\documentstyle[12pt,twoside]{report}
\newtheorem{defin}{Definition}[chapter]
\newtheorem{theorem}[defin]{Theorem}
\newtheorem{lemma}[defin]{Lemma}
\newtheorem{proposition}[defin]{Proposition}
\newtheorem{corollary}[defin]{Corollary}

\newcommand{\R}{{\rm l \! R}}

\newcommand{\C}{{\rm C \! \! \! l}}
\newcommand{\N}{{\rm l \! N}}

\title{Classical Representations of Quantum Mechanics Related to
Statistically Complete Observables}
\author{Werner Stulpe\\
\mbox{}\\
Aachen University of Applied Sciences, J\"ulich Campus\\
D-52428 J\"ulich, Germany\\
Email stulpe@fh-aachen.de\\
\mbox{}\\
\mbox{}\\
\mbox{}\\
\mbox{}\\
\mbox{}\\
\mbox{}\\
\mbox{}\\
\mbox{}\\
\fbox{\parbox{\linewidth}{\small This work was published as a book 1997
by Wissenschaft und Technik Verlag Berlin (Dresdener Str.\ 26,
D-10999 Berlin, Germany; email info@wt-verlag.de), ISBN 3-89685-438-0.}}}
\date{}

\begin{document}
\maketitle

\tableofcontents
\mbox{}\\
{\bf References \hfill 113}

\chapter{Introduction}

The problem of reformulating conventional Hilbert-space quantum mechanics
in terms of the classical phase space, i.e.\ the attempt to represent quantum
states and observables by probability densities and functions on phase space,
is almost as old as quantum mechanics itself and has been discussed by
many authors. In fact, the investigations on this subject originated with
E.~P.~Wigner's famous paper from 1932 and were continued by, for instance,
J.~E.~Moyal (1949), J. C. T. Pool (1966), L. Cohen (1966), E. P. Wigner
himself (1971), M. D. Srinivas and E. Wolf (1975), S. T. Ali and
E. Prugove\v{c}ki (1977a), F. E. Schroeck, Jr.\ (1982b), and W. Guz (1984). In
this paper, we present a reformulation of quantum mechanics in terms of
probability measures and functions on a general classical sample space
and in particular in terms of probability densities and functions on
phase space. The basis of our proceeding is the existence of so-called
statistically complete observables and the duality between the state spaces
and the spaces of the observables, the latter holding in the quantum
as well as in the classical case.

In order to give a preliminary impression of our subject as well as to review
some well-known facts, we tentatively define, for simplicity for spinless
particles with one-dimensional configuration space, a {\it classical
representation of quantum mechanics on phase space} to be a map
$ W \mapsto \rho_W $ that assigns to each density operator $W$
a probability density $ (q,p) \mapsto \rho_W (q,p) \geq 0 $ on phase space
and satisfies---as far as possible---the following postulates:
\begin{enumerate}
\item[{\rm (i)}] $ W \mapsto \rho_W $ is affine
\item[{\rm (ii)}] $ W \mapsto \rho_W $ is injective
\item[{\rm (iii)}] there exists a (possibly nonlinear) map $ A \mapsto f_A $
assigning to each (possibly unbounded) self-adjoint operator $A$ in Hilbert
space a real-valued measurable function $ f_A $ on phase space such that
for all density operators $W$ and all $A$ the quantum mechanical expectation
values can be represented according to
\begin{equation}
{\rm tr} \, WA = \int \rho_W(q,p) f_A(q,p) dqdp
\end{equation}
\item[{\rm (iv)}] the marginal densities of the probability densities
$ \rho_W $ coincide with the usual probability densities for position and
momentum, respectively; that is,
\begin{eqnarray*}
\langle q|Wq \rangle & = & \int \rho_W(q,p) dp   \\
\langle p|Wp \rangle & = & \int \rho_W(q,p) dq
\end{eqnarray*}
holds for all density operators $W$ or, equivalently,
\begin{eqnarray}
{\rm tr} \, WE^Q(b) & = & \int_b \left( \int \rho_W(q,p) dp \right) dq   \\
{\rm tr} \, WE^P(b) & = & \int_b \left( \int \rho_W(q,p) dq \right) dp
\end{eqnarray}
for all $W$ and all Borel subsets $b$ of $\R$ where $ E^Q $ and $ E^P $
are the spectral measures of the position operator $Q$, respectively,
the momentum operator $P$.
\end{enumerate}
We emphasize that we understand the densities $ \rho_W $ to be real
probability densities. In Sections 7.1 and 7.3 we shall show that there exist
maps $ W \mapsto \rho_W $ satisfying postulates (i) and (ii) and essentially
also (iii). In fact, from (i) and (ii) it follows the existence of
an assignment $ A \mapsto f_A $ such that Eq.~(1.1) holds exactly
for a large class of bounded self-adjoint operators $A$ and in
arbitrarily good physical approximation for all bounded self-adjoint
operators (cf.\ Theorem 7.5); we call such maps $ A \mapsto f_A $
{\it dequantizations}. Moreover, in some cases the dequantizations can
be extended to some unbounded self-adjoint operators (cf.\ Section 7.3).

We notice that postulate (iv) is not implied by (iii). Namely, if Eq.\ (1.1)
is valid for $ A = E^Q(b) $, respectively, $ A = E^P(b) $, then Eqs.\ (1.2)
and (1.3) contain the additional statement that
$ f_{E^Q(b)} = \chi_{b \times \R} $, respectively,
$ f_{E^P(b)} = \chi_{\R \times b} $ where $ \chi_B $ denotes
the characteristic function of a Borel set $ B \subseteq \R^2 $. According to
a theorem due to E. P. Wigner (1971), postulate (iv) cannot be satisfied;
we give a proof of that theorem below. However, according to
our reflections in Chapter 5, (iv) can be fulfilled approximately;
the better the approximation for (1.2) is, the worse the approximation
for (1.3), and conversely.

Summarizing, we shall show the existence of maps $ W \mapsto \rho_W $ that
satisfy postulates (i) and (ii), (iii) essentially, and (iv) approximately. In
contrast, the map $ W \mapsto \rho^W_W $ assigning to each density operator
the {\it Wigner function} $ \rho^W_W $ (the upper index is to indicate the
name ``Wigner'') fulfils all postulates (i) -- (iv); more precisely,
$ W \mapsto \rho^W_W $ satisfies (i), (ii), and (iv) exactly and (iii)
essentially in an analogous sense as before $ W \mapsto \rho_W $ where,
for $ W \mapsto \rho^W_W $, the map $ A \mapsto f_A $ is related to
the {\it Weyl correspondence} (cf.\ J. C. T. Pool, 1966; M. Hillery,
R. F. O'Connell, M. O. Scully, and E. P. Wigner, 1984). However, as is
well known, $ \rho^W_W $ is in general not nonnegative, i.e., $ \rho^W_W $
is in general only a pseudo-probability density (cf.\ R. L. Hudson, 1974).

We now prove Wigner's theorem stating that a map $ W \mapsto \rho_W $
cannot satisfy postulates (i) and (iv) if all $ \rho_W $ are real
probability densities. Instead of repeating E. P. Wigner's argumentation
from his paper from 1971, we proceed on our conceptual lines of
positive-operator-valued (POV-) measures etc. Assume there exists
an affine map $ W \mapsto \rho^W_W \geq 0 $ satisfying Eqs.\ (1.2)
and (1.3). It follows that, for each Borel set $ B \subseteq \R^2 $,
$ W \mapsto l_B(W) := \int_B \rho_W(q,p) dqdp $ is an affine functional
fulfilling $ 0 \leq l_B(W) \leq 1 $. This implies that $ l_B $ can
be represented by a bounded self-adjoint operator $ F(B) $,
$ 0 \leq F(B) \leq 1 $, according to $ l_B(W) = {\rm tr} \, WF(B) $. Hence,
\begin{equation}
{\rm tr} \, WF(B) = \int_B \rho_W(q,p) dqdp
\end{equation}
where $ B \mapsto F(B) $ is a normalized POV-measure. From (1.2), (1.3),
and (1.4) we obtain
\begin{eqnarray*}
{\rm tr} \, WE^Q(b) & = & {\rm tr} \, WF(b \times \R)     \\
{\rm tr} \, WE^P(b) & = & {\rm tr} \, WF(\R \times b)
\end{eqnarray*}
for all $W$ and all Borel sets $ b \subseteq \R $, respectively,
\begin{equation}
\begin{array}{ccc}
E^Q(b) & = & F(b \times \R)     \vspace{2mm}\\
E^P(b) & = & F(\R \times b)
\end{array}
\end{equation}
for all $b$. Thus, the projection-valued (PV-) measures $ E^Q $ and $ E^P $
appear as the marginal measures of the POV-measure $F$. As will be shown
immediately, from this it follows that $ [E^Q(b_1),E^P(b_2)] = 0 $ for all
Borel sets $ b_1,b_2 \subseteq \R $; i.e., the operators $Q$ and $P$ commute,
which is a contradiction. Hence, Wigner's theorem holds.

To show that the validity of Eqs.\ (1.5) would imply that all projections
$ E^Q(b) $ commute with all projections $ E^Q(b) $, we first prove the
following preparatory statement. If, for an orthogonal projection $E$ and
a bounded self-adjoint operator $ A \geq 0 $, $ A \leq E $ holds, then
$ A = EAE $ is valid. In fact, from $ 0 \leq A \leq E $ we obtain that,
for a vector $ \phi $ of Hilbert space, $ E\phi = 0 $ implies
$ A\phi = 0 $. In consequence, $ A(1 - E)\psi = 0 $ for all $ \psi $
since $ E(1 - E)\psi = 0 $. Hence, $ A(1 - E) = 0 $ or, equivalently,
$ A = EAE $.---Now assume that $F$ is a POV-measure on some measurable space
$ (M,\Xi) $ such that, for some set $ B_0 \in \Xi $, the operator $ F(B_0) $
is a projection. Then every positive operator $ F(B) $, $ B \in \Xi $,
commutes with $ F(B_0) $, as we are going to prove. We notice that
\begin{equation}
F(B) = F(B \setminus (B \cap B_0)) + F(B \cap B_0)
\end{equation}
and
\begin{equation}
F(B \setminus (B \cap B_0)) + F(B_0) = F(B \cup B_0) \leq F(M) = 1
\end{equation}
hold. From (1.7) it follows that
$ F(B \setminus (B \cap B_0)) \leq 1 - F(B_0) $; furthermore,
$ F(B \cap B_0) \leq F(B_0) $ is valid. The latter two inequalities and
the fact that $ F(B_0) $ is a projection imply
\begin{equation}
\begin{array}{rcl}
F(B \setminus (B \cap B_0)) & = & (1 - F(B_0)) \,
             F(B \setminus (B \cap B_0)) \, (1 - F(B_0))   \vspace{3mm}\\
F(B \cap B_0) & = & F(B_0) F(B \cap B_0) F(B_0) \ .
\end{array}
\end{equation}
Inserting (1.8) into (1.6) and multiplying with $ F(B_0) $ from the right,
respectively, from the left, we obtain
\[
F(B) F(B_0) = F(B_0) F(B \cap B_0) F(B_0) = F(B_0) F(B) \ .
\]
Hence, all $ F(B) $ commute with $ F(B_0) $, which finishes the proof of
Wigner's theorem.

In Chapter 2, we introduce the concept of statistical completeness of
observables and show the existence of such observables. In a very general
context, classical representations of quantum mechanics are introduced
in Chapter 3 and their relation to statistical complete observables is
discussed. A classical representation and a corresponding dequantization
enable the reformulation of the statistical scheme of quantum mechanics
on a classical sample space in the sense of Eq.\ (1.1). Moreover,
quantum dynamics can be reformulated on that sample space, as is shown
in Section 3.3. In Chapter 4 we discuss continuous resolutions
of the identity of Hilbert space. Such continuous resolutions,
in particular in the context of irreducible group representations, enable,
on the one hand, the introduction of physically interesting, possibly
statistically complete observables and, on the other hand, the construction
of Hilbert spaces consisting entirely of continuous square-integrable
functions; the latter gives a new aspect being interesting both from the
mathematical and the physical point of view.

In Chapter 5 a special class of continuous resolutions of the identity
is presented that gives rise to observables on phase space
that can be interpreted to describe joint position-momentum
measurements. Furthermore, such a continuous resolution on $ \R^{2N} $
gives rise to a representation of quantum mechanics on a Hilbert space of
continuous or even infinitely differentiable wave functions on the phase space
$ \R^{2N} $; we call such representations {\it phase-space representations of
quantum mechanics} and discuss them in Chapter 6. Some particular phase-space
representations are related to a well-known Hilbert space of entire functions,
which is pointed out in Section 6.4. Our phase-space representations are
conceptually different from our {\it classical representations of quantum
mechanics on phase space}. The latter ones are the subject of Chapter 7, they
are induced by statistically complete joint position-momentum observables and
concern the reformulation of the statistical scheme of quantum mechanics on
phase space in the sense of Eq.\ (1.1). Chapter 7 specifies the results of
Chapter 3 for the case that the phase space takes the role of the general
sample space; in particular, the reformulation of quantum dynamics on phase
space is presented in Section 7.4.

Chapters 2--4 have a general character whereas the more specific topics
of Chapters 5--7 are somehow related to the phase space. A further
specification is given by the consideration of spin systems (see
E. Prugove\v{c}ki, 1977b; F. E. Schroeck, Jr., 1982a; P. Busch 1986). In
this case the underlying Hilbert space is finite-dimensional, and the density
operators can classically be represented by probability measures on a finite
sample space, i.e.\ by finite-dimensional probability vectors (see P. Busch,
K.-E. Hellwig, and W.~Stulpe, 1993). For reasons of limitation, we do not
discuss this interesting specification in this paper.

We emphasize that our investigations on classical reformulations of quantum
mechanics are mainly motivated by our interest in structural aspects of
statistical physical theories, in particular, in insights into the relation
between quantum mechanics and classical probability theory, respectively,
classical statistical mechanics. Two important insights of this kind are
the following. For quantum mechanics based on a separable Hilbert space,
classical representations, i.e.\ injective affine maps from the density
operators into the probability measures on some sample space, always exist;
however, in this way the set of all density operators cannot be mapped onto
the set of all probability measures on that sample space. Physically, these
results can be interpreted in terms of hidden variables. Namely, the points
of the sample space are the hidden variables, and nature forbids that all
classically possible states can be prepared for microsystems; only those
states occur that are given as images of density operators under some
distinguished classical representation. A similar interpretation is possible
for the conceptually completely different framework of S. Gudder (1984, 1985,
1988) and I. Pitowsky (1983, 1989).

Besides the structural insights and the possible hidden-variables
interpretation, there are further consequences and several applications
of the topics presented here, as we indicate briefly. The existence of joint
position-momentum observables concerns fundamental aspects of the measurement
in quantum mechanics (cf.\ P. Busch and P. J. Lahti, 1984; P. Busch, 1985;
S.~Gudder, J. Hagler, and W. Stulpe, 1988; P. Busch, M. Grabowski, and
P.~J.~Lahti, 1995) and is a link to large quantum systems (cf.\ G. Ludwig,
1987); the existence of statistically complete observables has a fundamental
meaning for the determination of quantum states (cf.\ P. Busch and
P.~J.~Lahti, 1989) and perhaps for quantum information theory
(cf.\ E. Prugove\v{c}ki, 1977a; K.-E. Hellwig, 1993). Joint position-momentum
observables of many-particles systems have been proved to be a useful concept
in quantum statistical mechanics and especially for a derivation of the
Boltzmann equation (see E. Prugove\v{c}ki, 1984; L. Lanz, O. Melsheimer,
and E. Wacker, 1985; G. Ludwig, 1987). Moreover, the joint position-momentum
observables of one-particle systems have relativistic analogs (see
E. Prugove\v{c}ki, 1984; S.~T.~Ali, 1985); those observables give rise to
some kind of covariant relativistic position observables, for instance, and
are thus of fundamental importance in the context of the localization of
particles in relativistic quantum mechanics. Finally, phase-space formulations
of quantum mechanics have various applications in quantum optics.

It may be that our classical representations are, because of Eq.\ (1.1)
or some analog of it, also useful for calculational purposes. However,
our impression is that for calculational purposes the Wigner functions
are more suitable, whereas the classical representations on phase space
as presented here are more meaningful for fundamental interpretational
questions.

\chapter{Observables and Statistical Completeness}

Hilbert-space quantum mechanics is based on a complex separable Hilbert space
$ {\cal H} \neq \{0\} $. We denote the real Banach space of all bounded
self-adjoint operators in $ \cal H $ by $ {\cal B}_s({\cal H}) $ and
the Banach space of all self-adjoint trace-class operators by
$ {\cal T}_s({\cal H}) $. As is well known, $ {\cal B}_s({\cal H}) $
can be considered as the dual space $ ({\cal T}_s({\cal H}))' $ where
the duality is given by the trace functional. Let
$ K({\cal H}) \subset {\cal T}_s({\cal H}) $ be the convex set of all
positive trace-class operators $W$ with $ {\rm tr} \, W = 1 $, i.e.\ the
set of all density operators, and let $ [0,1] \subset {\cal B}_s({\cal H}) $
be the convex set of all bounded self-adjoint operators
$A$ satisfying $ 0 \leq A \leq 1 $. The density operators describe the
{\it statistical ensembles} of a sort of microsystems which we briefly call
{\it states}. The elements of $ [0,1] $ describe the {\it effects},
i.e.\ the classes of statistically equivalent realistic measurements with the
outcomes 0 and 1. For $ W \in K({\cal H}) $ and $ A \in [0,1] $, the
number $ {\rm tr} \, WA \in [0,1] $ is interpreted to be the
{\it probability for the outcome $1$ of the effect $A$ in the state $W$}.

The real Banach space $ {\cal B}_s({\cal H}) $ can be equipped with the
weak topology $ \sigma ({\cal B}_s({\cal H}),{\cal T}_s({\cal H})) $
which is the coarsest topology such that all linear functionals given by
the elements of $ {\cal T}_s({\cal H}) $ are continuous. We call this
topology briefly the {\it $ \sigma $-topology}. Since
$ \sigma ({\cal B}_s({\cal H}),{\cal T}_s({\cal H}))
= \sigma ({\cal B}_s({\cal H}),K({\cal H})) $ holds,
a neighborhood base of $ A \in {\cal B}_s({\cal H}) $
is given by the sets
\begin{equation}
U(A;W_1 ,\ldots,W_n ;\epsilon) := \{ \tilde{A} \in {\cal B}_s({\cal H})
\, | \, | {\rm tr} \, W_i \tilde{A} - {\rm tr} \, W_i A | < \epsilon
\ {\rm for} \ i= 1,\ldots,n \}
\end{equation}
where $ \epsilon > 0 $ and $ W_i \in K({\cal H}) $. An effect
$ A \in [0,1] $ is physically approximated by
$ \tilde{A} \in [0,1] $ if in many (but finitely many) states
$ W_1 ,\ldots,W_n \in K({\cal H}) $ the probabilities
$ {\rm tr} \, W_i \tilde {A} $ differ from $ {\rm tr} \, W_i A $
by an amount less than a small $ \epsilon > 0 $. This statement can be
tested experimentally and can be characterized mathematically by
$ \tilde {A} \in U(A;W_1 ,\ldots,W_n ;\epsilon) $. Hence, the
$ \sigma $-topology, respectively, its restriction to $ [0,1] $
describes the physical approximation of effects (cf.\ G.~Ludwig, 1970, 1983,
1985; R. Werner, 1983; R. Haag and D. Kastler, 1964).

An {\it observable} $F$ on some measurable space $(M, \Xi )$
is a normalized effect-valued measure on $\Xi$, i.e.\ a map
$ F \!\! : \Xi \rightarrow [0,1] $
satisfying $ F(\emptyset) = 0 $, $ F(M) = 1 $, and
$ F(\bigcup_{i=1}^{\infty} B_i) = \sum_{i=1}^{\infty} F(B_i) $ where
the sets $ B_i \in \Xi $ are mutually disjoint and the sum converges in
the $ \sigma $-topology, for instance. Thus, observables are normalized
{\it positive-operator-valued measures (POV-measures)}, whereas the
more common {\it projection-valued measures (PV-measures)} are special
cases. A state $ W \in K({\cal H}) $ and an observable $F$ define
a probability measure $ P^F_W $ on $ (M, \Xi ) $ by
\begin{equation}
P^{F}_{W} (B) := {\rm tr} \, WF(B) \ .
\end{equation}
We call $ P^{F}_{W} $ the {\it probability distribution of $F$ in the
state $W$}. Now let $F$ be an observable with real measuring values, i.e.,
let $ (M,\Xi) = (\R,\Xi (\R )) $ where $ \Xi (\R ) $ denotes the
$ \sigma $-algebra of Borel sets of $ \R $. The
{\it expectation value of $F$ in the state} $W$ is defined by
\begin{equation}
\langle F \rangle_{W} := \int \xi P^{F}_{W} (d\xi)
= \int {\rm id}_{\R} dP^{F}_{W} \ ,
\end{equation}
provided that the integral exists. If $ \langle F \rangle_W $ exists, then
the {\it variance of $F$ in the state $W$} is given by
\begin{equation}
{\rm var}_W F := \int (\xi - \langle F \rangle_W)^2 P^F_W(d\xi)
               = \int \xi^2 P^F_W(d\xi) - \langle F \rangle^2_W \ ;
\end{equation}
either $ {\rm var}_W F $ exists also or it is infinite. The existence of
$ {\rm var}_W F $ requires the existence of $ \langle F \rangle_W $ and
implies the existence of $ \int \xi^2 P^F_W(d\xi) $; conversely, if
$ {\rm id}^{2}_{\R} $ is $ P^F_W $-integrable, then $ \langle F \rangle_W $
as well as $ {\rm var}_W F $ exist.

Let $F$ be an observable on an arbitrary measurable space $ (M,\Xi) $ and
$ f \!\! : M \rightarrow \R $ a $ \Xi $-measurable function. If $f$ is
$ P^F_W $-integrable for all $ W \in K({\cal H}) $, then the integral
$ \int fdF $ exists, as a bounded self-adjoint operator, in the \linebreak
{\it $ \sigma $-weak} sense, i.e.,
\[
\int f dP^F_W = \int f d({\rm tr} \, WF(\, .\,))
              = {\rm tr} \left( W\int fdF \right)
\]
holds for all $ W \in K({\cal H}) $ (W. Stulpe, 1986, 1988). Equivalently,
\[
\int f d({\rm tr} \, VF(\, .\,)) = {\rm tr} \left( V\int fdF \right)
\]
is valid for all $ V \in {\cal B}_s({\cal H}) $. The existence of $ \int fdF $
in the {\it weak} sense is necessary and sufficient for the existence of
$ \int fdF $ in the $ \sigma $-weak sense. If $F$ is a PV-measure, then
$ \int fdF $ exists in the $ \sigma $-weak sense if and only if $f$ is
$F$-a.e.\ bounded, i.e.\ bounded almost everywhere with respect to $F$.

Now, let $F$ again be an observable on $ (\R,\Xi(\R)) $ and consider
the expectation value of $F$ in some state $W$. If $ {\rm id}_{\R} $
is $ P^F_W $-integrable even for all $ W \in K({\cal H}) $, then
$ \int {\rm id}_{\R} dF =: A \in {\cal B}_s({\cal H}) $ exists
in the $ \sigma $-weak sense, and we obtain
\begin{equation}
\langle F \rangle_{W} = \int {\rm id}_{\R} d({\rm tr} \, WF(\, .\,))
                      = {\rm tr} \left( W \int {\rm id}_{\R} dF \right) \ ,
\end{equation}
respectively,
\begin{equation}
\langle F \rangle_{W} = {\rm tr} \, WA \ .
\end{equation}
If, in addition, $F$ is a PV-measure on $ (\R,\Xi(\R)) $, i.e., if $F$ is
the spectral measure of some self-adjoint operator $ \tilde{A} $, then
$ \int {\rm id}_{\R} dF $ exists in the $ \sigma $-weak sense
if and only if $ \tilde{A} $ is bounded; in this case
$ A = \int {\rm id}_{\R} dF = \tilde{A} $ holds (for some further
discussion on $ \langle F \rangle_W $ and $ {\rm var}_W F $ when $F$ is
a spectral measure, see Section 7.2 and Eq.\ (7.31)).---In particular, for
any $ W \in K({\cal H}) $ and any $ A \in {\cal B}_s({\cal H}) $, one can
interpret the real number $ {\rm tr} \, WA $ as the expectation value
of some observable. According to (2.1) and (2.6), the $ \sigma $-topology
then describes the physical approximation of observables.

Following G. Ludwig (1970, 1983), we call a state $ W \in K({\cal H}) $
{\it effective} if $ {\rm tr} \, WA = 0 $ and $ A \in [0,1] $ imply
$ A = 0 $. Given an observable $F$ on $ (M,\Xi) $, the state $W$ is called
{\it effective with respect to $F$} if $ {\rm tr} \, WF(B) = 0 $ with
$ B \in \Xi $ implies $ F(B) = 0 $. To prove the existence of an effective
state, observe that, as a consequence of the separability of the Hilbert
space $ \cal H $, the space $ {\cal T}_s({\cal H}) $ as well as the set
$ K({\cal H}) $ are separable with respect to the trace norm. Let
$ \{ W_i \}_{i \in \N} $ be a dense sequence in $ K({\cal H}) $ and
$ \alpha_i > 0 $ with $ \sum_{i=1}^{\infty} \alpha_i = 1 $. Define
$ W := \sum_{i=1}^{\infty} \alpha_i W_i $, then $ W \in K({\cal H}) $ is
an effective state. Alternatively, choose a complete orthonormal system
$ \{ \phi_i \}_{i \in \N} $ in $ \cal H $ and define
$ \tilde{W} := \sum_{i=1}^{\infty} \alpha_i P_{\phi_i} $ where
$ \alpha_i > 0 $, $ \sum_{i=1}^{\infty} \alpha_i = 1 $, and
$ P_{\phi_i} := |\phi_i\rangle \langle\phi_i| $;
$ \tilde{W} \in K({\cal H}) $ is also effective.

For every observable $F$ on $ (M,\Xi) $ there exists a measure $ \lambda $
on $ \Xi $ such that for every $ W \in K({\cal H}) $ the probability measure
$ P^F_W $ is absolutely continuous with respect to $ \lambda $ (G. Ludwig,
1970). That is, $ P^F_W $ can be characterized by a probability density
$ \rho_W $ on the measure space $ (M,\Xi,\lambda) $ such that
\begin{equation}
P^F_W(B) = \int_B \rho_W d\lambda = \int_B \rho_W(x) \lambda(dx)
\end{equation}
holds for all $ B \in \Xi $. The measure $ \lambda $ can even be choosen as
a probability measure. In fact, let $ W_0 \in K({\cal H}) $ be effective
with respect to $F$ and define $ \lambda := P^{F}_{W_0} $. Since
$ \lambda(B) = {\rm tr} \, W_0F(B) = 0 $ implies $ F(B) = 0 $ and hence
$ P^F_W(B) = {\rm tr} \, WF(B) = 0 $ for every $ W \in K({\cal H}) $, every
$ P^F_W $ is absolutely continuous with respect to $ \lambda $.

We call a family $ \{ F_{\alpha} \}_{\alpha \in I} $ of observables on
$ (M_{\alpha},\Xi_{\alpha}) $ {\it statistically complete} if every state
is determined by the probability distributions (2.2) of all $ F_{\alpha} $,
i.e., if for any two states $ W_1,W_2 \in K({\cal H}) $,
$ P^{F_{\alpha}}_{W_1} = P^{F_{\alpha}}_{W_2} $ for all $ \alpha \in I $
implies $ W_1 = W_2 $. The concept of statistical completeness was introduced
by S.~T.~Ali and E. Prugove\v{c}ki (1977a,b; see also E. Prugove\v{c}ki,
1977a) who called it {\it informational completeness}. Next we present
a criterion for statistical completeness, and then we prove the remarkable
fact that there exist single statistically complete observables.

\begin{lemma}
A family $\{F_{\alpha} \}_{\alpha \in I}$ of observables on
$ (M_{\alpha},\Xi_{\alpha}) $ is statistically complete if
and only if the linear hull of the set
$\bigcup_{\alpha \in I} F_{\alpha}(\Xi_{\alpha})
    =  \{F_{\alpha}(B) \, | \, B \in \Xi_{\alpha} , \ \alpha \in I \}$
is $ \sigma $-dense in $ {\cal B}_s({\cal H}) $.
\end{lemma}
{\bf Proof}: Suppose $\{F_{\alpha} \}_{\alpha \in I}$ is statistically
complete, i.e., for any $ W_1,W_2 \in K({\cal H}) $,
\[
{\rm tr} \, W_1 A = {\rm tr} \, W_2 A
\]
for all $ A \in \bigcup_{\alpha \in I} F_{\alpha}(\Xi_{\alpha})$ implies
$ W_1 = W_2 $. Let $ V_1 $ and $ V_2 $ be arbitrary positive trace-class
operators and assume
\begin{equation}
{\rm tr} \, V_1 A = {\rm tr} \, V_2 A
\end{equation}
for all $ A \in \bigcup_{\alpha \in I} F_{\alpha}(\Xi_{\alpha})$. Setting
$ A = F_{\alpha} (M_{\alpha}) = 1 $, it follows that
$ {\rm tr} \, V_1 = {\rm tr} \, V_2 =: \gamma $. For $ \gamma = 0 $, we
obtain $ V_1 = V_2 = 0 $. For $ \gamma \neq 0 $,
divide (2.8) by $ \gamma $ and observe that $ \frac{1}{\gamma} V_1 $ and
$ \frac{1}{\gamma} V_2 $ are density operators. Consequently, $ V_1 = V_2 $
holds. Now, let $ V_1 ,V_2 \in {\cal T}_s({\cal H}) $ be arbitrary
and assume again the validity of (2.8) for all
$ A \in \bigcup_{\alpha \in I} F_{\alpha}(\Xi_{\alpha})$. Decomposing
$ V_1 $ and $ V_2 $ into positive operators, we obtain
\[
{\rm tr} \, (V^{+}_{1} - V^{-}_{1}) A = {\rm tr} \, (V^{+}_{2} - V^{-}_{2}) A
\]
or, equivalently,
\[
{\rm tr} \, (V^{+}_{1} + V^{-}_{2}) A
                               = {\rm tr} \, (V^{+}_{2} + V^{-}_{1}) A \ .
\]
It follows $ V^{+}_{1} + V^{-}_{2} = V^{+}_{2} + V^{-}_{1} $, respectively,
$ V_1 = V_2 $. Hence, the set
$ \bigcup_{\alpha \in I} F_{\alpha}(\Xi_{\alpha})$ separates
the elements of $ {\cal T}_s({\cal H}) $.

If the linear hull of $\bigcup_{\alpha \in I} F_{\alpha}(\Xi_{\alpha})$
were not $ \sigma $-dense in $ {\cal B}_s({\cal H}) $, then, according
to a well-known consequence of the Hahn--Banach theorem, there would
exist a $ \sigma $-continuous linear functional $ \Lambda  \neq 0 $ on
$ {\cal B}_s({\cal H}) $ such that $ \Lambda (A) = 0 $ for all
$ A \in \overline{{\rm lin} \bigcup_{\alpha \in I} F_{\alpha}(\Xi_{\alpha})}^
{\sigma} $. Since the $ \sigma $-continuous linear
functionals on $ {\cal B}_s({\cal H}) $ are just those ones that are
represented by the elements of $ {\cal T}_s({\cal H}) $,
\[
\Lambda (A) = {\rm tr} \, VA = 0
\]
would hold for some $ V \in {\cal T}_s({\cal H}) $, $ V \neq 0 $, and
all $ A \in \overline{{\rm lin} \bigcup_{\alpha \in I}
F_{\alpha}(\Xi_{\alpha})}^{\sigma} $. Hence,
$ \bigcup_{\alpha \in I} F_{\alpha}(\Xi_{\alpha}) $ would not
separate $ {\cal B}_s({\cal H}) $.

Conversely, suppose
$ \overline{{\rm lin} \bigcup_{\alpha \in I} F_{\alpha}(\Xi_{\alpha})}^
{\sigma} = {\cal B}_s({\cal H}) $. Let $ V_1 , V_2 \in {\cal T}_s({\cal H}) $
and assume Eq.\ (2.8) holds for all
$ A \in \bigcup_{\alpha \in I} F_{\alpha}(\Xi_{\alpha}) $. Considering
$ V_1 $ and $ V_2 $ as $ \sigma $-continuous linear functionals on
$ {\cal B}_s({\cal H}) $, one obtains the validity of (2.8) for all
$ A \in {\cal B}_s({\cal H}) $. In consequence, $ V_1 = V_2 $. Hence, the set
$ \bigcup_{\alpha \in I} F_{\alpha}(\Xi_{\alpha}) $ separates
$ {\cal T}_s({\cal H}) $, and the family $ \{ F_{\alpha} \}_{\alpha \in I} $
of observables is statistically complete. \hspace*{\fill} $ \Box $\\

\begin{theorem}
There exists a single statistically complete observable.
\end{theorem}
{\bf Proof:}  According to the Banach--Alaoglu theorem, the closed unit ball
$ D := \{ A \in {\cal B}_s({\cal H}) \, | \, \| A \| \leq 1 \} $ of
$ {\cal B}_s({\cal H}) = ({\cal T}_s({\cal H}))' $ is $ \sigma $-compact
(i.e.\ compact with respect to the $ \sigma $-topology). Furthermore,
because of the norm-separability of $ {\cal T}_s({\cal H}) $ which
is a consequence of the norm-separability of $ \cal H $, the
$ \sigma $-topology restricted to $D$ is metrizable (see, e.g.,
N. Dunford and J. T. Schwartz, 1958). Hence, the metrizable compact space
$ (D,\sigma \cap D) $ is separable. Likewise, the interval
$ [0,1] := \{ A \in {\cal B}_s({\cal H}) \, | \, 0 \leq A \leq 1 \} \subset D $
is $ \sigma $-separable. \vspace{1pt}

Let $ \{ \tilde{A}_{n} \}_{n \in \N} $ be a $ \sigma $-dense
sequence in $ [0,1] $ and define a further sequence by
\begin{eqnarray*}
A_{1} &:=& 1 -  \sum_{i=1}^{\infty} \frac{1}{2^{i}} \tilde{A}_{i} \\
A_{n} &:=& \frac{1}{2^{n-1}} \tilde{A}_{n-1} \ \ {\rm for} \ \ n \geq 2 \ .
\end{eqnarray*}
Observe that (i) the infinite sum is even norm-convergent,
(ii) $ A_n \in [0,1] $, \vspace{1pt}
(iii) $ \overline{{\rm lin} \{ A_{n} \, | \, n \in \N \}}^{\sigma}
= {\cal B}_s({\cal H}) $, and
(iv) $ \sum_{n=1}^{\infty} A_{n} = 1 $. Now define an observable $F$
on the power set of $\N$ by
\[
F(B) := \sum_{i \in B} A_{i}
\]
where $ B \subseteq \N $. Because of (iii) and the preceding lemma,
the observable  $F$ is statistically complete. \hfill $ \Box $\\

Lemma 2.1 and Theorem 2.2 were proved by M. Singer and W. Stulpe (1992)
within the more general context of statistical dualities; however, the
statement of Theorem 2.2 was already be concluded by G. Ludwig (1970). The
observable constructed in the proof of Theorem 2.2 is a discrete one,
another construction of a discrete statistically complete observable
within the framework of Hilbert-space quantum mechanics
was given by P. Busch and P.~J.~Lahti (1989). That construction
is perhaps more concrete than ours, however, our proof of
Theorem 2.2 shows that the existence of single statistically complete
quantum observables is possibly related to the norm-separability of the
state space. There exist also continuous statistically complete observables,
important examples of those will be discussed in Section~7.1.

The following proposition was proved by M. Singer and W. Stulpe (1992)
as follows, another proof had already been given by P. Busch and P. J. Lahti
(1989).

\begin{proposition}
Let $ \{ F_{\alpha} \}_{\alpha \in I} $ be a family of observables
on $ (M_{\alpha},\Xi_{\alpha}) $ such that
$ \bigcup_{\alpha \in I} F_{\alpha}(\Xi_{\alpha}) $ is a set of
commuting positive operators. Then $ \{ F_{\alpha} \}_{\alpha \in I} $
cannot be statistically complete, provided that $ \dim {\cal H} \geq 2 $. In
particular, one statistically complete observable cannot be a PV-measure.
\end{proposition}
{\bf Proof:}  Assume $ \{ F_{\alpha} \}_{\alpha \in I} $ is statistically
complete and the set $ \bigcup_{\alpha \in I} F_{\alpha}(\Xi_{\alpha}) $ is
commuting. Then it follows that
\begin{enumerate}
\item[{\rm (i)}] $ \bigcup_{\alpha \in I} F_{\alpha}(\Xi_{\alpha}) $ separates
the space $ {\cal T}({\cal H}) $ of all (not necessarily self- \linebreak
adjoint) trace-class operators in $ {\cal H} $ and, analogously to Lemma 2.1
and its proof, $ \overline{{\rm lin} \bigcup_{\alpha \in I}
F_{\alpha}(\Xi_{\alpha})}^{\sigma({\cal B}({\cal H}),{\cal T}({\cal H}))}
= {\cal B}({\cal H}) $ (here we consider the complex linear hull of
$ \bigcup_{\alpha \in I} F_{\alpha}(\Xi_{\alpha}) $;
$ \sigma({\cal B}({\cal H}),{\cal T}({\cal H})) $ is just
the ultraweak operator topology in $ {\cal B}({\cal H}) $)
\item[{\rm (ii)}] the $ \ast $-algebra
$ {\cal A} (\bigcup_{\alpha \in I} F_{\alpha}(\Xi_{\alpha})) $
generated by $ \bigcup_{\alpha \in I} F_{\alpha}(\Xi_{\alpha}) $
in $ {\cal B}({\cal H}) $ is Abelian.
\end{enumerate}
Statement (ii) implies that the von Neumann algebra
\[
{\cal A} := \overline{{\cal A}
\left( \bigcup_{\alpha \in I} F_{\alpha}(\Xi_{\alpha}) \right)}^
{\sigma({\cal B}({\cal H}),{\cal T}({\cal H}))}
\]
generated by $ \bigcup_{\alpha \in I} F_{\alpha}(\Xi_{\alpha}) $
is also Abelian, whereas from statement (i) we obtain
$ {\cal A} = {\cal B}({\cal H}) $. Hence, $ {\cal B}({\cal H}) $ is
Abelian, which is a contradiction if $ \dim {\cal H} \geq 2 $.
\hfill $\Box$\\

\chapter{The Representation of Quantum Mechanics on a Classical Sample Space}

\section{Classical Representations}

Most generally, classical statistical physics is based on a nontrivial
measurable space $ (\Omega,\Sigma) $, i.e.\ on a set $ \Omega \neq \emptyset $
and a $ \sigma $-algebra $ \Sigma $ in $ \Omega $. The set $ \Omega $ is
called a {\it sample space}, and the elements of $ \Sigma $ are called
{\it events}. We~denote the space of all bounded signed measures
on $ \Sigma $, i.e.\ the space of all $ \sigma $-additive real-valued
set functions on $ \Sigma $, by $ {\cal M}_{\R}(\Omega,\Sigma) $. By means of
$ \| \nu \| := |\nu|(\Omega) $ where $ |\nu| $ is the total variation of
$ \nu \in {\cal M}_{\R}(\Omega,\Sigma) $, $ {\cal M}_{\R}(\Omega,\Sigma) $
becomes a real Banach space. Let $ {\cal F}_{\R}(\Omega,\Sigma) $ be
the space of all real-valued bounded $ \Sigma $-measurable functions
on $ \Omega $. Defining $ \|f\| := \sup_{\omega \in \Omega} |f(\omega)| $
for $ f \in {\cal F}_{\R}(\Omega,\Sigma) $, $ {\cal F}_{\R}(\Omega,\Sigma) $
is also a real Banach space. By the integral, the spaces
$ {\cal M}_{\R}(\Omega,\Sigma) $ and $ {\cal F}_{\R}(\Omega,\Sigma) $
are placed in duality to each other; in particular, this means that
$ {\cal F}_{\R}(\Omega,\Sigma) $ can be considered as a closed subspace
of the dual space $ ({\cal M}_{\R}(\Omega,\Sigma))' $ and that
$ {\cal F}_{\R}(\Omega,\Sigma) $ separates the elements of
$ {\cal M}_{\R}(\Omega,\Sigma) $ (for more details, see W.~Stulpe, 1986;
M. Singer and W. Stulpe, 1992).

Let $ K(\Omega,\Sigma) \subset {\cal M}_{\R}(\Omega,\Sigma) $ be
the convex set of all probability measures on $ \Sigma $ and
$ [0,\chi_{\Omega}] \subset {\cal F}_{\R}(\Omega,\Sigma) $ the
convex set of all $ f \in {\cal F}_{\R}(\Omega,\Sigma) $ satisfying
$ 0 \leq f(\omega) \leq 1 $ for all $ \omega \in \Omega $. The probability
measures describe the {\it classical states}, i.e.\ the {\it classical
statistical ensembles}, and the elements of $ [0,\chi_{\Omega}] $
the {\it classical effects}. For $ \mu \in K(\Omega,\Sigma) $ and
$ f \in [0,\chi_{\Omega}] $, the number $ \int fd\mu \in [0,1] $ is
interpreted to be the {\it probability for the outcome $1$ of the effect $f$
in the state $ \mu $}. Particular effects are given by the characteristic
functions $ \chi_A $, $ A \in \Sigma $, these effects correspond to the
events. The probability for the outcome $1$ of an effect $ \chi_A $ in
the state $ \mu $ is $ \int \chi_A d\mu = \mu(A) $, $ \mu(A) $ is usually
interpreted as the {\it probability for the occurrence of the event $A$}.

Particular classical observables are given by the {\it random variables}
on $ \Omega $, i.e.\ by the $ \Sigma $-$ \Xi $-measurable maps
$ X \!\! : \Omega \rightarrow M $ where $ (M,\Xi) $ is a further
measurable space (for more details, see W. Stulpe, 1986; M. Singer and
W.~Stulpe, 1992). The {\it probability distribution of $X$ in the state
$ \mu $} is given by
\[
P^{X}_{\mu}(B) := \mu(X^{-1}(B)) =: (\mu \circ X^{-1})(B)
\]
where $ B \in \Xi $. Now let $X$ be a real-valued random variable, i.e.\
$ (M,\Xi) = (\R,\Xi(\R)) $. The {\it expectation value of $X$ in the state
$ \mu $} is
\begin{equation}
\langle X \rangle_{\mu} := \int \xi P^{X}_{\mu}(d\xi)
                         = \int {\rm id}_{\R} d(\mu \circ X^{-1})
                         = \int X d\mu
\end{equation}
(compare Eqs.\ (2.3), (2.5), (2.6), and (7.14)), and the {\it variance of $X$
in the state $ \mu $} is
\[
{\rm var}_{\mu} X := \int (\xi - \langle X \rangle_{\mu})^2 P^{X}_{\mu}(d\xi)
                   = \int \xi^2 P^{X}_{\mu}(d\xi) - \langle X \rangle^{2}_{\mu}
\]
(compare Eq.\ (2.4)), respectively,
\begin{equation}
{\rm var}_{\mu} X = \int (X - \langle X \rangle_{\mu})^2 d\mu
                  = \int X^2 d\mu - \langle X \rangle^{2}_{\mu}
                  = \langle X^2 \rangle_{\mu} - \langle X \rangle^{2}_{\mu}
\end{equation}
(compare Eq.\ (7.31)). Of course, we assume that at least one
of the integrals in (3.1) exists; $ {\rm var}_{\mu} X $ then either
exists also or is infinite. If $ {\rm id}^{2}_{\R} $ is
$ P^{X}_{\mu} $-integrable, respectively, if $ X^2 $ is $ \mu $-integrable,
then $ \langle X \rangle_{\mu} $ as well as $ {\rm var}_{\mu} X $
exist. Every function $ f \in {\cal F}_{\R}(\Omega,\Sigma) $ can be
interpreted as a bounded real-valued random variable, in this case
$ \langle f \rangle_{\mu} $ and $ {\rm var}_{\mu} f $ exist for all
$ \mu \in K(\Omega,\Sigma) $.

One often works only with those probability measures on $ \Sigma $ that are
absolutely continuous with respect to some distinguished positive measure,
respectively, with the corresponding probability densities. Accordingly, let
$ \lambda \neq 0 $ be a fixed $ \sigma $-finite (not necessarily finite)
positive measure on $ \Sigma $, i.e., $ (\Omega,\Sigma,\lambda) $ is
a nontrivial $ \sigma $-finite measure space. In classical statistical
mechanics, $ \lambda $ is the Lebesgue measure defined on the
$ \sigma $-algebra of Borel sets of the usual {\it phase space}. Denote
the set of all probability densities on $ (\Omega,\Sigma,\lambda) $, i.e.\
the set of all $ \rho \in L^1_{\R}(\Omega,\Sigma,\lambda) $ satisfying
$ \rho \geq 0 $ $ \lambda $-a.e.\ and $ \int \rho d\lambda = 1 $, by
$ K(\Omega,\Sigma,\lambda) $. The set $ K(\Omega,\Sigma,\lambda) $ is
a convex subset of the real Banach space $ L^1_{\R}(\Omega,\Sigma,\lambda) $,
the latter one can be considered as a closed subspace of
$ {\cal M}_{\R}(\Omega,\Sigma) $. Furthermore,
$ (L^1_{\R}(\Omega,\Sigma,\lambda))' = L^{\infty}_{\R}(\Omega,\Sigma,\lambda) $
holds. If $ \rho \in K(\Omega,\Sigma,\lambda) $,
$ f \in L^{\infty}_{\R}(\Omega,\Sigma,\lambda) $ with
$ 0 \leq f(\omega) \leq 1 $ for $ \lambda $-almost all $ \omega \in \Omega $,
and $ A \in \Sigma $, then $ \int \rho f d\lambda $ is again the probability
for the outcome $1$ of the effect $f$ in the state $ \rho $  and
$ \int \rho \chi_A d\lambda = \int_A \rho d\lambda $ the probability
for the occurrence of the event $A$. According to Eqs.\ (3.1) and (3.2),
the expectation value and the variance of a real-valued random variable $X$
in the state $ \rho $ are given by
\begin{equation}
\langle X \rangle_{\rho} := \langle X \rangle_{\mu}
                          = \int X d\mu = \int \rho X d\lambda
\end{equation}
and
\begin{eqnarray*}
{\rm var}_{\rho} X & := & {\rm var}_{\mu} X
  =   \int (X - \langle X \rangle_{\mu})^2 d\mu                       \\
& = & \int \rho (X - \langle X \rangle_{\rho})^2 d\lambda             \\
& = & \langle X^2 \rangle_{\rho} - \langle X \rangle^{2}_{\rho} \ ,
\end{eqnarray*}
respectively, where $ \mu $ is the probability measure corresponding to
the density $ \rho $. In particular, for bounded real-valued random variables
$ f \in L^{\infty}_{\R}(\Omega,\Sigma,\lambda) $, $ \langle f \rangle_{\rho} $
and $ {\rm var}_{\rho} f $ exist for all $ \rho \in K(\Omega,\Sigma,\lambda) $.

A {\it classical representation $ \tilde{T} $ of quantum mechanics} is
a map that assigns to every quantum state $ W \in K({\cal H}) $ injectively
a probability measure $ \mu \in K(\Omega,\Sigma) $. Since $ K({\cal H}) $
and $ K(\Omega,\Sigma) $ are convex sets, we furthermore assume that
$ \tilde{T} $ is affine. As one can prove easily, $ \tilde{T} $
can uniquely be extended to an injective positive linear map
$ T \!\! : {\cal T}_s({\cal H}) \rightarrow {\cal M}_{\R}(\Omega,\Sigma) $
with the property $ TK({\cal H}) \subseteq K(\Omega,\Sigma) $.

\begin{defin}
{\rm We call a linear map
$ T \!\! : {\cal T}_s({\cal H}) \rightarrow {\cal M}_{\R}(\Omega,\Sigma) $
a} classical representation of quantum mechanics on $ (\Omega,\Sigma) $
{\rm if}
\begin{enumerate}
\item[{\rm (i)}] $ TK({\cal H}) \subseteq K(\Omega,\Sigma) $
\item[{\rm (ii)}] $T$ {\rm is injective.}
\end{enumerate}
{\rm If $ (\Omega,\Sigma,\lambda) $ is a $ \sigma $-finite measure space,
then an injective linear map $ \hat{T} \!\! : {\cal T}_s({\cal H})
\rightarrow L^1_{\R}(\Omega,\Sigma,\lambda) $ with
$ \hat{T}K({\cal H}) \subseteq K(\Omega,\Sigma,\lambda) $ is called a}
classical representation of quantum mechanics on
{\rm $ (\Omega,\Sigma,\lambda) $.}
\end{defin}

Some simple properties of classical representations are stated in the next
lemma. In this context as well as in the following, we understand, for
simplicity, the dual map $T'$ of a bounded linear map
$ T \!\! : {\cal T}_s({\cal H}) \rightarrow {\cal M}_{\R}(\Omega,\Sigma) $
as the map
$ T' \!\! : {\cal F}_{\R}(\Omega,\Sigma) \rightarrow {\cal B}_s({\cal H}) $
that is the restriction of the Banach-space adjoint map of $T$ to
$ {\cal F}_{\R}(\Omega,\Sigma) $. For a bounded linear map $ \hat{T} \!\! :
{\cal T}_s({\cal H}) \rightarrow L^1_{\R}(\Omega,\Sigma,\lambda) $
where $ (\Omega,\Sigma,\lambda) $ is a $ \sigma $-finite measure space,
$ \hat{T}' $ is understood to be the Banach-space adjoint map
$ \hat{T}' \!\! : L^{\infty}_{\R}(\Omega,\Sigma,\lambda) \rightarrow
{\cal B}_s({\cal H}) $.

\begin{lemma}
A linear map
$ T \!\! : {\cal T}_s({\cal H}) \rightarrow {\cal M}_{\R}(\Omega,\Sigma) $
fulfilling $ TK({\cal H)} \subseteq K(\Omega,\Sigma) $ is positive
and bounded with $ \|T\| = 1 $. The property
$ TK({\cal H}) \subseteq K(\Omega,\Sigma) $ of a bounded linear map
$ T \!\! : {\cal T}_s({\cal H}) \rightarrow {\cal M}_{\R}(\Omega,\Sigma) $
is equivalent to $ T' \geq 0 $ and $ T'\chi_{\Omega} = 1 $. The latter
two conditions imply $ T'[0,\chi_{\Omega}] \subseteq [0,1] $. For linear maps
$ \hat{T} \!\! : {\cal T}_s({\cal H}) \rightarrow
L^1_{\R}(\Omega,\Sigma,\lambda) $, the analogous statements hold.
\end{lemma}
{\bf Proof:} Let
$ T \!\! : {\cal T}_s({\cal H}) \rightarrow {\cal M}_{\R}(\Omega,\Sigma) $
be linear with $ TK({\cal H}) \subseteq K(\Omega,\Sigma) $. Then $T$
is positive and $ \|T\| = \sup_{V \in K({\cal H})} \|TV\| = 1 $. The
map $T'$ is also positive; from $ TK({\cal H}) \subseteq K(\Omega,\Sigma) $
it follows for $ W \in K({\cal H}) $ that $ \|W\|_{\rm tr} = 1 = \|TW\|
= \int \chi_{\Omega} d(TW) = {\rm tr} \, W(T'\chi_{\Omega}) $, and
$ {\rm tr} \, W(T'\chi_{\Omega}) = 1 $ for all $ W \in K({\cal H}) $
implies $ T'\chi_{\Omega} = 1 $. The rest of the proof is clear.
\hfill $ \Box $\\

It is remarkable that injective affine maps from $ K({\cal H}) $ into
$ K(\Omega,\Sigma) $ do exist. In fact, the classical representations on
$ (\Omega,\Sigma) $ are in one-to-one correspondence with the statistically
complete observables on $ (\Omega,\Sigma) $.

\begin{theorem}
Every statistically complete observable $F$ on $ (M,\Xi) := (\Omega,\Sigma) $
defines a classical representation $T$ on $ (\Omega,\Sigma) $ by
\[
(TV)(B) := {\rm tr} \, VF(B)
\]
where $ V \in {\cal T}_s({\cal H}) $ and $ B \in \Sigma $. Conversely,
every classical representation
$ T \!\! : {\cal T}_s({\cal H}) \rightarrow {\cal M}_{\R}(\Omega,\Sigma) $
determines uniquely a statistically complete observable
$ F \!\! : \Sigma \rightarrow [0,1] $ such that
$ TV = {\rm tr} \, VF(\, . \,) $ holds. In particular, for
$ W \in K({\cal H}) $, $TW$ is just the probability distribution of $F$:
\[
TW = {\rm tr} \, WF(\, . \,) = P^F_W \ .
\]
Moreover, $T'$ and $F$ are related by
\begin{equation}
T'f = \int fdF \ ,
\end{equation}
respectively, by
\begin{equation}
T'\chi_{B} = F(B)
\end{equation}
where $ f \in {\cal F}_{\R}(\Omega,\Sigma) $, $ B \in \Sigma $, and the
integral is understood in the $ \sigma $-weak sense.
\end{theorem}
{\bf Proof:} Given an arbitrary observable
$ F \!\! : \Sigma \rightarrow [0,1] $, a linear map
$ T \!\! : {\cal T}_s({\cal H}) \rightarrow {\cal M}_{\R}(\Omega,\Sigma) $
with property (i) of Definition 3.1 can be defined by
$ (TV)(B) := {\rm tr} \, VF(B) $. If $F$ is statistically complete, then,
according to the proof of Lemma 2.1, its range $ F(\Sigma) $ separates
the elements of $ {\cal T}_s({\cal H}) $. Hence, $T$ is injective.

Now assume $T$ is a classical representation on $ (\Omega,\Sigma) $. Define
$ F(B) := T'\chi_{B} $. Then Lemma 3.2 implies $ F(B) \in [0,1] $ and
$ F(\Omega) = T'\chi_{\Omega} = 1 $, and from $ {\rm tr} \, VF(B)
= {\rm tr} \, V(T'\chi_{B}) = \int \chi_{B} d(TV) = (TV)(B) $ it follows
that $ TV = {\rm tr} \, VF(\, . \,) $. Next we show that $F$ is
$ \sigma $-additive. Taking a sequence of disjoint sets $ B_i \in \Sigma $,
we obtain
\begin{eqnarray*}
{\rm tr} \, VF \left( \bigcup_{i=1}^{\infty} B_i \right)
& = & (TV) \left( \bigcup_{i=1}^{\infty} B_i \right)
  = \sum_{i=1}^{\infty} (TV)(B_i) \\
& = & \lim_{n \rightarrow \infty} \sum_{i=1}^{n} (TV)(B_i)
  = \lim_{n \rightarrow \infty}
               {\rm tr} \left( V\sum_{i=1}^{n} F(B_i) \right)
\end{eqnarray*}
for all $ V \in {\cal T}_s({\cal H}) $. In consequence,
\[
F\left( \bigcup_{i=1}^{\infty} B_i \right)
= \sigma\mbox{--} \! \lim_{n \rightarrow \infty} \sum_{i=1}^{n} F(B_i)
= \sum_{i=1}^{\infty} F(B_i)
\]
holds. Hence, $F$ is an observable satisfying
$ TV = {\rm tr} \, VF(\, . \,) $. It is also statistically complete because
$T$ is injective.

Finally, for $ f \in {\cal F}_{\R}(\Omega,\Sigma) $ we have
\[
{\rm tr} \, V(T'F( \, . \, )) = \int fd(TV)
                              = \int f d({\rm tr} \, VF(\, . \,))
                              = {\rm tr} \left( V \int fdF \right) \ .
\]
Thus, $ T'f = \int fdF $. \hfill $ \Box $\\

Let $ \tau $ be the canonical embedding of $ L^1_{\R}(\Omega,\Sigma,\lambda) $
into $ {\cal M}_{\R}(\Omega,\Sigma) $, i.e.,
$ (\tau\rho)(B) := \int_{B} \rho d\lambda $ where
$ \rho \in L^1_{\R}(\Omega,\Sigma,\lambda) $ and $ B \in \Sigma $. The map
$ \tau $ and its inverse $ \tau^{-1} $ defined on
$ \tau L^1_{\R}(\Omega,\Sigma,\lambda) $ are linear, isometric, and
positive. Every classical representation $ \hat{T} $ on
$ (\Omega,\Sigma,\lambda) $ defines a classical representation $T$ on
$ (\Omega,\Sigma) $ by $ T := \tau \hat{T} $. Conversely, if $T$ is
a classical representation on $ (\Omega,\Sigma) $ such that all measures
$TV$ are absolutely continuous with respect to $ \lambda $, then a map
$ \hat{T} $ can be defined by $ \hat{T} := \tau^{-1}T $. According to
our reflections around Eq.\ (2.7), there exists always a $ \sigma $-finite
positive measure $ \lambda $ on $ \Sigma $ (even a probability measure)
and a classical representation $ \hat{T} $ on $ (\Omega,\Sigma,\lambda) $
such that $T$ can be written as $ T = \tau\hat{T} $. However, the general
construction of that $ \lambda $ seems to be somewhat artificial; in
Chapter 7 we shall discuss a class of physically important examples for
classical representations of the form $ T = \tau\hat{T} $ where the measure
$ \lambda $ arises quite natural as a consequence of the mathematical structure
of the underlying statistically complete observables.

Remember that, for classical representations $T$ on $ (\Omega,\Sigma) $
and $ \hat{T} $ on \linebreak $ (\Omega,\Sigma,\lambda) $,
we understand $T'$ and $ \hat{T}' $ to be the adjoint maps
$ T' \!\! : {\cal F}_{\R}(\Omega,\Sigma) \rightarrow {\cal B}_s({\cal H}) $
and $ \hat{T}' \!\! : L^{\infty}_{\R}(\Omega,\Sigma,\lambda)
\rightarrow {\cal B}_s({\cal H}) $, respectively, where
$ {\cal F}_{\R}(\Omega,\Sigma) \subseteq $ \linebreak
$ ({\cal M}_{\R}(\Omega,\Sigma))' $. Moreover, in contrast to
$ {\cal F}_{\R}(\Omega,\Sigma) $, the elements of
$ L^{\infty}_{\R}(\Omega,\Sigma,\lambda) $ are classes of
$ \lambda $-essentially bounded functions. We now derive the analog of (3.4)
for $ \hat{T} $. For every $ f \in L^{\infty}_{\R}(\Omega,\Sigma,\lambda) $
and all $ V \in {\cal T}_s({\cal H}) $, we have
\begin{eqnarray*}
{\rm tr} \, V(\hat{T}'f) & = & \int \rho f d\lambda = \int f d(\tau\rho) \\
& = & \int fd(TV) = \int f d({\rm tr} \, VF(\, . \,))
= {\rm tr} \left( V \int fdF \right)
\end{eqnarray*}
where $ \rho := \hat{T}V $ and $F$ is the observable corresponding to
$ T := \tau \hat{T} $. From this it follows that
\begin{equation}
\hat{T}'f = \int fdF
\end{equation}
holds where the $ \sigma $-weak integral does not depend on the representative
of $ f \in L^{\infty}_{\R}(\Omega,\Sigma,\lambda) $.

\section{Dequantizations}

By means of a classical representation, the quantum states can be identified
with probability measures and probability densities, respectively. The
\linebreak following theorem concerns the corresponding description of
effects and observables (cf.\ M. Singer and W. Stulpe, 1992).

\begin{theorem}
Let $ T \!\! : {\cal T}_s({\cal H}) \rightarrow {\cal M}_{\R}(\Omega,\Sigma) $
be a classical representation on the measurable space $ (\Omega,\Sigma) $
and $F$ the corresponding observable. Then the following statements are valid:
\begin{enumerate}
\item[(a)] To each bounded self-adjoint operator
$ A \in R(T') := T'{\cal F}_{\R}(\Omega,\Sigma) $ a function
$ f \in {\cal F}_{\R}(\Omega,\Sigma) $ can be assigned such that for all
states $ W \in K({\cal H}) $
\[
{\rm tr} \, WA = \int fd\mu
\]
holds where $ \mu := TW = P^F_W $.
\item[(b)] For every $ A \in {\cal B}_s({\cal H}) $, every $ \epsilon > 0 $,
and any finitely many states $ W_1,\ldots,W_n \in K({\cal H}) $ there exists
a function $ f \in {\cal F}_{\R}(\Omega,\Sigma) $ such that
\[
\left| {\rm tr} \, W_i A - \int fd\mu_i \right| < \epsilon
\]
holds where $ \mu_i := TW_i = P^{F}_{W_i} $ ($ i=1,\ldots,n $).
\end{enumerate}
\end{theorem}
{\bf Proof:} (a) If $ A \in R(T') $, choose $f$ as one of the functions
satisfying $ A = T'f $. We then obtain
\[
{\rm tr} \, WA = {\rm tr} \, W(T'f) = \int fd(TW) = \int fd\mu
\]
for all states $ W \in K({\cal H}) $.

(b) Since
$ T \!\! : {\cal T}_s({\cal H}) \rightarrow {\cal M}_{\R}(\Omega,\Sigma) $
is an injective linear map and the adjoint map
$ T' \!\! : {\cal F}_{\R}(\Omega,\Sigma) \rightarrow {\cal B}_s({\cal H}) $
exists, the range of $T'$ is a $ \sigma $-dense subspace of
$ {\cal B}_s({\cal H}) $. This is a general result in duality theory, but
in our case we can obtain it also from
\begin{equation}
R(T') = T'{\cal F}_{\R}(\Omega,\Sigma) \supseteq
{\rm lin} \, \{ T'\chi_B \, | \, B \in \Sigma \} = {\rm lin} \, F(\Sigma)
\end{equation}
and Lemma 2.1; the second equality sign in (3.7) is a consequence of (3.5).

From the $ \sigma $-denseness of $ R(T') $ in $ {\cal B}_s({\cal H}) $
and Eq.\ (2.1) it follows that, for every $ A \in {\cal B}_s({\cal H}) $,
every $ \epsilon > 0 $, and any $ W_1,\ldots,W_n \in K({\cal H}) $,
there exists a function $ f \in {\cal F}_{\R}(\Omega,\Sigma) $ satisfying
\[
| {\rm tr} \, W_i A - {\rm tr} \, W_i (T'f) | < \epsilon \ .
\]
Now, the assertion is implied by
$ {\rm tr} \, W_i (T'f) = \int fd(TW_i) = \int fd\mu_i $. \hfill $ \Box $\\

If $ \hat{T} \!\! : {\cal T}_s({\cal H}) \rightarrow
L^1_{\R}(\Omega,\Sigma,\lambda) $ is a classical representation on
the $ \sigma $-finite measure space $ (\Omega,\Sigma,\lambda) $, then,
for instance, the analog of statement (b) of the theorem reads as follows. For
every $ A \in {\cal B}_s({\cal H}) $, every $ \epsilon > 0 $, and any
$ W_1,\ldots,W_n \in K({\cal H}) $ there exists a function
$ f \in L^{\infty}_{\R}(\Omega,\Sigma,\lambda) $ such that
\begin{equation}
\left| {\rm tr} \, W_i A - \int \rho_i fd\lambda \right| < \epsilon
\end{equation}
holds where $ \rho_i := \hat{T} W_i $ ($ i=1,\ldots,n$). This result can be
concluded from part (b) of Theorem 3.4 or from duality theory (cf.\ W. Stulpe,
1992, 1994).

When a classical representation of quantum mechanics is given, the quantum
states $ W \in K({\cal H}) $ can be described like classical states. Moreover,
according to Theorem 3.4 and statement (3.8), respectively, the quantum
mechanical effects $ A \in [0,1] $ and observables $G$ on $ (\R,\Xi(\R)) $
for which the $ \sigma $-weak integral
$ \int {\rm id}_{\R} dG =: A \in {\cal B}_s({\cal H}) $
exists can be described like classical observables, namely, by
random variables on a classical sample space. That is, the quantum mechanical
probabilities and expectation values can, exactly or at least in arbitrarily
good physical approximation, be represented by the corresponding classical
expressions (compare Eqs.\ (2.6), (3.1), and (3.3)). The approximation
involved in Theorem 3.4 and statement (3.8) is physical in the sense that
probabilities and expectation values cannot be measured exactly and in
the laboratory physicists are not able to prepare more than finitely many
states. In particular, one can work with the same small $ \epsilon > 0 $
and the same many states $ W_1,\ldots,W_n $ for all observables.

We add some remarks. First, a function $ f \in {\cal F}_{\R}(\Omega,\Sigma) $
describing an effect $ A \in [0,1] $ need not satisfy
$ 0 \leq f \leq \chi_{\Omega} $. Second, writing
$ \int fd\mu = \int fd P^F_W $ and $ \int fd\mu_i = \int fd P^F_{W_i} $
in the respective statements of Theorem 3.4, we note that all quantum
mechanical probabilities and expectation values can be calculated from
the distributions of one single observable. Third, remembering that
$ R(T') = T'{\cal F}_{\R}(\Omega,\Sigma) $ is a $ \sigma $-dense subspace
of $ {\cal B}_s({\cal H}) $ and that the $ \sigma $-topology describes the
physical approximation of quantum observables $ A \in {\cal B}_s({\cal H}) $,
one can replace $ {\cal B}_s({\cal H}) $, as a space of observables, by
$ R(T') $. The classical description of observables then becomes exact.

One can show that, in the case of a finite-dimensional Hilbert space,
there~exist even bijective classical representations
$ T \!\! : {\cal T}_s({\cal H}) \rightarrow {\cal M}_{\R}(\Omega,\Sigma) $
where the sample space $ \Omega $ consists of $N$ elements, $ N := n^2 $,
$ n := \dim {\cal H} $, \linebreak and $ \Sigma := {\cal P} (\Omega) $ is the
power set of $ \Omega $ (see P. Busch, K.-E. Hellwig, and W.~Stulpe, 1993;
K.-E. Hellwig and W. Stulpe, 1993). Denoting the counting measure on
$ \Sigma $ by $ \kappa $ and identifying the respective spaces,
we have $ {\cal M}_{\R}(\Omega,\Sigma) $ \linebreak
$ = L^1_{\R}(\Omega,\Sigma,\kappa) = \R^N $,
$ T =: \hat{T} $, and $ \hat{T} \!\! : {\cal T}_s({\cal H}) \rightarrow
L^{1}_{\R}(\Omega,\Sigma,\kappa) $. For an infinite-dimensional Hilbert space,
it is an open question whether bijective classical representations $T$ or
$ \hat{T} $ on suitable measurable spaces $ (\Omega,\Sigma) $ and measure
spaces $ (\Omega,\Sigma,\lambda) $, respectively, do exist. Now let
$ \dim {\cal H} $ be arbitrary and assume that
$ T \!\! : {\cal T}_s({\cal H}) \rightarrow {\cal M}_{\R}(\Omega,\Sigma) $
is a bijective classical representation. Because of the different
geometrical structure of the convex sets $ K({\cal H}) $ and
$ K(\Omega,\Sigma) $, $T$ cannot map $ K({\cal H}) $ onto $ K(\Omega,\Sigma) $,
provided that $ \dim {\cal H} \geq 2 $ (for details, see M. Singer and
W. Stulpe, 1992; P. Busch, K.-E. Hellwig, and W.~Stulpe, 1993). That is, even
if $T$ is bijective, the quantum states can always be identified only with
a proper subset of $ K(\Omega,\Sigma) $. Finally, consider a bijective
classical representation $ \hat{T} \!\! : {\cal T}_s({\cal H}) \rightarrow
L^1_{\R}(\Omega,\Sigma,\lambda) $. Since the adjoint map $ \hat{T}' \!\! :
L^{\infty}_{\R}(\Omega,\Sigma,\lambda) \rightarrow {\cal B}_s({\cal H}) $
is also bijective, for every $ A \in {\cal B}_s({\cal H}) $
\vspace{1pt} there exists a uniquely determined function
$ f \in L^{\infty}_{\R}(\Omega,\Sigma,\lambda) $, namely
$ f := (\hat{T}')^{-1} A $, such that for all $ W \in K({\cal H}) $
\[
{\rm tr} \, WA = \int \rho f d\lambda
\]
holds where $ \rho := \hat{T}W $. Again, we have
$ \hat{T} K({\cal H}) \subset K(\Omega,\Sigma,\lambda) $. This proper
inclusion is related to $ \hat{T}' [0,\chi_{\Omega}] \subset [0,1] $,
respectively, to $ (\hat{T}')^{-1} [0,1] \supset [0,\chi_{\Omega}] $
(compare Lemma 3.2). Some further thoughts on these and similar problems
can be found in some papers of S. Bugajski (1993a,b) and a paper of
E.~G.~Beltrametti and S. Bugajski (1995).

For arbitrary classical representations $T$ and $ \hat{T} $, the adjoint maps
$ T' \!\! : {\cal F}_{\R}(\Omega,\Sigma) \rightarrow {\cal B}_s({\cal H}) $
and $ \hat{T}' \!\! : L^{\infty}_{\R}(\Omega,\Sigma,\lambda) \rightarrow
{\cal B}_s({\cal H}) $ assign self-adjoint operators $A$ to classical
random variables $f$. Therefore, we call the map $T'$ the {\it quantization
corresponding to the classical representation $T$} and the map $ \hat{T}' $
the {\it quantization corresponding to $ \hat{T} $}. Accordingly, the
assignments $ A \mapsto f $ that can be defined by Theorem 3.4 and
statement (3.8), respectively (if necessary, for given $ \epsilon > 0 $
and $ W_1,\ldots,W_n \in K({\cal H}) $), are {\it dequantizations}. The
particular case of a bijective classical representation \vspace{1pt}
$ \hat{T} \!\! : {\cal T}_s({\cal H}) \rightarrow
L^{1}_{\R}(\Omega,\Sigma,\lambda) $ induces canonically
exactly one dequantization $ A \mapsto f $, namely
$ (\hat{T}')^{-1} \!\! : {\cal B}_s({\cal H}) \rightarrow
L^{\infty}_{\R}(\Omega,\Sigma,\lambda) $; we call \vspace{2pt}
$ (\hat{T}')^{-1} = (\hat{T}^{-1})' $ the {\it dequantization
corresponding to $ \hat{T} $}.

Summarizing, a far-reaching classical reformulation of the statistical
\linebreak scheme of quantum mechanics is possible. In particular,
probabilities and expectation values which appear in reality as relative
frequencies and mean values can be calculated on the basis of Hilbert space
and in principle also on the basis of a classical sample space. Nevertheless,
the probabilistic structure of quantum mechanics differs essentially from that
of classical statistical physics, as reflected by the fact that the embedding
of the quantum states into the classical ones is proper.

\section{Quantum Dynamics on a Sample Space}

The reformulation of the statistical scheme of quantum mechanics in classical
terms can be supplemented by a corresponding reformulation of quantum
dynamics. To that end, consider first usual quantum dynamics which is given
by some Hamiltonian $H$ according to
\begin{equation}
W \mapsto W_t := \tau_t W := e^{-iHt}We^{iHt}
\end{equation}
where $ W \in K({\cal H}) $ and $ t \in \R $. Obviously,
$ \{ \tau_t \}_{t \in \R} $ is a one-parameter group of norm-automorphisms
of the Banach space $ {\cal T}_s({\cal H}) $ mapping $ K({\cal H}) $
onto $ K({\cal H}) $. To prove its strong continuity, i.e.,
$ \| \tau_tV - \tau_{t_0}V \|_{\rm tr} \rightarrow 0 $ for $t$ converging
to any $ t_0 \in \R $ and all $ V \in {\cal T}_s({\cal H}) $, we need
the following simple lemma.

\begin{lemma}
Let $ \phi,\psi \in {\cal H} $ with $ \|\phi\| = \|\psi\| = 1 $ and denote
the corresponding one-dimensional orthogonal projections by $ P_{\phi} $ and
$ P_{\psi} $. Then
\[
\| P_{\phi} - P_{\psi} \|_{\rm tr} \leq 4 \, \| \phi - \psi \|
\]
holds where $ \| \, . \, \|_{\rm tr} $ is the trace norm
in $ {\cal T}_s({\cal H}) $.
\end{lemma}
{\bf Proof:} Since the range of the operator $ P_{\phi} - P_{\psi} $ is a
two-dimensional subspace of $ \cal H $ (unless it is $ \{0\} $), we obtain
\begin{eqnarray*}
\| P_{\phi} - P_{\psi} \|_{\rm tr} & = & {\rm tr} \, | P_{\phi} - P_{\psi} |
        = \langle \chi_1 | \, | P_{\phi} - P_{\psi} | \chi_1 \rangle
        + \langle \chi_2 | \, | P_{\phi} - P_{\psi} | \chi_2 \rangle \\
        & \leq & 2 \, \| \, | P_{\phi} - P_{\psi} | \, \|
        = 2 \, \| P_{\phi} - P_{\psi} \|
\end{eqnarray*}
where $ \chi_1 $ and $ \chi_2 $ are suitably chosen vectors and
$ \| \, . \, \| $ is the usual operator norm. Taking account of
\[
\| P_{\phi} - P_{\psi} \|
\leq \| P_{\phi} - | \phi \rangle \langle \psi | \, \|
   + \| \, | \phi \rangle \langle \psi | - P_{\psi} \|
\leq 2 \, \| \phi - \psi \| \ ,
\]
it follows that
\[
\| P_{\phi} - P_{\psi} \|_{\rm tr} \leq 4 \, \| \phi - \psi \| \ .
\]
\hspace*{\fill} $ \Box $\\

\begin{proposition}
The {\it dynamical group} $ \{ \tau_t \}_{t \in \R} $ of automorphisms of
$ {\cal T}_s({\cal H}) $ defined by Eq.\ (3.9) is strongly continuous.
\end{proposition}
{\bf Proof:} It suffices to show
$ \| \tau_tW - \tau_{t_0}W \|_{\rm tr} \rightarrow 0 $ for
$ t \rightarrow t_0 $ and all $ W \in K({\cal H}) $. Writing
$ W = \sum_{i=1}^{\infty} \alpha_i P_{\chi_i} $ where $ \alpha_i \geq 0 $,
$ \sum_{i=1}^{\infty} \alpha_i = 1 $, $ \| \chi_i \| = 1 $, and
$ P_{\chi_i} := |\chi_i \rangle \langle \chi_i | $, we obtain, for
$ \epsilon > 0 $ and a sufficiently large $ N \in \N $,
\begin{eqnarray*}
\| \tau_tW - \tau_{t_0}W \|_{\rm tr} & \leq & \left\| (\tau_t - \tau_{t_0})
   \left( \sum_{i=1}^{N} \alpha_i P_{\chi_i} \right) \right\|_{\rm tr}
 + 2\left\| \sum_{i=N+1}^{\infty} \alpha_i P_{\chi_i} \right\|_{\rm tr} \\
& < & \sum_{i=1}^{N} \alpha_i
                     \| \tau_tP_{\chi_i} - \tau_{t_0}P_{\chi_i} \|_{\rm tr}
                                         + \frac{\epsilon}{2} \ .
\end{eqnarray*}
Taking account of Eq.\ (3.9) and applying Lemma 3.5, it follows with
$ e^{-iHt} =: U(t) $ that
\begin{eqnarray*}
\sum_{i=1}^{N} \alpha_i \| \tau_tP_{\chi_i} - \tau_{t_0}P_{\chi_i} \|_{\rm tr}
& = & \sum_{i=1}^{N} \alpha_i \| P_{U(t)\chi_i}
           - P_{U(t_0)\chi_i} \|_{\rm tr}  \\
& \leq & 4\sum_{i=1}^{N} \alpha_i \| U(t)\chi_i - U(t_0)\chi_i \|  \ .
\end{eqnarray*}
Now the strong continuity of the unitary one-parameter group
$ \{ U(t) \}_{t \in \R} $ implies $ \| U(t)\chi_i - U(t_0)\chi_i \|
\leq \| (U(t-t_0) - 1)\chi_i \| < \frac{\epsilon}{8} $ for
$ |t - t_0| < \delta (\frac{\epsilon}{8},\chi_i) $. Hence,
\[
\| \tau_tW - \tau_{t_0}W \|_{\rm tr} < \epsilon
\]
holds for $ |t - t_0| < \delta (\epsilon,W)
:= \min_{1 \leq i \leq N} \delta (\frac{\epsilon}{8},\chi_i) $.
\hfill $ \Box $\\

We mention that there is a converse of Proposition 3.6: For every strongly
continuous dynamical group $ \{ \tau_t \}_{t \in \R} $ of automorphisms of
$ {\cal T}_s({\cal H}) $ there exists a self-adjoint operator $H$ in $ \cal H $
such that $ \tau_t $ can be represented according to Eq.~(3.9); $H$ is
uniquely determined up to an additive constant $c1$ with $ c \in \R $. This
is in fact a deep result whose proof can be found in V.~S.~Varadarajan
(1970). Furthermore, the strongly continuous dynamical group
$ \{ \tau_t \}_{t \in \R} $ according to (3.9) has an infinitesimal
generator $Z$ which is a closed operator with domain $ D(Z) $ being dense
in $ {\cal T}_s({\cal H}) $. In particular, $ D(Z) $ is invariant under
$ \tau_t $, and
\begin{equation}
\dot{W}_t = ZW_t
\end{equation}
holds for $ t \mapsto W_t = \tau_t W $ with
$ W \in K({\cal H}) \cap D(Z) $. The domain $ D(Z) $ consists of all
$ V \in {\cal T}_s({\cal H}) $ satisfying $ VD(H) \subseteq D(H) $ for which
the operator $ HV - VH $ on $ D(H) $ is norm-bounded and can be extended to
a trace-class operator on $ \cal H $ (E. B. Davies, 1976). Since
$ P_{\psi} = |\psi \rangle \langle \psi | \in D(Z) $ for $ \psi \in D(H) $,
$ \| \psi \| = 1 $, it follows that $ K({\cal H}) \cap D(Z) $ is dense in
$ K({\cal H}) $, which is not obvious otherwise. The infinitesimal generator
$Z$ is given by
\begin{equation}
(ZV)\phi = -i(HV - VH)\phi
\end{equation}
where $ V \in D(Z) $ and $ \phi \in D(H) $, i.e., $ZV$ is just the extension
of $ -i(HV - VH) $. Eq.\ (3.10) then reads
\[
\dot{W}_t = -i(HW_t - W_tH)
\]
which is known as the von Neumann equation.

Now, let $ T \!\! : {\cal T}_s({\cal H}) \rightarrow
{\cal M}_{\R}(\Omega,\Sigma) $ be a classical representation
of quantum mechanics on $ (\Omega,\Sigma) $. Applying $T$ to Eq.\ (3.9),
we obtain a corresponding classical reformulation of quantum dynamics
according to
\begin{equation}
\mu \mapsto \mu_t := TW_t = T\tau_tW = T\tau_tT^{-1}\mu
\end{equation}
where $ W \in K({\cal H}) $, $ \mu := TW \in K(\Omega,\Sigma) $,
$ t \in \R $, and $ T^{-1} \!\! : R(T) \rightarrow {\cal T}_s({\cal H}) $,
$ R(T) := T{\cal T}_s({\cal H}) $ being the range of $T$. Defining
\begin{equation}
\delta_t := T\tau_tT^{-1} \ ,
\end{equation}
i.e.\ $ \mu_t = \delta_t \mu $ for $ \mu = TW $, one observes that
$ \{ \delta_t \}_{t \in \R} $ is a one-parameter group of linear isomorphisms
$ \delta_t \!\! : R(T) \rightarrow R(T) $ satisfying
$ \delta_t(TK({\cal H})) = TK({\cal H}) $. However, $ R(T) $, equipped with
the norm $ \| \, . \, \| $ of $ {\cal M}_{\R}(\Omega,\Sigma) $, need not be
a Banach space, and the maps $ T^{-1} $ and $ \delta_t $ need not be bounded.

The one-parameter group $ \{ \delta_t \}_{t \in \R} $ is strongly continuous
in the sense that $ \| \delta_t\nu - \delta_{t_0}\nu \| \rightarrow 0 $ for
$ t \rightarrow t_{0} $ and all $ \nu \in R(T) $. This, however, does not
imply that $ \{ \delta_t \}_{t \in \R} $ can be reconstructed from an
infinitesimal generator since $ (R(T),\| \, . \, \|) $ is not a Banach space
(if it is, then $ T^{-1} $ and $ \delta_t $ are necessarily bounded). To avoid
these difficulties, we equip $ R(T) $ with a new norm, namely
$ \|\nu\|' := \|V\|_{\rm tr} $ where $ \nu = TV $. Because of
$ \|\nu\| = \|TV\| \leq \|V\|_{\rm tr} = \|\nu\|' $, $ \| \, . \, \|' $
on $ R(T) $ is stronger than $ \| \, . \, \| $. Using $ \| \, . \, \|' $, one
can introduce an infinitesimal generator for $ \{ \delta_t \}_{t \in \R} $,
which is done in the following theorem due to the author.

\begin{theorem}
Consider $ R(T) $ with the norm $ \| \, . \, \|' $. Then
$ R(T) $ is a \mbox{Banach} space, and the map
$ T \!\! : {\cal T}_s({\cal H}) \rightarrow R(T) $ is a norm-isomorphism. In
particular, $ \{ \delta_t \}_{t \in \R} $ is a strongly continuous
one-parameter group of norm-automorphisms of $ R(T) $ mapping
$ TK({\cal H}) $ onto itself; $ \{ \delta_t \}_{t \in \R} $ can
be reconstructed from its infinitesimal generator $L$. For
$ t \mapsto \mu_t = \delta_t \mu $ with $ \mu \in TK({\cal H}) \cap D(L) $
and $ t \in \R $, the equation
\begin{equation}
\dot{\mu}_t = L\mu_t
\end{equation}
holds where the derivative can be taken with respect to $ \| \, . \, \|' $
as well as to $ \| \, . \, \| $. Moreover, $ D(L) $ is dense in $ R(T) $
and $ TK({\cal H}) \cap D(L) $ in $ TK({\cal H}) $, both with respect to
$ \| \, . \, \|' $ and $ \| \, . \, \| $; $L$ is related to $Z$ according to
$ D(L) = TD(Z) $ and
\[
L = TZT^{-1} \ .
\]
\end{theorem}
{\bf Proof:} Most of the statements of the theorem is obvious. By construction,
the derivative in (3.14) is understood with respect to $ \| \, . \, \|' $,
but, since $ \| \, . \, \|' $ is stronger than $ \| \, . \, \| $, it can
also be taken in $ \| \, . \, \| $. Clearly, $ D(L) $ is dense
in $ R(T) $. For $ \nu \in TD(Z) $, i.e.\ $ T^{-1}\nu \in D(Z) $,
$ \left. \frac{d}{dt} (\tau_tT^{-1}\nu) \right|_{t=0} $ exists. It follows
that
\[
\left. T\frac{d}{dt} (\tau_tT^{-1}\nu) \right|_{t=0}
= \left. \frac{d}{dt} (T\tau_tT^{-1}\nu) \right|_{t=0}
= \left. \frac{d}{dt} \delta_t\nu \right|_{t=0}
\]
where the latter two derivatives exist in $ \| \, . \, \|' $. This implies
$ \nu \in D(L) $. Conversely, if $ \nu \in D(L) $, then
$ \left. \frac{d}{dt} \delta_t\nu \right|_{t=0} $ exists
in $ \| \, . \, \|' $. In consequence,
\[
\left. T^{-1} \frac{d}{dt} \delta_t\nu \right|_{t=0}
= \left. \frac{d}{dt} (T^{-1}\delta_t\nu) \right|_{t=0}
= \left. \frac{d}{dt} (\tau_tT^{-1}\nu) \right|_{t=0}
\]
and $ T^{-1}\nu \in D(Z) $, respectively, $ \nu \in TD(Z) $. Hence,
$ D(L) = TD(Z) $.

Since $ K({\cal H}) \cap D(Z) $ is dense in $ K({\cal H}) $
and $T$ is a norm-isomorphism,
$ TK({\cal H}) \cap D(L) = TK({\cal H}) \cap TD(Z) = T(K({\cal H}) \cap D(Z)) $
is $ \| \, . \, \|' $-dense in $ TK({\cal H}) $ and consequently also
$ \| \, . \, \| $-dense. Finally, for $ \nu \in D(L) $ we obtain
\begin{eqnarray*}
\left. \dot{\nu}_t \right|_{t=0}
             & = & \left. \frac{d}{dt} \delta_t\nu \right|_{t=0}
               =   \left. \frac{d}{dt} (T\tau_tT^{-1}\nu) \right|_{t=0}  \\
             & = & T\left. \frac{d}{dt} (\tau_tT^{-1}\nu) \right|_{t=0}  \\
             & = & TZ(T^{-1}\nu)                                         \\
             & = & TZT^{-1}\nu  \ ,
\end{eqnarray*}
and from $ \left. \frac{d}{dt} \delta_t\nu \right|_{t=0} = L\nu $ we
conclude $ L = TZT^{-1} $. \hfill $ \Box $\\

As a remark, we notice the obvious relations
\begin{equation}
TK({\cal H}) \subseteq K(\Omega,\Sigma) \cap R(T)
\end{equation}
and
\begin{eqnarray}
B' & = & {\rm conv} (TK({\cal H}) \cup (-TK({\cal H})))  \nonumber\\
   & \subseteq & {\rm conv} (K(\Omega,\Sigma)
                  \cup (-K(\Omega,\Sigma))) \cap R(T)             \\
   & = & B \cap R(T)                                     \nonumber
\end{eqnarray}
where $B'$ is the closed unit ball of $ (R(T),\| \, . \, \|') $, $B$ the
unit ball of \linebreak $ ({\cal M}_{\R}(\Omega,\Sigma),\| \, . \, \|) $, and
$ B \cap R(T) $ that of $ (R(T),\| \, . \, \|) $. The fact that the inclusion
$ TK({\cal H}) \subseteq K(\Omega,\Sigma) $ is always proper for
$ \dim {\cal H} \geq 2 $ (see the end of Section 3.2) suggests that
the inclusions (3.15) and (3.16) are also proper. This helps to understand
why the linear map $ T^{-1} $ may be unbounded on $ (R(T),\| \, . \, \|) $
although it is bounded on $ (R(T),\| \, . \, \|') $.

Since $ R(T) $ is a space of signed measures, $ \| \, . \, \|' $ is no
natural norm on $ R(T) $. It was introduced for technical reasons,
however, some statements of Theorem~3.7 can be derived without use of
$ \| \, . \, \|' $ as indicated now. From Eq.\ (3.10) it follows that
$ t \mapsto \mu_t = \delta_t \mu $ satisfies $ \dot{\mu}_t = TZT^{-1} \mu_t $
for all $ \mu \in TK({\cal H}) \cap TD(Z) $. The derivative $ \dot{\mu}_t $
is taken in $ \| \, . \, \| $, and the $ \| \, . \, \| $-continuity of $T$
implies that $ TD(Z) $ is dense in $ R(T) $ and $ TK({\cal H}) \cap TD(Z) $
in $ TK({\cal H}) $.

We have shown that, by means of a classical representation $T$ on
$ (\Omega,\Sigma) $, quantum dynamics can completely be reformulated in terms
of measures evolving in time on the classical sample space $ \Omega $. An
analogous reformulation can be obtained by means of a classical representation
of quantum mechanics on some $ \sigma $-finite measure space
$ (\Omega,\Sigma,\lambda) $. If $ \hat{T} \!\! : {\cal T}_s({\cal H})
\rightarrow L^{1}_{\R}(\Omega,\Sigma,\lambda) $ is such a classical
representation, we define, analogously to (3.12) and (3.13),
$ \rho_t := \hat{\delta}_t \rho $ and
$ \hat{\delta}_t := \hat{T}\tau_t\hat{T}^{-1} $
where $ \rho \in \hat{T}K({\cal H}) $ is a probability density and
$ \hat{T}^{-1} \!\! : R(\hat{T}) \rightarrow {\cal T}_s({\cal H}) $. Again,
$ R(\hat{T}) $ can be equipped with two norms, $ \| \, . \, \| $ and
$ \| \, . \, \|' $. Using $ \| \, . \, \|' $, one can introduce
an infinitesimal generator $ \hat{L} $ for the strongly continuous
one-parameter group $ \{ \hat{\delta}_t \}_{t \in \R} $. The equation
of motion corresponding to (3.14) then reads
\begin{equation}
\dot{\rho}_t = \hat{L} \rho_t
\end{equation}
where $ \rho_t = \hat{\delta}_t \rho $ with
$ \rho \in \hat{T}K({\cal H}) \cap D(\hat{L}) $ and $ t \in \R $. Moreover,
$ D(\hat{L}) $ \vspace{2pt} is dense in $ R(\hat{T}) $ and
$ \hat{T}K({\cal H}) \cap D(\hat{L}) $ in $ \hat{T}K({\cal H}) $;
finally, $ D(\hat{L}) = \hat{T}D(Z) $ and
$ \hat{L} = \hat{T}Z\hat{T}^{-1} $. In Section 7.4, we shall discuss
Eq.\ (3.17) in the context of phase space and a particular classical
representation $ \hat{T} $ where, essentially, it becomes equivalent to
a partial differential equation, as demonstrated by some special instances. It
turns out that Eq.\ (3.17) is related to the classical Liouville equation
and $ \hat{L} $ to the classical Liouville operator $ -\{ H, \, . \, \} $.

\chapter{Generalized Coherent States}

Important instances for observables and in particular statistically complete
observables can be obtained from the so-called continuous resolutions of the
identity of $ \cal H $. In Section 4.1, we introduce that concept and study
some remarkable consequences of the existence of continuous resolutions;
in Section~4.2 we show how they can be obtained from irreducible group
representations.

\section{Continuous Resolutions of the Identity}

In this section, let $M$ be a locally compact Hausdorff space satisfying
the axiom of second countability and $ \lambda $ some Borel measure on $M$. In
particular, the Borel sets of $M$ coincide with its Baire sets, and $ \lambda $
is a $ \sigma $-finite measure. We denote the $ \sigma $-algebra of the Borel
sets of $M$ by $ \Xi(M) $ and write briefly $ dx := \lambda (dx) $.

A family $ \{ u_x \}_{x \in M} $ of unit vectors in $ \cal H $ is called
a {\it continuous resolution of the identity of $ \cal H $} if the map
$ x \mapsto u_x $ is norm-continuous and there exists a number
$ \alpha > 0 $ such that
\begin{equation}
1 = \frac{1}{\alpha} \int | u_x \rangle \langle u_x | \, dx
\end{equation}
holds where the integral is understood to exist in the weak sense. As we
shall see in the next section, continuous resolutions of the identity often
arise from irreducible, strongly continuous projective unitary representations
of groups; in fact, if $ g \mapsto U_g $ is such a representation of
a group $ \cal G $ on $ \cal H $ and $ u \in {\cal H} $ a unit vector,
then the vectors $ u_g := U_g u $ may constitute a continuous resolution
$ \{ u_g \}_{g \in {\cal G}} $ of the identity of $ \cal H $. A very important
instance of a continuous resolution of type $ \{ u_g \}_{g \in {\cal G}} $
will be discussed in the next chapter; it is given by
$ \{ u_{qp} \}_{(q,p) \in \R^2} $ where
$ u_{qp} := U_{qp}u $, $ u \in {\cal H} $, $ \| u \| = 1 $,
and $ (q,p) \mapsto U_{qp} := e^{ipQ}e^{-iqP} $ is an irreducible,
strongly continuous projective unitary representation of the additive
group $ \R^2 $ (essentially, the operators $ U_{qp} $ are the Weyl
operators). Choosing $ {\cal H} := L^2_{\C} \, (\R,d\xi) $ and
\[
u(\xi) := \frac{1}{\sqrt[4]{2\pi\sigma^2}} e^{-\frac{\xi^2}{4\sigma^2}}
        =: u^{\sigma} (\xi) \ ,
\]
$ \{ u^{\sigma}_{qp} \}_{(q,p) \in \R^2} $ is, for each $ \sigma > 0 $,
a continuous resolution consisting of coherent states. For this reason,
the vectors $ u_x $ of a general continuous resolution are called
{\it generalized coherent states}.---Our account on this topic is close
to that in the book of E. B. Davies (1976), other interesting references
are A.~M.~Perelomov (1986) and S. T. Ali (1985).

A continuous resolution of the identity of some Hilbert space can be used
to construct a Hilbert space of continous functions. This remarkable property
is stated in the context of the next theorem (cf.\ E. B. Davies, 1976).

\begin{theorem}
A continuous resolution $ \{ u_x \}_{x \in M} $ of the identity of $ \cal H $
defines bounded continuous square-integrable functions $ \Psi $ and an
isometry $ V \!\! : {\cal H} \rightarrow L^{2}_{\C} \, (M,dx) $ by
\begin{equation}
(V\psi)(x) := \frac{1}{\sqrt{\alpha}} \langle u_x | \psi \rangle =: \Psi(x)
\end{equation}
where $ \psi \in {\cal H} $ and
$ L^{2}_{\C} \, (M,dx) := L^{2}_{\C} \, (M,\Xi(M),\lambda) $. In particular,
$ R(V) = V{\cal H} $ is a Hilbert space consisting entirely of bounded
continuous functions. Conversely, a norm-continuous family
$ \{ u_x \}_{x \in M} $ for which the linear operator $V$ defined by (4.2)
is an isometry from $ \cal H $ into $ L^{2}_{\C} \, (M,dx) $, is a continuous
resolution of the identity of $ \cal H $.
\end{theorem}
{\bf Proof:} If $ \{ u_x \}_{x \in M} $ is a continuous resolution, then
\[
\int |\Psi|^2 d\lambda
     = \frac{1}{\alpha} \int |\langle \psi | u_x \rangle|^2 dx
     = \frac{1}{\alpha} \int \langle \psi | P_{u_x}\psi \rangle \, dx
     = \| \psi \|^2
\]
holds where $ P_{u_x} := |u_x \rangle \langle u_x | $. Consequently, $ \Psi $
is a bounded continuous square-integrable function, and the linear operator
$ V \!\! : {\cal H} \rightarrow L^{2}_{\C} \, (M,dx) $ is isometric. Hence,
its range $ V{\cal H} $ is a Hilbert space of bounded continuous functions. If,
conversely, $ x \mapsto u_x $ is norm-continuous and $V$ defined by (4.2) is
an isometry into $ L^{2}_{\C} \, (M,dx) $, then it follows that
\[
\| \psi \|^2 = \| V\psi \|^2
     = \frac{1}{\alpha} \int \langle \psi | P_{u_x}\psi \rangle \, dx
\]
and hence $ \frac{1}{\alpha} \int P_{u_x} dx = 1 $. \hfill $ \Box $\\

We remark that, since the $ \sigma $-algebra $ \Xi(M) $ is countably
generated and the measure $ \lambda $ is $ \sigma $-finite, the Hilbert space
$ L^{2}_{\C} \, (M,dx) $ is separable. The following proposition gives further
information on the situation described in Theorem 4.1.

\begin{proposition}
The orthogonal projection $P$ of $ L^{2}_{\C} \, (M,dx) $ onto $ V{\cal H} $
can be represented by
\[
(P\Phi)(x) = \int K(x,y) \Phi(y) dy = \langle K( \, . \, ,x)|\Phi\rangle
\]
where $ \Phi \in L^{2}_{\C} \, (M,dx) $ and
\[
K(x,y) := \frac{1}{\alpha} \, \langle u_x | u_y \rangle \ .
\]
The integral kernel $K$ is bounded and continuous and satisfies
\begin{enumerate}
\item[(i)] $ K( \, . \, ,y), \, K(x, \, . \, ) \in L^{2}_{\C} \, (M,dx) $
\item[(ii)] $ K(x,y) = \overline{K(y,x)} $
\item[(iii)] $ K(x,y) = \int K(x,z)K(z,y) dz $ \ ,
\end{enumerate}
i.e., $K$ is a reproducing kernel. The function
$ x \mapsto \langle K( \, . \, ,x)|\Phi \rangle $ is a continuous
representative of $ P\Phi $.
\end{proposition}
{\bf Proof:} Because of
\[
|K(x,y)| = \frac{1}{\alpha} \, |\langle u_x | u_y \rangle|
         \leq \frac{1}{\alpha} \, \| u_x \| \, \| u_y \| = \frac{1}{\alpha}
\]
$K$ is bounded, and because of
\begin{eqnarray*}
|K(x,y) - K(x_0,y_0)| & \leq & \frac{1}{\alpha} \,
    (|\langle u_x | u_y \rangle - \langle u_{x_0} | u_y \rangle|
     + |\langle u_{x_0} | u_y \rangle - \langle u_{x_0} | u_{y_0} \rangle|) \\
  & \leq & \frac{1}{\alpha} \, (\| u_x - u_{x_0} \| \, \| u_y \|
                              + \| u_{x_0} \| \, \| u_y - u_{y_0} \|)
\end{eqnarray*}
$K$ is continuous. Obviously,
$ K( \, . \, ,y) = \frac{1}{\sqrt{\alpha}} Vu_y \in L^{2}_{\C} \, (M,dx) $,
$ K(x,y) = \overline{K(y,x)} $, and $ K(x, \, . \, ) \in L^{2}_{\C} \, (M,dx) $
hold. Property (iii) of $K$ follows from
$ K(x,y) = \frac{1}{\alpha} \, \langle u_x | 1u_y \rangle $ and
Eq.\ (4.1) or from
\begin{eqnarray*}
K(x,y) & = & \frac{1}{\alpha} \, \langle u_x | u_y \rangle
         =   \frac{1}{\alpha} \, \langle Vu_x | Vu_y \rangle \\
       & = & \frac{1}{\alpha} \int \overline{(Vu_x)(z)} (Vu_y)(z) \, dz \\
       & = & \frac{1}{\alpha^2} \int \langle u_x | u_z \rangle
                                     \langle u_z | u_y \rangle \, dz \\
       & = & \int K(x,z) K(z,y) dz \ .
\end{eqnarray*}

For $ \Phi \in L^{2}_{\C} \, (M,dx) $ and $ \psi \in {\cal H} $ we have
\begin{eqnarray*}
\langle V^{\ast}\Phi | \psi \rangle & = & \langle \Phi | V\psi \rangle
   =   \langle \Phi | PV\psi \rangle \\
 & = & \langle P\Phi | V\psi \rangle
   =   \langle V^{-1}P\Phi | V^{-1}V\psi \rangle
   =   \langle V^{-1}P\Phi | \psi \rangle
\end{eqnarray*}
where $ V^{-1} $ is defined on $ V{\cal H} $. That is, $ V^{\ast} = V^{-1}P $
or, equivalently, $ VV^{\ast} = P $. Hence,
\begin{eqnarray*}
(P\Phi)(x) & = & (VV^{\ast}\Phi)(x)
             =   \frac{1}{\sqrt{\alpha}} \langle u_x | V^{\ast}\Phi \rangle
             =   \left\langle \left. \frac{1}{\sqrt{\alpha}} Vu_x \,
                              \right| \, \Phi \right\rangle       \\
           & = & \langle K( \, . \, ,x) | \Phi \rangle
             =   \int K(x,y) \Phi(y) dy \ .
\end{eqnarray*}
In particular, $ x \mapsto \langle K( \, . \, ,x) | \Phi \rangle $ is
continuous. \hfill $ \Box $\\

\begin{corollary}
The kernel $K$ is square-integrable with respect to both of its arguments,
i.e.\
\[
\int |K|^2 d\lambda^2 = \int |K(x,y)|^2 dxdy < \infty \ ,
\]
respectively, $ K \in L^{2}_{\C} \, (M^2,dxdy) $, if and only if $ \cal H $
is a finite-dimensional Hilbert space. In particular, $ \cal H $ must be
finite-dimensional if the Borel measure $ \lambda $ is finite; also,
$ \cal H $ is finite-dimensional if $M$ is compact.
\end{corollary}
{\bf Proof:} If and only if $ K \in L^{2}_{\C} \, (M^2,dxdy) $, the
projection $P$ is a Hilbert-Schmidt operator. Equivalently, its range
$ R(P) = PL^{2}_{\C} \, (M,dx) $ is finite-dimensional, respectively,
$ \dim {\cal H} = \dim V{\cal H} = \dim PL^{2}_{\C} \, (M,dx) < \infty $. If
$ \lambda(M) < \infty $, then $ K \in L^{2}_{\C} \, (M^2,dxdy) $; if
$M$ is compact, then, as a Borel measure on $M$, $ \lambda $
is finite. \hfill $ \Box $\\

The second statement of the corollary, $ \dim {\cal H} < \infty $ if
$ \lambda < \infty $, can be obtained more directly. Namely, assuming
$ \dim {\cal H} = \infty $ and using a complete orthonormal system
$ \{ \phi_i \}_{i \in \N} $ of $ \cal H $, it follows from Eq.\ (4.1) that
\begin{eqnarray}
\alpha n & = & \alpha \sum_{i=1}^{n} \langle \phi_i | 1\phi_i \rangle
  =   \sum_{i=1}^{n} \int |\langle \phi_i | u_x \rangle|^2 dx  \nonumber\\
& = & \int \sum_{i=1}^{n} |\langle \phi_i | u_x \rangle|^2 dx           \\
& \leq & \lambda(M) < \infty                                   \nonumber
\end{eqnarray}
holds for every $ n \in \N $. But $ \alpha n \leq \lambda(M) $ for every $n$ is
a contradiction. Hence, $ \cal H $ must be finite-dimensional.---Consequently,
a continuous resolution $ \{ u_x \}_{x \in M} $ of the identity of an
infinite-dimensional Hilbert space is possible only with an infinite measure
$ \lambda $ and a noncompact space $M$.

The spaces $ \cal H $ and $ V{\cal H} = PL^{2}_{\C} \, (M,dx) $ are
isomorphic where, by restriction of its range, the isometry $V$ can
be regarded as a unitary operator. Therefore, using the unitary operator
$ V \!\! : {\cal H} \rightarrow V{\cal H} $, the quantum states
and observables can be represented by operators in the Hilbert space
$ V{\cal H} $ which is a space of bounded continuous,
square-integrable functions on $M$, in general a proper subspace of
$ L^{2}_{\C} \, (M,dx) $. Thus, on the one hand, a continuous resolution
of the identity of $ \cal H $ gives rise to a particular representation
of quantum mechanics on $ V{\cal H} $ which we call an
{\it $M$-representation of quantum mechanics}. On the other hand, continuous
resolutions can be used to define observables as we discuss now.

\begin{proposition}
Given a continuous resolution $ \{ u_x \}_{x \in M} $ of the identity of
$ \cal H $, then, for every Borel set $ B \in \Xi(M) $, the integral
\begin{equation}
F(B) := \frac{1}{\alpha} \int_B |u_x \rangle \langle u_x | \, dx
\end{equation}
exists in the weak sense and defines an observable $F$ on $ (M,\Xi(M)) $.
\end{proposition}
{\bf Proof:} From (4.1) it follows that the integral
\[
\Phi_{B} (\phi,\psi) := \frac{1}{\alpha} \int_{B} \langle \phi |u_x \rangle
                                         \langle u_x |\psi \rangle \, dx
\]
exists for all $ \phi,\psi \in {\cal H} $ and $ B \in \Xi(M) $. The
sesquilinear functional $ \Phi_B $ has the property
$ | \Phi_{B}(\psi,\psi) | \leq \| \psi \|^2 $. By polarization,
this implies $ | \Phi_{B}(\phi,\psi) | \leq 2 $ for
$ \| \phi \|, \| \psi \| \leq 1 $. Hence, $ \Phi_{B} $ is a bounded
sesquilinear functional, and there exists a bounded linear operator $ F(B) $
satisfying
\[
\Phi_{B} (\phi,\psi) = \langle \phi | F(B) \psi \rangle \ ;
\]
in particular, $ F(B) $ is just the weak integral
$ \frac{1}{\alpha} \int_{B} | u_x \rangle \langle u_x | \, dx $. Since
$ \Phi_B (\phi,\psi) $ $ = \overline{\Phi_B (\psi,\phi)} $ and
$ 0 \leq \Phi_B (\psi,\psi) \leq \| \psi \|^2 $, $ F(B) $ is self-adjoint
and $ 0 \leq F(B) \leq 1 $. Moreover, $ B \mapsto F(B) $ is
an observable. \hfill $ \Box $\\

We next show that the integrals (4.1) and (4.4) do exist also
in the $ \sigma $-weak sense. Consider the integral (4.4)
which contains the other one as a special case, and let
$ V = \sum_{i=1}^{\infty} \alpha_{i} P_{\chi_i} $ where $ \alpha_i \geq 0 $,
$ \| \chi_i \| = 1 $, and $ P_{\chi_i} = | \chi_i \rangle \langle \chi_i | $
be any positive trace-class operator. Writing
$ | u_x \rangle \langle u_x | = P_{u_x} $, we then have
\begin{eqnarray*}
{\rm tr} \, VF(B)
& = & \sum_{i=1}^{\infty} \alpha_{i} \langle \chi_{i} | F(B)\chi_{i} \rangle \\
& = & \frac{1}{\alpha} \sum_{i=1}^{\infty} \alpha_{i} \int_{B}
      \langle \chi_{i} | P_{u_x} \chi_{i} \rangle \, dx                      \\
& = & \frac{1}{\alpha} \int_{B} {\rm tr} \, VP_{u_x} \, dx \ .
\end{eqnarray*}
For an arbitrary self-adjoint trace-class operator
$ V \in {\cal T}_s({\cal H}) $, we write $ V = V^{+} - V^{-} $
and obtain also
\begin{equation}
{\rm tr} \, VF(B) = \frac{1}{\alpha} \int_{B} {\rm tr} \, VP_{u_x} \, dx \ ,
\end{equation}
showing that the integral (4.4) exists in the $ \sigma $-weak sense.

From (4.5) it follows that the probability distribution of the observable $F$
in a state $ W = \sum_{i=1}^{\infty} \alpha_i P_{\phi_i} \in K({\cal H}) $,
$ P_{\phi_i} = |\phi_i \rangle \langle \phi_i | $, is given by
\[
P^F_W (B) = {\rm tr} \, WF(B)
          = \frac{1}{\alpha} \int_{B} \langle u_x | W u_x \rangle \, dx \ ;
\]
$ P^F_W $ has the bounded continuous probability density
\[
x \mapsto \rho(x) := \frac{1}{\alpha} \, \langle u_x | Wu_x \rangle
                   = \sum_{i=1}^{\infty} \alpha_i |\Phi_i(x)|^2
\]
where $ \Phi_i(x) := (V\phi_i)(x)
= \frac{1}{\sqrt{\alpha}} \langle u_x | \phi_i \rangle $. If
$ W = P_{\psi} = |\psi\rangle \langle \psi | $ is a pure state, then
$ \rho(x) = |(V\psi)(x)|^2 = |\Psi(x)|^2 $. Although $V$ is an isometry,
the map $ P_{\psi} \mapsto \rho $ need not be injective. Still less, the map
$ W \mapsto \rho =: \hat{T}W $ need be injective; that is, $ \hat{T} $ need
not be a classical representation on $ (M,\Xi(M),\lambda) $, reflecting the
fact that the observable $F$ need not be statistically complete.

A trivial example of a continuous resolution of the identity is given
by a complete orthonormal system $ \{ \phi_n \}_{n \in \N} $ in an,
e.g., infinite-dimensional Hilbert space $ \cal H $ where $ M := \N $
is equipped with the discrete topology and $ \lambda := \kappa $
is the counting measure defined on the power set $ \Xi(\N) $ of $ \N $;
Eq.\ (4.1) reads $ 1 = \sum_{n=1}^{\infty} |\phi_n \rangle \langle \phi_n |
= \int |\phi_n \rangle \langle \phi_n | \, dn $ where
$ dn := \kappa(dn) $. The isometry (4.2) is given by
$ (V\psi)(n) := \langle \phi_n | \psi \rangle $ and is a unitary map
from $ \cal H $ onto the sequence space $ l^{2}_{\C} = L^{2}_{\C} \, (\N,dn)
:= L^{2}_{\C} \, (\N,\Xi(\N),\kappa) $. Consider the observable (4.4), i.e.,
$ B \mapsto F(B) := \sum_{n \in B} |\phi_n \rangle \langle \phi_n | $
where $B$ is any subset \vspace{1pt} of $ \N $, and define a linear map
$ \hat{T} \!\! : {\cal T}_s({\cal H}) \rightarrow l^{1}_{\R}
= L^{1}_{\R}(\N,dn) $ by assigning to each
$ W \in K({\cal H}) $ the probability vector
$ p := (p_1,p_2,\ldots) \in K(l^{1}_{\R}) := K(\N,\Xi(\N),\kappa) $
with $ p_n := \langle \phi_n | W\phi_n \rangle $; $p$ is just the density
of the probability distribution $ P^F_W $ with respect to the counting measure
on $ \N $. If $ p \in K(l^{1}_{\R}) $ is any \vspace{1pt} probability vector,
then $ \psi := \sum_{n=1}^{\infty} \sqrt{p_n} \phi_n \in {\cal H} $,
$ \|\psi\| = 1 $, and $ \hat{T}P_{\psi} = p $ where
$ P_{\psi} = |\psi \rangle \langle \psi | $. Hence, $ \hat{T} $ maps
$ K({\cal H}) $ onto $ K(l^{1}_{\R}) $; $ \hat{T} $ is surjective, but
\vspace{2pt} not injective (the latter can be seen directly by
$ \tilde{\psi} := \sum_{n=1}^{\infty} (-1)^n \sqrt{p_n} \phi_n $,
for instance, and $ \hat{T}P_{\tilde{\psi}} = p) $. Thus, $ \hat{T} $
is not a classical representation on $ (\N,\Xi(\N),\kappa) $, corresponding
to the fact that the observable $F$ is not statistically complete. The
two crucial properties of this example, namely,
$ V{\cal H} = L^{2}_{\C} \, (\N,dn) $ and $F$ being not statistically
complete, are not independent of each other, as the next proposition
states more precisely.

\begin{proposition}
Let $ \{ u_x \}_{x \in M} $ be a continuous resolution of the identity
of $ \cal H $, $V$ the isometry (4.2), and $F$ the observable (4.4). If
$ V{\cal H} = L^{2}_{\C} \, (M,dx) $, then $F$ cannot be statistically
complete.
\end{proposition}
{\bf Proof:} Let
$ \hat{T} \!\! : {\cal T}_s({\cal H}) \rightarrow L^{1}_{\R}(M,dx) $
be the linear map that assigns to each $ W \in K({\cal H}) $
the probability density
$ x \mapsto \rho(x) := \frac{1}{\alpha} \, \langle u_x | Wu_x \rangle $
of $ P^F_W $. For any probability density
$ \rho \in K(M,dx) := K(M,\Xi(M),\lambda) $, define functions
\[
\Psi_f(x) := e^{if(x)} \sqrt{\rho(x)}
\]
where $f$ is an arbitrary real-valued measurable function on $M$; we have
$ \Psi_f \in L^{2}_{\C} \, (M,dx) $ and $ \|\Psi\| = 1 $. By means of
$ \psi_f := V^{-1}\Psi_f $, we obtain
$ \hat{T}P_{\psi_f} = |V\psi_f|^2 = |\Psi_f|^2 = \rho $. Hence,
$ \hat{T}K({\cal H}) = K(M,dx) $, $ \hat{T} $ is surjective, but not
injective. In particular, $F$ is not statistically complete. \hfill $ \Box $\\

As we shall see later, if $ V{\cal H} $ is a proper subspace of
$ L^{2}_{\C} \, (M,dx) $, then both is possible, i.e., $F$ may be
statistically complete or not.

Finally, we remark that the observable (4.4), based on $ \cal H $,
is related to an observable that is based on the entire space
$ L^2_{\C} \, (M,dx) $. Namely, if $G$ is the observable defined by
\[
G(B)\Phi := \chi_B \Phi
\]
where $ B \in \Xi(M) $ and $ \Phi \in L^2_{\C} \, (M,dx) $, then it follows
from Eq.\ (4.2) that
\begin{eqnarray*}
\langle \psi | F(B)\psi \rangle
& = & \frac{1}{\alpha} \int_B |\langle u_x | \psi \rangle |^2 dx
  =   \int_B |\Psi(x)|^2 dx \\
& = & \langle \Psi | G(B)\Psi \rangle
  =   \langle V\psi | G(B)V\psi \rangle
  =   \langle \psi | V^{\ast}G(B)V\psi \rangle
\end{eqnarray*}
holds for all $ \psi \in {\cal H} $. Therefore,
\begin{equation}
F(B) = V^{\ast}G(B)V = V^{-1}PG(B)V \ .
\end{equation}
Introducing an observable $ \tilde{F} $ on $ V{\cal H} $ by
\[
\tilde{F}(B) = VF(B)V^{-1}
\]
which is unitarily equivalent to $F$, we obtain
\begin{equation}
\tilde{F}(B) = \left. PG(B) \right|_{V{\cal H}}
             = \left. PG(B)P \right|_{V{\cal H}} \ .
\end{equation}
Eqs.\ (4.6) and (4.7) show that the POV-measures $F$ and $ \tilde{F} $ are
related to the PV-measure $G$ on a larger Hilbert space. These are particular
cases of a theorem due to M. A. Naimark which states that every POV-measure
defined on a suitable measurable space can be obtained from a PV-measure on
a larger Hilbert space by means of a projection (see E. B. Davies, 1976,
and the references given there).

\section{\sloppy Continuous Resolutions Induced by Group Representations}

Specifying the space $M$ and the Borel measure $ \lambda $, we now
replace $M$ by a locally compact group $ \cal G $ satisfying the axiom
of second countability and $ \lambda $ by a left-invariant Haar measure
$ \mu $ on $ \cal G $. Again, we briefly write $ dg := \mu(dg) $.

The following theorem is due to the author, however, it is related to some
theorem in E. B. Davies' account (1976).

\begin{theorem}
Let $ g \mapsto U_g $ be an irreducible, strongly continuous projective
unitary representation of $ \cal G $ on $ \cal H $, $ u \in {\cal H} $
with $ \| u \| = 1 $, and $ u_g := U_g u $. If
$ \int |\langle u_g | \psi \rangle|^2 dg $ exists for all
$ \psi \in {\cal H} $ and is bounded in $ \psi $, i.e., if
\begin{equation}
\int |\langle u_g | \psi \rangle|^2 dg \leq C
\end{equation}
for $ \| \psi \| \leq 1 $, then the integral
\[
A := \int | u_g \rangle \langle u_g | \, dg
\]
exists in the weak sense, and $ A = \alpha 1 $ holds with some
$ \alpha > 0 $, i.e.\
\begin{equation}
1 = \frac{1}{\alpha} \int | u_g \rangle \langle u_g | \, dg \ .
\end{equation}
In particular, the family $ \{ u_g \}_{g \in {\cal G}} $ is then a continuous
resolution of the identity of $ \cal H $.
\end{theorem}
{\bf Proof:} By polarization, we obtain from (4.8) that the integral
\[
\Phi(\phi,\psi) := \int \langle \phi | u_g \rangle
                        \langle u_g | \psi \rangle \, dg
\]
exists and that $ \Phi $ is a bounded sesquilinear functional. Hence,
there exists a bounded linear operator $A$ satisfying
\[
\Phi(\phi,\psi) = \langle \phi | A\psi \rangle \ ;
\]
in particular, $A$ is just the weak integral
$ \int | u_g \rangle \langle u_g | \, dg $. Moreover, $A$ is a positive
self-adjoint operator.

For any $ h \in {\cal G} $, we have
\begin{eqnarray*}
\langle \psi | U_{h}AU^{\ast}_{h} \psi \rangle
& = & \langle U^{\ast}_{h} \psi | AU^{\ast}_{h} \psi \rangle
  =   \int |\langle \psi | U_{h}U_{g}u \rangle|^2 dg \\
& = & \int |\langle \psi | U_{hg}u \rangle|^2 \mu(dg)
  =   \int |\langle \psi | u_{hg} \rangle|^2 (\mu \circ H^{-1}) (dg)
\end{eqnarray*}
where $ H(g) := h^{-1}g $ and the invariance of $ \mu $ under left
translations has been used. It follows that
\[
\langle \psi | U_{h}AU^{\ast}_{h} \psi \rangle
 = \int |\langle \psi | u_{hH(g)} \rangle|^2 \mu(dg)
 = \int |\langle u_g | \psi \rangle|^2 dg
 = \langle \psi | A \psi \rangle
\]
holds for all $ \psi \in {\cal H} $. In consequence,
\[
U_h A = AU_h
\]
is valid for all $ h \in {\cal G} $. Since the projective representation
$ h \mapsto U_h $ is irreducible and $ A \geq 0 $, we obtain $ A = \alpha 1 $
with $ \alpha \geq 0 $. Since $ g \mapsto |\langle u_g | \psi \rangle|^2 $
is continuous, nonnegative, and positive for some $g$ and some $ \psi $,
$ \langle \psi | A\psi \rangle > 0 $ holds for some $ \psi $. This
implies $ \alpha > 0 $. \hfill $ \Box $\\

We remark that, for a fixed representation $ g \mapsto U_g $, it possibly
depends on the unit vector $u$ whether condition (4.8) is fulfilled or not. If
condition (4.8) is fulfilled for different unit vectors $u$, the number
$ \alpha $ may depend on $u$. In the case of the irreducible, strongly
continuous projective unitary representation $ (q,p) \mapsto U_{qp} $ of
the additive group $ \R^2 $ that will be discussed in the next chapter,
condition (4.8) is satisfied for all unit vectors $u$, and the number
$ \alpha $ does not depend on $u$; however, the observable (4.4) does.

Suppose that $ \cal G $ is even a compact second countable group, i.e.,
$ \cal G $ is a compact metrizable group. Then the Haar measure $ \mu $
is finite, and we can assume $ \mu({\cal G}) = 1 $. For every irreducible,
strongly continuous projective unitary representation $ g \mapsto U_g $
of $ \cal G $ on some Hilbert space $ \cal H $ and every unit vector $u$,
condition (4.8) is automatically fulfilled with $ C = 1 $, and
$ \{ u_g \}_{g \in {\cal G}} $ is a continuous resolution of the
identity. Assume $ \dim {\cal H} = \infty $. Taking account of Eq.\ (4.9)
and using a complete orthonormal system $ \{ \phi_i \}_{i \in \N} $ of
$ \cal H $, we obtain, for every $ n \in \N $,
\[
\alpha n = \sum_{i=1}^{n} \langle \phi_i | A\phi_i \rangle
         = \int \sum_{i=1}^{n} |\langle \phi_i | u_g \rangle|^2 \, dg
         \leq 1 \ ,
\]
which is a contradiction. Hence, $ \dim {\cal H} < \infty $ (compare
Corollary 4.3 and the reasoning (4.3)). In particular, we have shown
the well-known fact that the Hilbert space on which an irreducible,
strongly continuous projective unitary representation of a compact
second countable group is based must be finite-dimensional. Furthermore,
with $ \dim {\cal H} =: N $ and $ |u_g\rangle \langle u_g| = P_{u_g} $
it follows that
\[
\alpha N = {\rm tr} \, A = \int {\rm tr} \, P_{u_g} \, dg = 1
\]
and consequently $ \alpha = \frac{1}{N} $. Hence, Eq.\ (4.9) reads
\[
1 = N\int |u_g\rangle \langle u_g| \, dg \ ;
\]
in particular, for a compact second countable group, $ \alpha $ does not
depend on $u$.

Returning to the general situation of Theorem 4.6, we finally consider
the observable defined by $ \{ u_g \}_{g \in {\cal G}} $ according to
Proposition 4.4. This observable on $ ({\cal G},\Xi({\cal G})) $ is given by
\[
F(B) := \frac{1}{\alpha} \int_B | u_g \rangle \langle u_g | \, dg \ .
\]
An easy calculation using the left-invariance of $ \mu $ shows that
\begin{equation}
F(gB) = U_{g}F(B)U^{-1}_{g}
\end{equation}
holds for every $ g \in {\cal G} $ and every $ B \in \Xi({\cal G}) $. That is,
the observable $F$ transforms {\it covariant under the representation
$ g \mapsto U_g $ of $ \cal G $}.

\chapter{Joint Position-Momentum Observables}

According to Theorem 2.2, there exist single statistically complete
quantum observables, and according to Theorem 3.3, there exist classical
representations of quantum mechanics on some sample space $ \Omega $. The
special case of $ \Omega $ being the phase space is significant. An important
class of observables on phase space can, according to Proposition 4.4,
be obtained from a particular class of continuous resolutions of the
identity. Those so-called joint position-momentum observables are the
subject of this chapter; as will be discussed later, they are often
statistically complete, thus giving rise to classical representations
of quantum mechanics on phase space.

\section{\sloppy Approximate Observables for Position and Momentum}

Let $E$ be a PV-measure defined on the Borel sets $ \Xi(\R^N) $ of $ \R^N $. We
interpret $E$ as an ``ideal'' observable that does not take the imprecision
of real measurements into account. To involve this imprecision, we assume
that a real measuring apparatus for $E$ has an intrinsic inaccuracy which
can be described by a probability density $ \eta $ on $ \R^N $ in the
following sense. The probability that the apparatus indicates a value in
the set $ B \in \Xi(\R^N) $ for systems in the state $ W \in K({\cal H}) $
is
\begin{equation}
\int_B \int \eta(y-x) \, {\rm tr} \, WE(dx) \, d^N \! y
= \int \int \chi_B (y) \eta(y-x) \, d^N \! y \, {\rm tr} \, WE(dx)
\end{equation}
where $ d^N \! y := \lambda^N(dy) $ and $ \lambda^N $ is the $N$-dimensional
Lebesgue measure on $ \Xi(\R^N) $. Note that $ \eta \geq 0 $ a.e.\ and
$ \int \int \eta(y-x) \, d^N \! y \, {\rm tr} \, WE(dx) = 1 $. We call
$ \eta $ the {\it confidence function} of the measuring apparatus. The smaller
the variance of $ \eta $, the better the probability (5.1) approximates
$ {\rm tr} \, WE(B) $. The second integral of (5.1) can be written as
\[
\int (\chi_B * \eta(- \, . \, ))(x) \, {\rm tr} \, WE(dx)
= {\rm tr} \left( W \int \chi_B * \eta(- \, . \, ) \, dE \right)
\]
where $ \chi_B * \eta(- \, . \, ) $ is the convolution of the
characteristic function $ \chi_B $ and the reflection of $ \eta $,
$ 0 \leq \chi_B * \eta(- \, . \, ) \leq 1 $, and
$ \int \chi_B * \eta(- \, . \, ) \, dE $ exists in the $ \sigma $-weak
sense. Moreover, $G$ defined by
\begin{equation}
G(B) := \int \chi_B * \eta(- \, . \, ) \, dE
\end{equation}
is a normalized POV-measure on $ (\R^N,\Xi(\R^N)) $.

By construction, the probability distribution of the observable $G$ in
the state $W$ is given by
\begin{equation}
P^G_W(B) = {\rm tr} \, WG(B)
= \int_B \int \eta(y-x) \, {\rm tr} \, WE(dx) \, d^N \! y \ .
\end{equation}
Since
\[
P^G_W (B) \approx {\rm tr} \, WE(B) = P^E_W(B)
\]
holds where the approximation is good for a confidence function with small
variance, we call $G$ an {\it approximate $E$-observable with confidence
function $ \eta $} and interpret it as a realistic substitute for $E$.---The
idea of approximate observables in the sense of Eq.\ (5.2) is due to
E. B. Davies (1970, 1976); confidence functions in order to describe
``unsharp measurements'' were introduced by S. T. Ali and E. Prugove\v{c}ki
(1977a; cf.\ also E. Prugove\v{c}ki, 1984). The interpretation of approximate
observables in the sense of Eq.\ (5.1) was given in the paper of S. Gudder,
J. Hagler, and W. Stulpe (1988).

Now let $ N = 1 $. If the expectation values
$ \langle E \rangle_W = \int x P^E_W (dx) $ and
$ \langle \eta \rangle = \int y \eta (y)dy $ as well as the variances
$ {\rm var}_W E = \int (x - \langle E \rangle_W )^2 P^E_W (dx)
= \int x^2 P^E_W (dx) - \langle E \rangle^2_W $ and
$ {\rm var} \, \eta = \int (y - \langle \eta \rangle )^2 \eta (y)dy
= \int y^2 \eta (y)dy - \langle \eta \rangle^2 $ exist, then
$ \langle G \rangle_W = \int y P^G_W (dy) $ and
$ {\rm var}_W G = \int (y - \langle G \rangle_W )^2 P^G_W (dy)
= \int y^2 P^G_W (dy) - \langle G \rangle^2_W $ exist also where
\begin{equation}
\langle G \rangle_W = \langle E \rangle_W + \langle \eta \rangle
\end{equation}
and
\begin{equation}
{\rm var}_W G = {\rm var}_W E + {\rm var} \, \eta
\end{equation}
hold. Namely, using (5.3), we obtain
\begin{eqnarray*}
\langle G \rangle_W & = & \int y \int \eta (y-x) P^E_W (dx) dy    \\
                    & = & \int \int (x+y) \eta (y) dy P^E_W (dx)  \\
                    & = & \int x P^E_W (dx) + \int y \eta (y)dy   \\
                    & = & \langle E \rangle_W + \langle \eta \rangle
\end{eqnarray*}
and
\begin{eqnarray*}
{\rm var}_W G & = & \int y^2 \int \eta (y-x) P^E_W (dx) dy
                                                    - \langle G \rangle^2_W \\
              & = & \int \int (x+y)^2 \eta (y) dy P^E_W (dx)
                                                    - \langle G \rangle^2_W \\
              & = & \int x^2 P^E_W (dx) + \int y^2 \eta (y)dy
                         + 2 \langle E \rangle_W \langle \eta \rangle
                         - ( \langle E \rangle_W + \langle \eta \rangle )^2 \\
              & = & {\rm var}_W E + {\rm var} \, \eta \ .
\end{eqnarray*}

Finally, consider the Hilbert space
$ {\cal H} := L^{2}_{\C} \, (\R^N,d^N \! x)
= L^2_{\C} \, (\R^N,\Xi(\R^N), $ $ \lambda^N) $. Then an
{\it $N$-dimensional position observable $ E^Q $}, as a PV-measure on
$ \Xi(\R^N) $, can be defined by
\begin{equation}
E^Q (B) \psi := \chi_B \psi
\end{equation}
where $ B \in \Xi(\R^N) $ and $ \psi \in {\cal H} $. An {\it $N$-dimensional
momentum observable $ E^P $} is given by
\begin{equation}
E^P (B) := F^{-1}E^Q(B)F
\end{equation}
where $F$ denotes the Fourier transformation in
$ {\cal H} = L^{2}_{\C} \, (\R^N,d^N \! x) $. In the case of $ E^Q $, $ \R^N $
is interpreted as {\it configuration space}, whereas in the case of $ E^P $,
$ \R^N $ is interpreted as {\it momentum space}. According to
\begin{equation}
G^Q (B) := \int \chi_B * \eta^Q (- \, . \, ) \, dE^Q
\end{equation}
an {\it approximate $N$-dimensional position observable with confidence
function $ \eta^Q $} is defined and according to
\begin{equation}
G^P (B) := \int \chi_B * \eta^P (- \, . \, ) \, dE^P
\end{equation}
an {\it approximate $N$-dimensional momentum observable with confidence
function $ \eta^P $}.

\section{Joint Position-Momentum Observables Generated by Wave Functions}

The following theorem is the basis for the definition of joint
position-momen-tum observables; we presuppose that
$ {\cal H} = L^{2}_{\C} \, (\R^N,d^N \! x) $.

\begin{theorem}
For each $ q = (q_1,\ldots,q_N) \in \R^N $ and each
$ p = (p_1,\ldots,p_N) \in \R^N $, define a unitary operator by
\begin{equation}
U_{qp} := \left( \prod_{j=1}^{N} e^{ip_j Q_j} \right)
          \left( \prod_{j=1}^{N} e^{-iq_j P_j} \right)
        = \prod_{j=1}^{N} e^{ip_j Q_j} e^{-iq_j P_j}
\end{equation}
where $ Q_j $ and $ P_j $ are the usual operators for position and
momentum. Let $ u \in {\cal H} $ be any function of norm $1$ and define
$ u_{qp} := U_{qp}u $, i.e.
\[
u_{qp} (x) := e^{ip \cdot x} u(x-q) \ .
\]
Then $ (q,p) \mapsto u_{qp} $ is a norm-continuous map on $ \R^{2N} $, and
\begin{equation}
1 = \frac{1}{(2\pi)^N} \int |u_{qp}\rangle \langle u_{qp} |
                                                 \, d^N \! q \, d^N \! p
\end{equation}
holds where the integral is understood in the weak sense. That is,
for every $ u \in {\cal H} $ with $ \|u\| = 1 $, the family
$ \{ u_{qp} \}_{(q,p) \in \R^{2N}} $ is a continuous resolution of
the identity of $ \cal H $ with normalization constant $ \alpha = (2\pi)^N $.
\end{theorem}
{\bf Proof:} The norm-continuity of $ (q,p) \mapsto u_{qp} = U_{qp}u $
follows easily from the fact that $ q_j \mapsto e^{-iq_j P_j} $ and
$ p_j \mapsto e^{-ip_j Q_j} $ are strongly continuous one-parameter
groups of unitary operators. To prove Eq.\ (5.11), let $ \psi \in {\cal H} $
be arbitrary. By Fubini's theorem, we obtain
\begin{equation}
\begin{array}{ccl}
\langle \psi | 1\psi \rangle & = & \|\psi\|^2 \|u\|^2
  =   \displaystyle{\int |\psi(q)|^2 \, d^N \! q \int |u(x)|^2 \, d^N \! x}
                                            \vspace{2.4mm}\\
& = & \displaystyle{\int |u(x)|^2 \int |\psi(x+q)|^2 \, d^N \! q \, d^N \! x}
                                            \vspace{2.4mm}\\
& = & \displaystyle{\int \int |u(x)|^2 |\psi(x+q)|^2 \, d^N \! x \, d^N \! q}
                                            \vspace{2.4mm}\\
& = & \displaystyle{\int \int |\overline{u}(x-q)\psi(x)|^2 \,
                                                    d^N \! x \, d^N \! q \ ;}
\end{array}
\end{equation}
in particular,
$ \overline{u}( \, . \, - q)\psi \in L^{2}_{\C} \, (\R^N,d^N \! x) $
for almost all $ q \in \R^N $. Furthermore,
$ \overline{u}( \, . \, - q)\psi \in L^{1}_{\C} \, (\R^N,d^N \! x) $
for all $ q \in \R^N $. Denoting the unitary operator of Fourier
transformation by $F$, it follows that
\begin{eqnarray}
F(\overline{u}( \, . \, - q)\psi)(p)
& = & \frac{1}{\sqrt{(2\pi)^N}} \int e^{-ip \cdot x} \overline{u}(x-q)\psi(x)
                                                    \, d^N \! x  \nonumber \\
& = & \frac{1}{\sqrt{(2\pi)^N}} \int \overline{u_{qp}(x)}\psi(x) \, d^N \! x \\
& = & \frac{1}{\sqrt{(2\pi)^N}} \langle u_{qp} | \psi \rangle    \nonumber
\end{eqnarray}
and
\begin{equation}
\int |\overline{u}(x-q)\psi(x)|^2 \, d^N \! x
 = \int |F(\overline{u}( \, . \, -q)\psi)(p)|^2 \, d^N \! p
\end{equation}
hold for almost all $q$. Now, (5.12), (5.14), and (5.13) imply
\begin{eqnarray*}
\langle \psi | 1\psi \rangle
& = & \int \int \frac{1}{(2\pi)^N} \, |\langle u_{qp} | \psi \rangle |^2
                                               \, d^N \! p \, d^N \! q \\
& = & \frac{1}{(2\pi)^N} \int |\langle u_{qp} | \psi \rangle |^2
                                               \, d^N \! q \, d^N \! p \\
& = & \left\langle \psi \, \left| \, \frac{1}{(2\pi)^N} \int | u_{qp}
      \rangle \langle u_{qp} | \, d^N \! q \, d^N \! p \right. \, \psi
                                                       \right\rangle \ .
\end{eqnarray*}
\hspace*{\fill} $ \Box $\\

Since $ (q,p) \mapsto U_{qp} $ is an irreducible, strongly continuous
projective unitary representation of the additive group $ \R^{2N} $ with
Haar measure $ \lambda^{2N} $, one could, without using Theorem 5.1, try
to apply Theorem 4.6 to the family $ \{ u_{qp} \}_{(q,p) \in \R^{2N}} $. Then
one had to prove condition (4.8). Actually, Theorem~5.1 states that, for
every unit vector $u$, (4.8) is satisfied and (4.9) is valid with
$ \alpha = (2\pi)^N $. We remark further that the statement of Theorem 5.1
holds whenever $ U_{qp} $ is defined according to (5.10) by operators
$ Q'_j $ and $ P_j' $ in $ {\cal H}' $ being unitarily equivalent to $ Q_j $
and $ P_j $ in $ \cal H $, i.e., whenever $ Q'_j $ and $ P'_j $ give rise to
an irreducible representation of Weyl's commutation relations. Finally, the
operators $ U_{qp} $ are essentially the Weyl operators which are usually
defined according to
\begin{equation}
U^{W}_{qp} := e^{i(q \cdot P + p \cdot Q)}
  = e^{i\frac{q \cdot p}{2}} \prod_{j=1}^{N} e^{ip_jQ_j} e^{iq_jP_j} \ ;
\end{equation}
in fact,
\[
U^{W}_{qp} = e^{i\frac{q \cdot p}{2}} U_{-q,p} \ ,
\]
respectively,
\[
U_{qp} = e^{i\frac{q \cdot p}{2}} U^{W}_{-q,p} \ .
\]
Analogously to the operators $ U_{qp} $, $ (q,p) \mapsto U^{W}_{qp} $ is an
irreducible, strongly continuous projective representation of $ \R^{2N} $, and
$ \{ u^{W}_{qp} \}_{(q,p) \in \R^{2N}} $ with $ u^{W}_{qp} := U^{W}_{qp}u $
and $ \|u\| = 1 $ is a continuous resolution of the identity.

According to Proposition 4.4, the continuous resolution \vspace{1pt}
$ \{ u_{qp} \}_{(q,p) \in \R^{2N}} $ introduced in Theorem 5.1 defines
an observable $F$ on $ (\R^{2N},\Xi(\R^{2N})) $ by
\begin{equation}
F(B) := \frac{1}{(2\pi)^N} \int_{B} |u_{qp}\rangle \langle u_{qp} |
                                                   \, d^N \! q \, d^N \! p \ .
\end{equation}
The probability distribution of $F$ in a state $ W \in K({\cal H}) $
is given by
\[
P^F_W (B) = \frac{1}{(2\pi)^N} \int_{B} \langle u_{qp} | W u_{qp} \rangle
                                                  \, d^N \! q \, d^N \! p \ ;
\]
$ P^F_W $ has the bounded continuous probability density $ (q,p) \mapsto
\rho(q,p) := \frac{1}{(2\pi)^N} \, \langle u_{qp} | W u_{qp} \rangle $.

The physical interpretation of the observable (5.16) is based on the fact
that its marginal observables are approximate observables for position and
momentum. The marginal observables of $F$ are defined by
\begin{eqnarray}
F^Q (b) & := & F(b \times \R^N) \\
F^P (b) & := & F(\R^N \times b)
\end{eqnarray}
where $ b \in \Xi(\R^N) $. From (5.17) and (5.16) it follows by means of
(5.13) and (5.14) that
\begin{eqnarray*}
\langle \psi | F^{Q} (b) \psi \rangle
& = & \frac{1}{(2\pi)^N} \int_{b \times \R^N} |\langle u_{qp} | \psi \rangle|^2
                                              \, d^N \! q \, d^N \! p       \\
& = & \int_{b} \int | \overline{u}(x-q)\psi(x) |^2 \, d^N \! x \, d^N \! q  \\
& = & \int \int \chi_{b} (q) |u(x-q)|^2 \, d^N \! q \, |\psi(x)|^2
                                                          \, d^N \! x       \\
& = & \int (\chi_{b} * |u|^2)(x) \, \langle \psi | E^Q (dx) \psi \rangle    \\
& = & \left\langle \psi \, \left| \, \int \chi_{b} * |u|^2 \, dE^Q
                           \right. \, \psi \right\rangle \ .
\end{eqnarray*}
Thus,
\begin{equation}
F^{Q} (b) = \int \chi_{b} * |u|^2 \, dE^Q \ ,
\end{equation}
and, according to (5.8), $ F^Q $ is an approximate position observable
with confidence function $ |u(- \, . \, )|^2 $. To derive the corresponding
result for $ F^P $, let $ \tilde{u} $ and $ \tilde{\psi} $ be the Fourier
transforms of $u$ and $ \psi $, respectively. A calculation analogous to
(5.12) shows that $ \overline{\tilde{u}} ( \, . \, -p) \tilde{\psi} \in
L^{1}_{\C} \, (\R^N,d^N \! k) \cap L^{2}_{\C} \, (\R^N,d^N \! k) $ for
almost all $ p \in \R^N $. Then
\begin{eqnarray}
\frac{1}{\sqrt{(2\pi)^N}} \langle u_{qp} | \psi \rangle
& = & \frac{1}{\sqrt{(2\pi)^N}} \langle \widetilde{u_{qp}} |
                                        \tilde{\psi} \rangle     \nonumber\\
& = & \frac{1}{\sqrt{(2\pi)^N}} \int e^{iq \cdot k}
               \overline{\tilde{u}} (k-p) \tilde{\psi} (k) \, d^N \! k    \\
& = & F( \overline{\tilde{u}} ( \, . \, -p) \tilde{\psi})(-q)    \nonumber
\end{eqnarray}
holds for almost all $p$ where
$ \widetilde{u_{qp}}(k) = e^{-iq \cdot k} \tilde{u}(k-p) $
and $F$ is the Fourier transformation in
$ L^{2}_{\C} \, (\R^N,d^N \! k) $. Furthermore,
\begin{equation}
\int |F(\overline{\tilde{u}}( \, . \, -p)\tilde{\psi})(q)|^2 \, d^N \! q
= \int |\overline{\tilde{u}}(k-p)\tilde{\psi}(k)|^2 \, d^N \! k
\end{equation}
is valid for almost all $p$. Now, using (5.18), (5.16), (5.20), and
(5.21), we obtain
\begin{eqnarray*}
\langle \psi | F^{P} (b) \psi \rangle
& = & \frac{1}{(2\pi)^N} \int_{\R^N \times b} |\langle u_{qp} | \psi \rangle|^2
                                                 \, d^N \! q \, d^N \! p \\
& = & \int_{b} \int | F(\overline{\tilde{u}}( \, . \, -p)\tilde{\psi})(q) |^2
                                                 \, d^N \! q \, d^N \! p \\
& = & \int_{b} \int | \overline{\tilde{u}}(k-p)\tilde{\psi}(k) |^2
                                                 \, d^N \! k \, d^N \! p \\
& = & \int \int \chi_{b} (p) |\tilde{u}(k-p)|^2 \, d^N \! p \,
                                         |\tilde{\psi}(k)|^2 \, d^N \! k \\
& = & \int (\chi_{b} * |\tilde{u}|^2)(k) \, \langle \psi |
                                           E^P (dk) \psi \rangle \ .
\end{eqnarray*}
Hence,
\begin{equation}
F^{P} (b) = \int \chi_{b} * |\tilde{u}|^2 \, dE^P \ ,
\end{equation}
and, according to (5.9), $ F^P $ is an approximate momentum observable
with confidence function $ |\tilde{u}(- \, . \, )|^2 $.

The approximate position observable $ F^Q $ and the approximate
momentum observable $ F^P $ can be measured jointly (simultaneously) by
one measuring apparatus representing the observable $F$. We call $F$
a {\it joint position-momentum observable} and interpret $ \R^{2N} $ as
{\it phase space}. According to the terminology of G. Ludwig (1983), $ F^Q $
and $ F^P $ are {\it coexistent} approximate observables for position and
momentum. However, the better $ F^Q $ approximates $ E^Q $, the worse
$ F^P $ approximates $ E^P $, and conversely.

Joint position-momentum observables were, on the basis of Theorem 5.1,
introduced by E. B. Davies (1970, 1976) and A. S. Holevo (1973). Although
attempts to describe joint position-momentum measurements in quantum mechanics
had already been made earlier, the aproaches of E. B. Davies and A. S. Holevo
have been the first rigorous ones.

\section{Joint Position-Momentum Observables Generated by Density Operators}

The type of joint position-momentum observables introduced in the preceding
section can be generalized. Namely, in Theorem (5.1) and Eq.\ (5.16),
the unit vector $u$ can be replaced by an arbitrary density operator
$ a \in K({\cal H}) $. Defining
\begin{equation}
a_{qp} := U_{qp}aU^{\ast}_{qp}
        = \sum_{i=1}^{\infty} \lambda_{i} | u_{i,qp} \rangle \langle u_{i,qp} |
\end{equation}
with $ U_{qp} $ according to (5.10),
$ a = \sum_{i=1}^{\infty} \lambda_i | u_i \rangle \langle u_i | $,
$ \lambda_i \geq 0 $, $ \sum_{i=1}^{\infty} \lambda_i = 1 $,
$ \| u_i \| = 1 $, and $ u_{i,qp} := U_{qp}u_i $, we have
that $ (q,p) \mapsto a_{qp} $ is a trace-norm continuous map
on $ \R^{2N} $ satisfying
\begin{equation}
1 = \frac{1}{(2\pi)^N} \int a_{qp} \, d^N \! q \, d^N \! p \ .
\end{equation}
Moreover, an observable $F$ on $ (\R^{2N},\Xi(\R^{2N})) $ is defined by
the weak integral
\begin{equation}
F(B) := \frac{1}{(2\pi)^N} \int_{B} a_{qp} \, d^N \! q \, d^N \! p \ .
\end{equation}
This definition of an observable is a consequence of Eq.\ (5.24) which
itself follows immediately from (5.11). The statement on the trace-norm
continuity, however, needs a proof.

To that end, consider any $ (q,p),(q_0,p_0) \in \R^{2N} $. Then, for
a sufficiently large $ n \in \N $,
\begin{eqnarray*}
\| a_{qp} - a_{q_{0}p_{0}} \|_{\rm tr}
& = & \left\| \sum_{i=1}^{\infty} \lambda_{i}
                                    (| u_{i,qp} \rangle \langle u_{i,qp} |
  - | u_{i,q_{0}p_{0}} \rangle \langle u_{i,q_{0}p_{0}} |) \right\|_{\rm tr} \\
& < & \left\| \sum_{i=1}^{n} \lambda_{i} (| u_{i,qp} \rangle \langle u_{i,qp} |
  - | u_{i,q_{0}p_{0}} \rangle \langle u_{i,q_{0}p_{0}} |) \right\|_{\rm tr}
      + \frac{\epsilon}{2} \\
& \leq & 4 \sum_{i=1}^{n} \lambda_{i} \| u_{i,qp} - u_{i,q_{0}p_{0}} \|
      + \frac{\epsilon}{2}
\end{eqnarray*}
holds where Lemma 3.5 has been used. The norm-continuity of
$ (q,p) \mapsto u_{qp} $ implies the existence of
a $ \delta (\epsilon;q_0,p_0) $ such that
$ \| u_{i,qp} - u_{i,q_{0}p_{0}} \| < \frac{\epsilon}{8} $ for
$ \| (q,p) - (q_0,p_0) \| < \delta (\epsilon;q_0,p_0) $ and
each $ i = 1,\ldots,n $. Hence,
\[
\| a_{qp} - a_{q_{0}p_{0}} \|_{\rm tr} < \frac{\epsilon}{2} \sum_{i=1}^{n}
                           \lambda_{i} + \frac{\epsilon}{2} \leq \epsilon
\]
for $ \| (q,p) - (q_0,p_0) \| < \delta (\epsilon;q_0,p_0) $, and
$ (q,p) \mapsto a_{qp} $ is trace-norm continuous.

Analogously to the integrals (5.11) and (5.16), the integrals (5.24) and
(5.25) do exist also in the $ \sigma $-weak sense. This follows from the
conclusion leading to Eq.\ (4.5) when the one-dimensional projection
$ P_{u_x} $, respectively, $ P_{u_{qp}} $ is replaced by the density operator
$ a_{qp} $. In particular, the probability distribution of the observable
(5.25) in a state $ W \in K({\cal H}) $ is given by
\[
P^F_W (B) = \frac{1}{(2\pi)^N} \int_{B} {\rm tr} \, Wa_{qp}
                                                 \, d^N \! q \, d^N \! p
\]
where $ (q,p) \mapsto \rho(q,p) := \frac{1}{(2\pi)^N} \, {\rm tr} \, Wa_{qp} $
is a bounded continuous probability density. Note that, independently of
$ W \in K({\cal H}) $, $ \rho(q,p) \leq \frac{1}{(2\pi)^N} $ for all
$ (q,p) \in \R^{2N} $.

From Eqs.\ (5.23), (5.19), and (5.22), we obtain that the marginal
observables of the observable (5.25) are given by
\begin{equation}
\begin{array}{ccc}
F^Q (b) & = & {\displaystyle \int \chi_b * \sum_{i=1}^{\infty}
                             \lambda_i | u_i |^2 \, dE^Q} \vspace{0.3cm}\\
F^P (b) & = & {\displaystyle \int \chi_b * \sum_{i=1}^{\infty}
                             \lambda_i | \tilde{u}_i |^2 \, dE^P}
\end{array}
\end{equation}
where $ b \in \Xi(\R^N) $. Again, $ F^Q $ and $ F^P $ are approximate
observables for position and momentum, respectively, and $F$ can be
interpreted as a joint position-momentum observable.

Next we consider the transformation properties of the observable (5.25) and
the special case (5.16). According to Eq.\ (4.10), the observable (5.16)
transforms covariant under the projective representation
$ (q,p) \mapsto U_{qp} $ of $ \R^{2N} $, i.e.,
\begin{equation}
F(B + (q_0,p_0)) = U_{q_0p_0}F(B)U^{-1}_{q_0p_0}
\end{equation}
holds for every $ (q_0,p_0) \in \R^{2N} $ and every
$ B \in \Xi(\R^{2N}) $. Eq.\ (5.27) is also valid for the general observable
(5.25) and is in particular a consequence of the invariance of the Lebesgue
measure $ \lambda^{2N} $ under translations. Since $ \lambda^{2N} $ is also
invariant under rotations and reflections, there is a further covariance
property of (5.25), provided that the density operator $a$ is spherically
symmetric. The latter means that
\[
a = U_R aU^{\ast}_{R}
\]
holds for all orthogonal matrices $ R \in O(N) $ where $ R \mapsto U_R $
is the unitary representation of the orthogonal group $ O(N) $ defined by
\[
(U_R \psi)(x) := \psi(R^{-1}x) \ ,
\]
$ \psi \in {\cal H} = L^{2}_{\C} \, (\R^N,d^N \! x) $. The spherical symmetry
of $a$ and the obvious relation
\[
U_{Rq,Rp} = U_R U_{qp}U^{-1}_{R}
\]
imply that
\begin{equation}
a_{Rq,Rp} = U_R a_{qp}U^{\ast}_{R}
\end{equation}
for all $ (q,p) \in \R^{2N} $ and all $ R \in O(N) $. From the invariance
of $ \lambda^{2N} $ under $ O(N) $ and Eq.\ (5.28) it then follows that
the observable (5.25) transforms covariant under the representation
$ R \mapsto U_R $, i.e.,
\begin{equation}
F(RB) = U_R F(B)U^{-1}_{R}
\end{equation}
is valid for every $ R \in O(N) $ and every $ B \in \Xi(\R^{2N}) $ where
$ RB := \{ (Rq,Rp) \, | $ \linebreak $ (q,p) \in B \} $.

We remark that the spherical symmetry of the unit vectors $ u_i $ is
sufficient for the spherical symmetry of the density operator
$ a = \sum_{i=1}^{\infty} \lambda_i |u_i\rangle \langle u_i| $, but
not necessary. A vector $ u \in {\cal H} = L^{2}_{\C} \, (\R^N,d^N \! x) $
is spherically symmetric if $ u = U_Ru $ for all $ R \in O(N) $;
equivalently, $ u(x) = u(Rx) $ for all $ x \in \R^N $ and all $ R \in O(N) $,
respectively, $ u(x) = \hat{u}(\|x\|) $ where $ \|x\| $ denotes the Euclidean
norm of $x$. Analogously to (5.28), a spherical symmetric vector $u$ satisfies
$ u_{Rq,Rp} = U_R u_{qp} $.---Summarizing, joint position-momentum observables
generated by spherically symmetric density operators have the transformation
properties (5.27) and (5.29), that is, they transform covariant under
Galilei transformations.

Our joint position-momentum observables are particular instances of
more general {\it phase-space observables}, i.e., of observables on
$ (\R^{2N},\Xi(\R^{2N})) $ having the covariance property (5.27). The paper
of S. T. Ali and E. Prugove\v{c}ki (1977a) provides information on such more
general observables; we give some remarks. The covariance property (5.27)
of a general phase-space observable $F$ implies that its probability
distributions $ P^F_W $ are absolutely continuous with respect to
Lebesgue measure, i.e., each $ P^F_W $ has a probability density
$ \rho $. An arbitrary observable $F$ on $ (\R^{2N},\Xi(\R^{2N})) $
is said to have an {\it operator-valued density}
$ (q,p) \mapsto \alpha(q,p) \in {\cal B}_s({\cal H}) $ if
\[
F(B) = \int_B \alpha(q,p) \, d^N \! q \, d^N \! p
\]
holds for all $ B \in \Xi(\R^{2N}) $ where $ \alpha(q,p) \geq 0 $ for almost
all $ (q,p) \in \R^{2N} $ and the map $ \alpha $ is weakly integrable. If
a phase-space observable possesses an operator-valued density, then $ \alpha $
is necessarily of the form $ \alpha(q,p) = U_{qp}AU^{-1}_{qp} $ with $A$
being a positive bounded self-adjoint operator; in particular, $ \alpha $
is strongly as well as $ \sigma $-continuous. Conversely, if for some
$ A \in {\cal B}_s({\cal H}) $ with $ A \geq 0 $,
$ (q,p) \mapsto \alpha(q,p) := U_{qp}AU^{-1}_{qp} $ is
the operator-valued density of an observable $F$, then $F$ is
a phase-space observable. Furthermore, a phase-space observable with
an operator-valued density has the covariance property (5.29) if and only if
$A$ is a spherically symmetric operator, i.e. $ A = U_R AU^{-1}_{R} $ for
all $ R \in O(N) $. Finally, a phase-space observable $F$ possesses an
operator-valued density if and only if all its probability distributions
$ P^F_W $ can be represented by continuous functions.

There are phase-space observables with an operator-valued density which
are generated by an operator $ A \in {\cal B}_s({\cal H}) $ that is not
a positive trace-class operator. The property $ A = \frac{1}{(2\pi)^N} \, a $
with $ a \in K({\cal H}) $ guarantees \vspace{2pt} that the generated
phase-space observable is a joint position-momentum observable, i.e.,
the marginal observables are approximate observables for position and
momentum. Thus, the observables (5.25) play a distinguished role under
all phase-space observables. For a broad discussion of the physical aspects
of joint position-momentum measurements based on Eqs.\ (5.25) and (5.26), it
seems to be mandatory to mention also the work of F. E. Schroeck, Jr.\ (1981),
P. Busch and P. J. Lahti (1984), P. Busch (1985, 1987), and
P. Busch, M. Grabowski, and P. J. Lahti (1995).

\section{An Uncertainty Relation}

We conclude this chapter with the discussion of an uncertainty relation
for joint position-momentum measurements on the basis of Eqs.\ (5.25) and
(5.26) for $ N = 1 $ (cf.\ S. Gudder, J. Hagler, and W. Stulpe, 1988). We
assume that the expectation values and variances of the ideal observables
$ E^Q $ and $ E^P $ in the state $ W \in K({\cal H}) $ as well as
the expectation values and variances of the confidence functions
$ \eta^Q := \sum_{i=1}^{\infty} \lambda_i | u_i (- \, . \, ) |^2 $ and
$ \eta^P := \sum_{i=1}^{\infty} \lambda_i | \tilde{u}_i (- \, . \, ) |^2 $
exist. From
\begin{eqnarray*}
{\rm var} \, \eta^Q
& = & \int y^2 \sum_{i=1}^{\infty} \lambda_i | u_i (y) |^2 \, dy - \left(
      \int y \sum_{i=1}^{\infty} \lambda_i | u_i (y) |^2 \, dy \right)^2  \\
& = & \int y^2 \, {\rm tr} \, aE^Q(dy) - \left(
      \int y \, {\rm tr} \, aE^Q(dy) \right)^2                            \\
& = & {\rm var}_a E^Q \ ,
\end{eqnarray*}
\[
{\rm var} \, \eta^P = {\rm var}_a E^P \ ,
\]
and the usual uncertainty relation it follows that
\begin{equation}
{\rm var} \, \eta^Q \, {\rm var} \, \eta^P \geq \frac{1}{4} \ .
\end{equation}
Moreover,
\begin{equation}
{\rm var}_W E^Q \, {\rm var}_W E^P \geq \frac{1}{4}
\end{equation}
holds. For numbers $ a_1,a_2,b_1,b_2 \geq 0 $, an application of the Schwarz
inequality yields
\[
(a_1 + a_2) (b_1 + b_2)
        \geq \left( \sqrt{a_1}\sqrt{b_1} + \sqrt{a_2}\sqrt{b_2} \right)^2 \ .
\]
If $ a_1 b_1 \geq \frac{1}{4} $ and $ a_2 b_2 \geq \frac{1}{4} $, we have
\begin{equation}
(a_1 + a_2) (b_1 + b_2) \geq 1 \ .
\end{equation}
Thus, from Eq.\ (5.5) and (5.30) -- (5.32) we obtain
\begin{equation}
{\rm var}_W F^Q \, {\rm var}_W F^P \geq 1
\end{equation}
which is an uncertainty relation for the coexistent approximate observables
for position and momentum that are related by the joint position-momentum
observable $F$.

The usual uncertainty relation
\begin{equation}
\Delta_W E^Q \, \Delta_W E^P \geq \frac{1}{2}
\end{equation}
for the complementary observables $ E^Q $ and $ E^P $ states that it is
impossible to prepare a state (i.e.\ a statistical ensemble)
$ W \in K({\cal H}) $ such that the product of the standard deviations
of $ E^Q $ and $ E^P $ is smaller than $ \frac{1}{2} $. In particular,
this lower bound can be attained. However, (5.34) refers to ideal
measurements of $ E^Q $ and $ E^P $ and does not take into account any
measuring inaccuracy. The imprecision of real measuring apparata for position
and momentum is involved in the approximate observables $ G^Q $ and $ G^P $
according to (5.8) and (5.9). From (5.5) it follows that
\[
\Delta_W G^Q \, \Delta_W G^P > \frac{1}{2} \ .
\]
This inequality refers to real separate measurements of position and
momentum. Since the variances of the independent confidence functions
$ \eta^Q $ and $ \eta^P $ can be made arbitrarily small, the uncertainty
product $ \Delta_W G^Q \, \Delta_W G^P $ can, for suitable states $W$,
approach its lower bound arbitrarily closely. For joint position-momentum
measurements, the confidence functions become dependent, and their variances
cannot simultaneously be made arbitrarily small; inequality (5.33) implies
\[
\Delta_W F^Q \, \Delta_W F^P \geq 1 \ .
\]
That is, the lower bound of the uncertainty product for joint position-
\linebreak momentum measurements is twice as large as that for separate
ideal measurements. This fact seems to be fundamental since it is due to
the approximation of complementary observables by coexistent ones.

\chapter{Phase-Space Representations of Quantum Mechanics}

For systems with configuration space $ \R^N $, the continuous resolutions
of the identity introduced in Theorem 5.1 give rise to two formulations
of quantum mechanics that are related to the classical phase space
$ \R^{2N} $. The first formulation is based on Theorem 4.1, it refers to
a Hilbert space of wave functions on phase space and is unitarily equivalent
to the usual configuration space representation. The second formulation
is based on statistically complete joint position-momentum observables and
concerns the reformulation of quantum mechanics in terms of probability
densities and functions on phase space. In this chapter, we discuss the
first formulation; the second one will be discussed in the next chapter.

\section{\sloppy Hilbert Spaces of Continuous Wave Functions on Phase Space}

Let $ {\cal H} = L^{2}_{\C} \, (\R^N,d^N \! x) $. According to Theorem 5.1,
every unit vector $ u \in {\cal H} $ defines a continuous resolution
$ \{ u_{qp} \}_{(q,p) \in \R^{2N}} $ of the identity of $ \cal H $ by
$ u_{qp} := U_{qp}u $. Moreover, according to Theorem 4.1, an isometry
$ V \!\! : {\cal H} \rightarrow L^{2}_{\C} \, (\R^{2N},d^N \! q \, d^N \! p) $
is defined by
\begin{equation}
(V\psi)(q,p) := \frac{1}{\sqrt{(2\pi)^N}} \langle u_{qp} | \psi \rangle
              =: \Psi(q,p) \ .
\end{equation}
The Hilbert space $ V{\cal H} $ consists of bounded continuous,
square-integrable functions on the phase space $ \R^{2N} $ and is isomorphic
to $ \cal H $. Corresponding to a terminology introduced in Section 4.1,
$ V{\cal H} $ gives rise to an {\it $ \R^{2N} $-representation} or
{\it phase-space representation of quantum mechanics}.

In physical terms, $ V{\cal H} $ is a Hilbert space of wave functions
on phase space, the normalized wave functions describing pure states. To
obtain an interpretation for the probability density $ |\Psi|^2 $ of
a wave function $ \Psi \in V{\cal H} $, $ \|\Psi\| = 1 $, consider the joint
position-momentum observable $F$ introduced in Section 5.2. This observable
is based on $ \cal H $ and is given by
\begin{equation}
F(B) = \frac{1}{(2\pi)^N} \int_B | u_{qp} \rangle \langle u_{qp} |
                                          \, d^N \! q \, d^N \! p
\end{equation}
where $ B \in \Xi(\R^{2N}) $. In a state $ W \in K({\cal H}) $, the
probability distribution of $F$ reads
\[
P^F_W (B) = \frac{1}{(2\pi)^N} \int_B \langle u_{qp} | Wu_{qp} \rangle
                                          \, d^N \! q \, d^N \! p
          = \int_B \rho (q,p) \, d^N \! q \, d^N \! p \ ,
\]
$ (q,p) \mapsto \rho(q,p)
= \frac{1}{(2\pi)^N} \, \langle u_{qp} | Wu_{qp} \rangle $
being a bounded continuous probability density. For a pure state
$ W = P_{\psi} = |\psi\rangle \langle\psi| $ we obtain
\[
\rho(q,p) = \frac{1}{(2\pi)^N} \, |\langle u_{qp} | \psi \rangle|^2
          = |\Psi(q,p)|^2
\]
where $ \Psi = V\psi $. That is, $ |\Psi|^2 $ is the probability density
of the joint position-momentum observable $F$ in the pure state
$ \Psi = V\psi $.

We stress two properties of our phase-space representations. First,
the Hilbert space $ V{\cal H} $ of wave functions on phase space is
a proper subspace of the Hilbert space
$ L^{2}_{\C} \, (\R^{2N},d^N \! q \, d^N \! p) $
of all square-integrable functions on phase space. Second, for
$ \Psi \in V{\cal H} $ with $ \|\Psi\| = 1 $, $ |\Psi|^2 $ is
a probability density on phase space whose marginal densities are only
approximations for the quantum probability densities of position and
momentum. The first property is a consequence of the continuity of the
wave functions $ \Psi \in V{\cal H} $ (and also, if the observable (6.2)
is statistically complete, a consequence thereof; cf.\ Proposition 4.5),
whereas the second one is related to the fundamental fact that in quantum
mechanics position and momentum are non-coexistent observables.

\section{\sloppy Hilbert Spaces of Infinitely Differentiable Wave Functions
on Phase Space}

In the next section we investigate the operators of position and momentum
in phase-space representation, i.e., we derive explicit expressions for the
operators $ VQ_j V^{-1} $ and $ VP_j V^{-1} $ where $ Q_j $ and $ P_j $ are
the usual operators of position and momentum, respectively, and $V$
defined by (6.1) is understood as the unitary operator
$ V \!\! : {\cal H} \rightarrow V{\cal H} $. To that end, the following
two lemmata are needed which, moreover, enable us to discover a class of
Hilbert spaces of infinitely differentiable functions. The results of
this section, in particular Theorem 6.3, are due to the author.

\begin{lemma}
Let $ v \in S_{\C} \, (\R^N) $ be a Schwartz function, i.e.\ an
infinitely differentiable function of rapid decrease, and let
$ v_{qp}(x) := e^{ip \cdot x} v(x-q) $, $ q,p \in \R^N $. Then,
for every $ \psi \in {\cal H} $,
\[
\int \frac{\partial}{\partial q_j} \overline{v_{qp}(x)} \psi(x) \, d^N \! x
= \frac{\partial}{\partial q_j} \int \overline{v_{qp}(x)} \psi(x) \, d^N \! x
\]
($ j = 1,\ldots,N $) holds.
\end{lemma}
{\bf Proof:} We have
\begin{eqnarray}
\frac{\partial}{\partial q_j} \overline{v_{qp}(x)}
& = & \frac{\partial}{\partial q_j} (e^{-ip \cdot x} \overline{v}(x-q))
  =   -e^{-ip \cdot x} \frac{\partial \overline{v}}{\partial x_j} (x-q)
                                                                 \nonumber \\
& = & -e^{-ip \cdot x} \lim_{h \rightarrow 0}
              \frac{\overline{v}(x-q+he_j) - \overline{v}(x-q)}{h}         \\
& = & \lim_{h \rightarrow 0} f_h(x)                              \nonumber
\end{eqnarray}
where $ e_j $ is the $j$-th vector of the canonical basis of $ \R^N $ and
\begin{equation}
f_h(x) := -e^{-ip \cdot x}
                  \frac{\overline{v}(x-q+he_j) - \overline{v}(x-q)}{h} \ ;
\end{equation}
$q$ and $p$ are fixed. We show that the measurable functions $ f_h $,
$ h \neq 0 $, are dominated by some square-integrable function. Using the
mean value theorem of differential calculus, we obtain
\begin{equation}
f_h(x) = -e^{-ip \cdot x}
                 \frac{\partial \overline{v}}{\partial x_j} (x-q+ \xi e_j)
       = -e^{-ip \cdot x} w(x-q+ \xi e_j)
\end{equation}
where $ \xi $ depends on $ x-q $ and $h$, $ 0 < |\xi| < |h| $, and
$ w := \frac{\partial \overline{v}}{\partial x_j} \in S_{\C} \, (\R^N) $. The
latter implies that
\begin{equation}
|w(y) \|y\|^{2N}| < 1
\end{equation}
holds for $ \|y\| > R $, $ \|y\| $ denoting the Euclidean norm
of $ y \in \R^N $. Without loss of generality, we can assume
$ |h| \leq 1 $. Thus, $ |\xi| < 1 $ and $ \|q - \xi e_j\| < \|q\| + 1 $. Then,
if $ \|x\| > R + \|q\| + 1 $,
\begin{eqnarray*}
\|x - q + \xi e_j\| & \geq & | \, \|x\| - \|q - \xi e_j\| \, |
                        =    \|x\| - \|q - \xi e_j\|     \\
                    &   >  & \|x\| - (\|q\| + 1) > R
\end{eqnarray*}
holds. From this and (6.6) it follows that
\begin{equation}
|w(x-q+ \xi e_j)| < \frac{1}{\|x - q + \xi e_j \|^{2N}}
                  < \frac{1}{(\|x\| - \|q\| - 1)^{2N}}
\end{equation}
is valid for $ \|x\| > R + \|q\| + 1 $.

Define
\[
g(x) := \left\{
\begin{array}{ccl}
{\displaystyle \frac{1}{(\|x\| - \|q\| - 1)^{2N}}} & {\rm for} &
                                 \|x\| > R + \|q\| + 1 \vspace{0.3cm}\\
{\displaystyle \max_{y \in \R^N} |w(y)|} & {\rm for} &
                                 \|x\| \leq R + \|q\| + 1 \ .
\end{array}
\right.
\]
Then, by (6.5) and (6.7), $ |f_h (x)| \leq g(x) $ holds for $ |h| \leq 1 $
and all $ x \in \R^N $. Furthermore, $g$ is square-integrable with respect
to the Lebesgue measure on $ \R^N $ (and also integrable). Hence, for any
$ \psi \in L^{2}_{\C} \, (\R^N,d^N \! x) $ we have
$ g\psi \in L^{1}_{\C} \, (\R^N,d^N \! x) $ and
$ |f_h \psi| \leq g|\psi| $. Now, from Eq.\ (6.3), Lebesgue's
dominated convergence theorem, and (6.4) we conclude
\begin{eqnarray*}
\int \frac{\partial}{\partial q_j} \overline{v_{qp}(x)} \psi(x) \, d^N \! x
& = & \int \lim_{h \rightarrow 0} (f_h (x) \psi(x)) \, d^N \! x \\
& = & \lim_{h \rightarrow 0} \int f_h (x) \psi(x) \, d^N \! x \\
& = & -\lim_{h \rightarrow 0} \int e^{-ip \cdot x}
       \frac{\overline{v}(x - q + he_j) - \overline{v}(x - q)}{h}
       \psi(x) \, d^N \! x \\
& = & \lim_{h \rightarrow 0} \frac{1}{-h} \left(
      \int e^{-ip \cdot x} \overline{v}(x - (q-he_j)) \psi(x) \, d^N \! x
                                                              \right. \\
&   & \mbox{} - \left. \int e^{-ip \cdot x} \overline{v}(x-q)
                                          \psi(x) \, d^N \! x \right) \\
& = & \frac{\partial}{\partial q_j}
      \int e^{-ip \cdot x} \overline{v}(x-q) \psi(x) \, d^N \! x \\
& = & \frac{\partial}{\partial q_j}
      \int \overline{v_{qp}(x)} \psi(x) \, d^N \! x \ .
\end{eqnarray*}
\hspace*{\fill} $ \Box $\\

\begin{lemma}
If $ v \in S_{\C} \, (\R^N) $ and $ \psi \in {\cal H} $, then
\[
\int \frac{\partial}{\partial p_j} \overline{v_{qp}(x)} \psi(x) \, d^N \! x
= \frac{\partial}{\partial p_j} \int \overline{v_{qp}(x)} \psi(x) \, d^N \! x
\]
($ j = 1,\ldots,N $) holds.
\end{lemma}
{\bf Proof:} Write
\begin{equation}
\frac{\partial}{\partial p_j} \overline{v_{qp}(x)}
                            = \lim_{h \rightarrow 0} f_h (x)
\end{equation}
where
\begin{equation}
f_h (x) := \frac{\overline{v_{q,p+he_j}(x)} - \overline{v_{qp}(x)}}{h} \ .
\end{equation}
Using the mean value theorem of differential calculus, we obtain
\[
f_h (x) = \left. \frac{\partial}{\partial p_j} \overline{v_{qp}(x)} \,
          \right|_{(q,p+ \xi e_j ,x)}
        = -ix_j e^{-i(p+ \xi e_j ) \cdot x} \overline{v}(x-q)
\]
where $ 0 < |\xi| < |h| $. It follows that
\[
|f_h(x)| \leq |x_jv(x-q)| =: g(x)
\]
with $ g \in L^{2}_{\C} \, (\R^N,d^N \! x) $ holds. Now, from Eq.\ (6.8),
the dominated convergence theorem, and (6.9) we conclude
\begin{eqnarray*}
\int \frac{\partial}{\partial p_j} \overline{v_{qp}(x)} \psi(x) \, d^N \! x
& = & \int \lim_{h \rightarrow 0} (f_h (x) \psi(x)) \, d^N \! x \\
& = & \lim_{h \rightarrow 0} \int f_h (x) \psi(x) \, d^N \! x \\
& = & \lim_{h \rightarrow 0} \frac{1}{h} \left(
              \int \overline{v_{q,p+he_j}(x)} \psi(x) \, d^N \! x
            - \int \overline{v_{qp}(x)} \psi(x) \, d^N \! x \right) \\
& = & \frac{\partial}{\partial p_j}
      \int \overline{v_{qp}(x)} \psi(x) \, d^N \! x \ . \\
&   & \hspace{8.63cm} \Box
\end{eqnarray*}

Proceeding similarly as in the proofs of Lemmata 6.1 and 6.2, one \linebreak
can prove that, for $ v \in S_{\C} \, (\R^N) $ and $ \psi \in {\cal H} $,
the \vspace{1pt} function $ (q,p) \mapsto $ \linebreak
$ \int \overline{v_{qp}(x)} \psi(x) \, d^N \! x $
is continuous. However, it is already \vspace{2pt} clear that
$ (q,p) \mapsto \int \overline{v_{qp}(x)} \psi(x) \, d^N \! x
= \langle v_{qp} | \psi \rangle $ is continuous even for all
$ v \in {\cal H} $ since $ v_{qp} = U_{qp}v $ with $ U_{qp} $ according to
Theorem 5.1 and $ (q,p) \mapsto U_{qp} $ is strongly continuous. Using
Hilbert-space arguments, we have also alternative, shorter proofs for
Lemmata (6.1) and (6.2). Namely, for $ v \in S_{\C} \, (\R^N) $ we obtain,
since $ S_{\C} \, (\R^N) $ is contained in the domain $ D(P_j) $ of the
momentum operator $ P_j $,
\begin{eqnarray*}
\int \frac{\partial}{\partial q_j} \overline{v_{qp}(x)} \psi(x) \, d^N \! x
& = & \int \overline{\left( -\frac{\partial}{\partial x_j} v(x - q) \right)}
                                   e^{-ip \cdot x} \psi(x) \, d^N \! x  \\
& = & \langle (-i)P_j e^{-iq \cdot P}v | e^{-ip \cdot Q}\psi \rangle    \\
& = & \left\langle \left. \frac{\partial}{\partial q_j} e^{-iq \cdot P}v \,
                        \right| \, e^{-ip \cdot Q}\psi \right\rangle    \\
& = & \frac{\partial}{\partial q_j} \, \langle e^{ip \cdot Q} e^{-iq \cdot P}v
                                                      | \psi \rangle    \\
& = & \frac{\partial}{\partial q_j} \, \langle U_{qp}v | \psi \rangle   \\
& = & \frac{\partial}{\partial q_j} \int \overline{v_{qp}(x)} \psi(x)
                                                              \, d^N \! x \ .
\end{eqnarray*}
Analogously, one can prove Lemma 6.2.

\begin{theorem}
Let $ v \in S_{\C} \, (\R^N) $ and $ \psi \in {\cal H} $. Then the function
$ x \mapsto \frac{\partial^{l_{1}+\ldots+l_{2N}}}
{\partial q^{l_1}_{1} \ldots \partial p^{l_{2N}}_{N}}
\overline{v_{qp}(x)} \psi(x) $
is Lebesgue-integrable,
$ (q,p) \mapsto \int \overline{v_{qp}(x)} \psi(x) \, d^N \! x $
is \vspace{1pt} an infinitely differentiable function, and
\begin{equation}
\int \frac{\partial^{l_{1}+\ldots+l_{2N}}}
{\partial q^{l_1}_{1} \ldots \partial p^{l_{2N}}_{N}}
\overline{v_{qp}(x)} \psi(x) \, d^N \! x
= \frac{\partial^{l_{1}+\ldots+l_{2N}}}
{\partial q^{l_1}_{1} \ldots \partial p^{l_{2N}}_{N}}
\int \overline{v_{qp}(x)} \psi(x) \, d^N \! x
\end{equation}
($ l_{1},\ldots,l_{2N} = 0,1,2,\ldots $) holds. In particular, if
$ u \in S_{\C} \, (\R^N) $ and $ \|u\| = 1 $, then
$ V\psi \in C^{\infty}_{\C} (\R^{2N}) $ with $V$ according to (6.1).
\end{theorem}
{\bf Proof:} We denote any differential operator
$ \frac{\partial^{l_{1}+\ldots+l_{2N}}}
{\partial q^{l_1}_{1} \ldots \partial p^{l_{2N}}_{N}} $
of order $ n := l_{1} + \ldots + l_{2N} $, $ n = 0,1,2,\ldots \ $, briefly
by $ \partial^n $. By induction over $n$, we prove \vspace{1pt} that, for all
$ v \in S_{\C} \, (\R^N) $ and all $ \psi \in {\cal H} $, the function
$ (q,p) \mapsto \int \overline{v_{qp}(x)} \psi(x) \, d^N \! x $ belongs to
$ C^{n}_{\C} \hspace{0.2mm} (\R^{2N}) $ where (6.10) is satisfied. For
$ n = 0 $, this statement is obviously valid. Now suppose that it holds
for all $ n \leq m $. Taking account of
\[
\frac{\partial}{\partial q_j} v_{qp}(x)
               = -e^{ip \cdot x} \frac{\partial v}{\partial x_j} (x-q)
               = w_{qp}(x)
\]
where $ w := -\frac{\partial v}{\partial x_j} \in S_{\C} \, (\R^N) $, we
obtain from the induction hypothesis and Lemma 6.1 that
\begin{eqnarray*}
\int \partial^m \frac{\partial}{\partial q_j} \overline{v_{qp}(x)}
                                              \psi(x) \, d^N \! x
& = & \int \partial^m \overline{w_{qp}(x)} \psi(x) \, d^N \! x        \\
& = & \partial^m \int \overline{w_{qp}(x)} \psi(x) \, d^N \! x        \\
& = & \partial^m \int \frac{\partial}{\partial q_j}
                      \overline{v_{qp}(x)} \psi(x) \, d^N \! x        \\
& = & \partial^m \frac{\partial}{\partial q_j} \int
                      \overline{v_{qp}(x)} \psi(x) \, d^N \! x \ .
\end{eqnarray*}
The continuity of
$ (q,p) \mapsto \partial^m \int \overline{w_{qp}(x)} \psi(x) \, d^N \! x $
implies that of $ (q,p) \mapsto \partial^m \frac{\partial}{\partial q_j}
\int \overline{v_{qp}(x)} \psi(x) \, d^N \! x $; in consequence,
\begin{equation}
\int \partial^m \frac{\partial}{\partial q_j} \overline{v_{qp}(x)}
                                              \psi(x) \, d^N \! x
= \frac{\partial}{\partial q_j} \partial^m \int
                      \overline{v_{qp}(x)} \psi(x) \, d^N \! x \ .
\end{equation}

To prove the analog of (6.11) for $ \frac{\partial}{\partial p_j} $,
notice that
\[
\frac{\partial}{\partial p_j} v_{qp}(x)
   = i(x_j - q_j) e^{ip \cdot x} v(x-q) + iq_j e^{ip \cdot x} v(x-q)
   = \tilde{w}_{qp}(x) + iq_j v_{qp}(x)
\]
where $ \tilde{w}(x) := ix_jv(x) $ and $ \tilde{w} \in S_{\C} \, (\R^N) $. For
two functions $f$ and $g$ of one real variable being $k$-times differentiable
and satisfying $ f'' = 0 $, the differentiation rule
\[
(fg)^{(k)} = fg^{(k)} + kf'g^{(k-1)}
\]
holds. Applying some analogous rule, we obtain
\begin{eqnarray*}
\partial^m \frac{\partial}{\partial p_j} v_{qp}(x)
& = & \partial^m (\tilde{w}_{qp}(x) + iq_j v_{qp}(x))                  \\
& = & \partial^m \tilde{w}_{qp}(x)
              + i\frac{\partial^{l_j}}{\partial q^{l_j}_{j}}
                  (q_j \partial^{m-l_j} v_{qp}(x))                     \\
& = & \partial^m \tilde{w}_{qp}(x) + iq_j \partial^m v_{qp}(x)
              + il_j \frac{\partial^{l_j-1}}{\partial q^{l_j-1}_{j}}
                           \partial^{m-l_j} v_{qp}(x)                  \\
& = & \partial^m \tilde{w}_{qp}(x) + iq_j \partial^m v_{qp}(x)
              + il_j \partial^{m-1} v_{qp}(x)
\end{eqnarray*}
where $ l_j $ is the number of differentiations with respect to $ q_j $
contained in $ \partial^m $. Using the latter result, it follows from the
induction hypothesis and Lemma~6.2 that
\begin{eqnarray*}
\lefteqn{\int \partial^m \frac{\partial}{\partial p_j} \overline{v_{qp}(x)}
                                       \psi(x) \, d^N \! x}\hspace{2cm}    \\
& = & \int (\partial^m \overline{\tilde{w}_{qp}(x)}
      - iq_j \partial^m \overline{v_{qp}(x)}
      - il_j \partial^{m-1} \overline{v_{qp}(x)}) \psi(x) \, d^N \! x      \\
& = & \partial^m \int \overline{\tilde{w}_{qp}(x)} \psi(x) \, d^N \! x
      - iq_j \partial^m \int \overline{v_{qp}(x)} \psi(x) \, d^N \! x      \\
&   & \mbox{} - il_j \partial^{m-1} \int \overline{v_{qp}(x)}
                                       \psi(x) \, d^N \! x                 \\
& = & \partial^m \int \overline{\tilde{w}_{qp}(x)} \psi(x) \, d^N \! x
      - i\partial^m \left( q_j \int \overline{v_{qp}(x)}
                                       \psi(x) \, d^N \! x \right)         \\
& = & \partial^m \int \overline{(\tilde{w}_{qp}(x) +iq_j v_{qp}(x))}
                                       \psi(x) \, d^N \! x                 \\
& = & \partial^m \int \frac{\partial}{\partial p_j} \overline{v_{qp}(x)}
                                       \psi(x) \, d^N \! x                 \\
& = & \partial^m \frac{\partial}{\partial p_j} \int \overline{v_{qp}(x)}
                                       \psi(x) \, d^N \! x \ .
\end{eqnarray*}
Again, $ (q,p) \mapsto \partial^m \frac{\partial}{\partial p_j}
\int \overline{v_{qp}(x)} \psi(x) \, d^N \! x $ is continuous; in consequence,
\begin{equation}
\int \partial^m \frac{\partial}{\partial p_j} \overline{v_{qp}(x)}
                                              \psi(x) \, d^N \! x
= \frac{\partial}{\partial p_j} \partial^m \int
                      \overline{v_{qp}(x)} \psi(x) \, d^N \! x \ .
\end{equation}

From (6.11) and (6.12) it follows that
$ (q,p) \mapsto \partial^m \int \overline{v_{qp}(x)} \psi(x) \, d^N \! x $
\vspace{1pt} has continuous partial derivatives, i.e.,
$ (q,p) \mapsto \int \overline{v_{qp}(x)} \psi(x) \, d^N \! x $
is $ (m+1) $-times differentiable. Moreover, by (6.11) and (6.12), Eq.\ (6.10)
holds for $ n = m+1 $. Hence, for every $ n = 0,1,2,\ldots \ $, the function
$ (q,p) \mapsto \int \overline{v_{qp}(x)} \psi(x) \, d^N \! x $ belongs to
$ C^{n}_{\C} \hspace{0.2mm} (\R^{2N}) $ and satisfies (6.10). In particular,
this function belongs to $ C^{\infty}_{\C} (\R^{2N}) $. \hfill $ \Box $\\

By Theorem 6.3, the space $ V{\cal H} $ is a Hilbert space of bounded
infinitely differentiable, square-integrable functions on phase space,
provided that $u$ is a Schwartz function. Of course, if $ \Psi \in V{\cal H} $,
then its partial derivatives do in general not belong to $ V{\cal H} $ and need
even not be square-integrable. The following proposition gives some information
about that.

\begin{proposition}
Let $ u \in S_{\C} \, (\R^N) $ and $ \|u\| = 1 $, and let $V$ be
the corresponding isometry according to Eq.\ (6.1). Then, for each
$ \Psi \in V{\cal H} $, $ \frac{\partial \Psi}{\partial q_j} \in
L^{2}_{\C} \, (\R^{2N},d^N \! q \, d^N \! p) $ holds ($ j = 1,\ldots,N $).
\end{proposition}
{\bf Proof:} From
\[
\Psi(q,p) = (V\psi)(q,p)
 = \frac{1}{\sqrt{(2\pi)^N}} \int \overline{u_{qp}(x)} \psi(x) \, d^N \! x
\]
we obtain, using Lemma 6.1,
\[
\frac{\partial \Psi}{\partial q_j}(q,p) = \frac{1}{\sqrt{(2\pi)^N}} \int
\overline{e^{ip \cdot x} \left( -\frac{\partial u}{\partial x_j}(x-q) \right)}
                                                     \psi(x) \, d^N \! x \ .
\]
Defining $ v_j = -\frac{\partial u}{\partial x_j} $, we have
$ v \in S_{\C} \, (\R^N) $, $ v_j \neq 0 $, and thus
$ \| v_j \| \neq 0 $. Hence,
\begin{eqnarray}
\frac{\partial \Psi}{\partial q_j}(q,p)
& = & \frac{1}{\sqrt{(2\pi)^N}} \langle U_{qp}v_j | \psi \rangle   \nonumber\\
& = & \frac{\| v_j \|}{\sqrt{(2\pi)^N}} \left\langle U_{qp} \left.
   \left( \frac{v_j}{\| v_j \|} \right) \, \right| \, \psi \right\rangle    \\
& = & \| v_j \| (\tilde{V}_j \psi)(q,p)                            \nonumber
\end{eqnarray}
where, according to (6.1), $ \tilde{V}_j \!\! : {\cal H} \rightarrow
L^{2}_{\C} \, (\R^{2N},d^N \! q \, d^N \! p) $ is \vspace{1pt} the isometry
corresponding to $ \frac{v_j}{\| v_j \|} $. In particular,
$ \frac{\partial \Psi}{\partial q_j} \in
L^{2}_{\C} \, (\R^{2N},d^N \! q \, d^N \! p) $. \hfill $ \Box $\\

We are going to consider the analog of Eq.\ (6.13) for
$ \frac{\partial \Psi}{\partial p_j} $. Using Lem- \linebreak ma 6.2,
we obtain
\[
\frac{\partial \Psi}{\partial p_j}(q,p) = \frac{1}{\sqrt{(2\pi)^N}} \int
\overline{ix_j e^{ip \cdot x} u(x-q)} \psi(x) \, d^N \! x \ .
\]
Now the definition $ w_j (x) := ix_j u(x) $, $ w_j \in S_{\C} \, (\R^N) $,
$ w_j \neq 0 $, yields
\begin{equation}
\begin{array}{ccl}
\displaystyle{\frac{\partial \Psi}{\partial p_j}(q,p)}
& = & \displaystyle{\frac{1}{\sqrt{(2\pi)^N}} (\langle U_{qp}w_j |
      \psi \rangle - iq_j \langle u_{qp} | \psi \rangle)}     \vspace{3mm}\\
& = & \displaystyle{\frac{\| w_j \|}{\sqrt{(2\pi)^N}} \left\langle U_{qp}
      \left. \left( \frac{w_j}{\| w_j \|} \right) \, \right| \,
      \psi \right\rangle - \frac{iq_j}{\sqrt{(2\pi)^N}} \langle u_{qp} |
      \psi \rangle}                                           \vspace{3mm}\\
& = & \| w_j \| (\tilde{\tilde{V}}_j \psi)(q,p) - iq_j (V\psi)(q,p)
                                                              \vspace{3mm}\\
& = & \| w_j \| (\tilde{\tilde{V}}_j \psi)(q,p) - iq_j \Psi(q,p)
\end{array}
\end{equation}
where $ \tilde{\tilde{V}}_j $ is the isometry corresponding to
$ \frac{w_j}{\| w_j \|} $. In particular, because of the factor $ q_j $
in (6.14), one cannot conclude that $ \frac{\partial \Psi}{\partial p_j} \in
L^{2}_{\C} \, (\R^{2N},d^N \! q \, d^N \! p) $. However, we shall see that,
as a consequence of Proposition 6.5, even
$ \frac{\partial \Psi}{\partial p_j} \in V{\cal H} $ holds if $ \Psi $
belongs to a suitable dense subspace of $ V{\cal H} $.

\section{Position and Momentum in Phase-Space Representation}

The next proposition now states how the operators of position and momentum
act in phase-space representation (cf., e.g., E. Prugove\v{c}ki, 1984).

\begin{proposition}
Let $ u \in S_{\C} \, (\R^N) $ and $ \|u\| = 1 $, and consider
the isometry according to Eq.\ (6.1) as unitary operator
$ V \!\! : {\cal H} \rightarrow V{\cal H} $. Then the operators $ VQ_jV^{-1} $
and $ VP_jV^{-1} $, $ j = 1,\ldots,N $, are self-adjoint on the dense domains
$ VD(Q_j) $ and $ VD(P_j) $, they can be represented by
\begin{equation}
VQ_jV^{-1} \Psi = i\frac{\partial \Psi}{\partial p_j}
\end{equation}
where $ \Psi \in VD(Q_j) $, respectively, by
\begin{equation}
(VP_jV^{-1} \Psi)(q,p) = p_j \Psi(q,p)
                          - i\frac{\partial \Psi}{\partial q_j}(q,p)
\end{equation}
where $ \Psi \in VD(P_j) $; briefly
\begin{equation}
\begin{array}{ccc@{\:}c@{\:}l}
VQ_jV^{-1} & = &     &   & {\displaystyle i\frac{\partial}{\partial p_j}}
                                                           \vspace{0.3cm}\\
VP_jV^{-1} & = & p_j & - & {\displaystyle i\frac{\partial}{\partial q_j} \ .}
\end{array}
\end{equation}
\end{proposition}
{\bf Proof:} If $ \Psi \in VD(Q_j) $ and $ \psi := V^{-1}\Psi $, we have
\begin{eqnarray*}
(VQ_jV^{-1} \Psi)(q,p) & = & (VQ_j \psi)(q,p)
          = \frac{1}{\sqrt{(2\pi)^N}} \langle u_{qp} | Q_j \psi \rangle \\
& = & \frac{1}{\sqrt{(2\pi)^N}} \int \overline{x_j u_{qp}(x)} \psi(x) \,
                                                               d^N \! x \\
& = & \frac{1}{\sqrt{(2\pi)^N}} \int
\overline{\frac{1}{i} \frac{\partial}{\partial p_j} e^{ip \cdot x} u(x-q)}
                                                    \psi(x) \, d^N \! x \\
& = & \frac{i}{\sqrt{(2\pi)^N}} \frac{\partial}{\partial p_j}
                          \int \overline{u_{qp}(x)} \psi(x) \, d^N \! x \\
& = & i\frac{\partial \Psi}{\partial p_j}(q,p)
\end{eqnarray*}
where Lemma 6.2 or Theorem 6.3 has been used. Similarly, for
$ \Psi \in VD(P_j) $ we obtain, using Lemma 6.1 or Theorem 6.3,
\begin{eqnarray*}
(VP_jV^{-1} \Psi)(q,p) & = & (VP_j \psi)(q,p)
  =   \frac{1}{\sqrt{(2\pi)^N}} \langle u_{qp} | P_j \psi \rangle \\
& = & \frac{1}{\sqrt{(2\pi)^N}} \langle P_j u_{qp} | \psi \rangle \\
& = & \frac{e^{-iq \cdot p}}{\sqrt{(2\pi)^N}} \int
      \overline{\frac{1}{i} \frac{\partial}{\partial x_j}
                     (e^{ip \cdot (x-q)} u(x-q))} \psi(x) \, d^N \! x \\
& = & \frac{e^{-iq \cdot p}}{\sqrt{(2\pi)^N}} \int
      \overline{\left( -\frac{1}{i} \right) \frac{\partial}{\partial q_j}
                     (e^{ip \cdot (x-q)} u(x-q))} \psi(x) \, d^N \! x \\
& = & \frac{-ie^{-iq \cdot p}}{\sqrt{(2\pi)^N}} \frac{\partial}{\partial q_j}
      \left( e^{iq \cdot p} \int \overline{u_{qp}(x)}
                                          \psi(x) \, d^N \! x \right) \\
& = & -ie^{-iq \cdot p} \left( ip_j e^{iq \cdot p} \Psi(q,p)
       + e^{iq \cdot p} \frac{\partial \Psi}{\partial q_j}(q,p) \right) \\
& = & p_j \Psi(q,p) - i\frac{\partial \Psi}{\partial q_j}(q,p) \ .
\end{eqnarray*}
\hspace*{\fill} $ \Box $\\

\begin{corollary}
For $ \Psi \in VD(Q_j) $, $ \frac{\partial \Psi}{\partial p_j} \in V{\cal H} $
and \vspace{1pt} $ q_j \Psi \in L^{2}_{\C} \, (\R^{2N},d^N \! q \, d^N \! p) $
hold; for $ \Psi \in VD(P_j) $,
$ p_j \Psi \in L^{2}_{\C} \, (\R^{2N},d^N \! q \, d^N \! p) $ holds.
\end{corollary}
{\bf Proof:} Let $ \Psi \in VD(Q_j) $. From Eq.\ (6.15) it follows that
$ \frac{\partial \Psi}{\partial p_j} \in V{\cal H} $. Eq.~(6.14) implies
\[
q_j \Psi(q,p) = i\frac{\partial \Psi}{\partial p_j}(q,p)
                 - i\| w_j \| (\tilde{\tilde{V}}_j \psi)(q,p) \ .
\]
Since $ \frac{\partial \Psi}{\partial p_j} \in V{\cal H} $ and
$ \tilde{\tilde{V}}_j \psi \in \tilde{\tilde{V}}_j {\cal H} $, we obtain
$ q_j \Psi \in L^{2}_{\C} \, (\R^{2N},d^N \! q \, d^N \! p) $. Now let
$ \Psi \in VD(P_j) $. Eq.\ (6.16) can be rewritten as
\[
p_j \Psi(q,p) = (VP_jV^{-1}\Psi)(q,p)
                + i\frac{\partial \Psi}{\partial q_j}(q,p) \ .
\]
Since, according to Proposition 6.4, $ \frac{\partial \Psi}{\partial q_j}
\in L^{2}_{\C} \, (\R^{2N},d^N \! q \, d^N \! p) $, it follows that
$ p_j \Psi \in L^{2}_{\C} \, (\R^{2N},d^N \! q \, d^N \! p) $ also.
\hfill $ \Box $\\

Eqs.\ (6.13) -- (6.17) show that in our phase-space representations the
variables $ q_j $ and $ p_j $ as well as the operators $ Q_j $ and $ P_j $
occur rather asymmetric. This can be corrected by a gauge transformation;
gauge transformations related to our phase-space representations are
introduced now (cf.\ E. Prugove\v{c}ki, 1984).

Let $ \Theta $ be a real-valued function on the phase space $ \R^{2N} $ which,
for instance, is infinitely differentiable. Then the functions
\[
x \mapsto \hat{u}_{qp}(x) := e^{-i\Theta(q,p)} u_{qp}(x)
                           = e^{i(p \cdot x - \Theta(q,p))} u(x-q)
\]
differ from the wave functions $ u_{qp} $ only by a phase factor. In
particular, \linebreak $ \{ \hat{u}_{qp} \}_{(q,p) \in \R^{2N}} $ is
also a continuous resolution of the identity of $ \cal H $, and
according to Eq.\ (6.2) it defines the same observable as
$ \{ u_{qp} \}_{(q,p) \in \R^{2N}} $. If, in Eq.\ (6.1), we replace
$ u_{qp} $ by $ \hat{u}_{qp} = e^{-i\Theta(q,p)} u_{qp} $,
we obtain new wave functions
\[
(q,p) \mapsto \hat{\Psi}(q,p) := e^{i\Theta(q,p)} \Psi(q,p)
\]
on phase space differing from the old ones by $ e^{i\Theta(q,p)} $. Now,
however, $ e^{i\Theta(q,p)} $ is in general not a phase factor. According to
\begin{equation}
(U_{\Theta}\Psi)(q,p) := e^{i\Theta(q,p)} \Psi(q,p)
\end{equation}
where $ \Psi \in V{\cal H} $, an isometry $ U_{\Theta} \!\! : V{\cal H}
\rightarrow L^{2}_{\C} \, (\R^{2N},d^N \! q \, d^N \! p) $
is defined, respectively, a unitary operator
$ U_{\Theta} \!\! : V{\cal H} \rightarrow U_{\Theta} V{\cal H} $. Because of
$ | \hat{\Psi}(q,p) |^2 = | \Psi(q,p) |^2 $, the transformation
\[
\Psi \mapsto \hat{\Psi} = U_{\Theta} \Psi
\]
can be considered as a {\it gauge transformation}.

Defining $ V_{\Theta} := U_{\Theta}V $, we obtain from Eqs.\ (6.17) and
(6.18) that
\[
\begin{array}{ccc@{\:}c@{\:}c@{\:}c@{\:}c}
V_{\Theta}Q_jV^{-1}_{\Theta} & = &     &   &
 {\displaystyle \frac{\partial \Theta}{\partial p_j}} & + &
 {\displaystyle i\frac{\partial}{\partial p_j}}  \vspace{0.3cm}\\
V_{\Theta}P_jV^{-1}_{\Theta} & = & p_j & - &
 {\displaystyle \frac{\partial \Theta}{\partial q_j}} & - &
 {\displaystyle i\frac{\partial}{\partial q_j}}
\end{array}
\]
holds where $ p_j $, $ \frac{\partial \Theta}{\partial p_j} $, and
$ \frac{\partial \Theta}{\partial q_j} $ are understood as multiplication
operators. The choice $ \Theta(q,p) := q \cdot p $ gives
\[
\begin{array}{ccc@{\:}c@{\:}l}
V_{\Theta}Q_jV^{-1}_{\Theta} & = & q_j & + &
 {\displaystyle i\frac{\partial}{\partial p_j}}  \vspace{0.3cm}\\
V_{\Theta}P_jV^{-1}_{\Theta} & = &     & - &
 {\displaystyle i\frac{\partial}{\partial q_j}} \ ,
\end{array}
\]
and the choice $ \Theta(q,p) := \frac{1}{2} q \cdot p $ gives the
symmetric representation
\[
\begin{array}{ccc@{\:}c@{\:}l}
V_{\Theta}Q_jV^{-1}_{\Theta} & = & \displaystyle{\frac{q_j}{2}} & + &
 \displaystyle{i\frac{\partial}{\partial p_j}}  \vspace{0.3cm}\\
V_{\Theta}P_jV^{-1}_{\Theta} & = & \displaystyle{\frac{p_j}{2}} & - &
 \displaystyle{i\frac{\partial}{\partial q_j}} \ .
\end{array}
\]
Furthermore, for the latter choice of $ \Theta $ we have
$ \hat{\Psi}(q,p) = \Psi^W(-q,p) $ where $ \hat{\Psi} = V_{\Theta}\psi $
and $ \Psi^W $ is defined by means of the Weyl operators (5.15) according to
$ \Psi^W(q,p) := \frac{1}{\sqrt{(2\pi)^N}} \langle u^{W}_{qp} | \psi \rangle $
and $ u^{W}_{qp} = U^{W}_{qp}u $.

Next we consider the phase-space representation of quantum mechanics
given by $V$ for a distinguished choice of $u$, namely, for a coherent
state. For simplicity, let $ N = 1 $; let $ u \in S_{\C} \, (\R) $,
$ \|u\| = 1 $, given by
\begin{equation}
u(x) = u^{\sigma}(x) := \frac{1}{\sqrt[4]{2\pi\sigma^2}}
                                      e^{-\frac{x^2}{4\sigma^2}}
\end{equation}
where $ \sigma > 0 $. In this situation, Propositions 6.4 and 6.5 can be
supplemented by the following one.

\begin{proposition}
Let $ N = 1 $, and let $u$ be given by the coherent state $ u^{\sigma} $
according to (6.19). Then, for each $ \Psi \in V{\cal H} $,
$ \frac{\partial \Psi}{\partial q} \in L^{2}_{\C} \, (\R^2,dqdp) $ and
\begin{equation}
\left\langle \Phi \, \left| \, \frac{\partial}{\partial q} \right. \Psi
                                                           \right\rangle = 0
\end{equation}
for all $ \Phi \in V{\cal H} $ hold, i.e.\
$ \frac{\partial \Psi}{\partial q} \in (V{\cal H})^{\bot} $. Furthermore,
on $ VD(Q) $ and $ VD(P) $, respectively, the representations
\begin{eqnarray}
VQV^{-1} & = & \hspace{7.2mm} i\frac{\partial}{\partial p}
           =   q + 2\sigma^2 \frac{\partial}{\partial q} \\
VPV^{-1} & = &          p - i\frac{\partial}{\partial q}
           =   p + \frac{i}{2\sigma^2}q
                 + \frac{1}{2\sigma^2} \frac{\partial}{\partial p}
\end{eqnarray}
are valid.
\end{proposition}
{\bf Proof:} By Proposition 6.4,
$ \frac{\partial \Psi}{\partial q} \in L^{2}_{\C} \, (\R^2,dqdp) $
holds. Since $ u^{\sigma} $ is a bounded square-integrable function,
we have $ u^{\sigma}( \, . \, -q)\psi \in
L^{1}_{\C} \, (\R,dx) \cap L^{2}_{\C} \, (\R,dx) $ for
$ \psi \in {\cal H} = L^{2}_{\C} \, (\R,dx) $ and all
$ q \in \R $. Therefore,
\[
\Psi(q,p) = (V\psi)(q,p)
          = \frac{1}{\sqrt{2\pi}} \int e^{-ipx} u^{\sigma}(x-q) \psi(x) dx
          = F(u^{\sigma}( \, . \, -q)\psi)(p)
\]
where $F$ denotes the Fourier transformation in
$ L^{2}_{\C} \, (\R,dx) $. Defining $ v^{\sigma} := -(u^{\sigma})' $
and taking account of the boundedness of $ v^{\sigma} $, we obtain
\[
\frac{\partial \Psi}{\partial q}(q,p)
          = \frac{1}{\sqrt{2\pi}} \int e^{-ipx} v^{\sigma}(x-q) \psi(x) dx
          = F(v^{\sigma}( \, . \, -q)\psi)(p) \ .
\]
Hence, for any $ \Phi,\Psi \in V{\cal H} $ with $ \Phi = V\phi $ and
$ \Psi = V\psi $ it follows that
\begin{eqnarray*}
\left\langle \Phi \, \left| \, \frac{\partial}{\partial q} \right. \Psi
                                                           \right\rangle
& = & \int \overline{\Phi(q,p)} \frac{\partial \Psi}{\partial q}(q,p) dqdp \\
& = & \int\int \overline{F(u^{\sigma}( \, . \, -q)\phi)(p)}
                         F(v^{\sigma}( \, . \, -q)\psi)(p) dpdq \\
& = & \int\int u^{\sigma}(x-q) v^{\sigma}(x-q)
                  \overline{\phi(x)} \psi(x) dxdq \\
& = & \int u^{\sigma}(x) v^{\sigma}(x)
                  \int \overline{\phi(x+q)} \psi(x+q) dqdx \\
& = & \int \frac{x}{2\sigma^2} (u^{\sigma}(x))^2 dx
                  \int \overline{\phi(q)} \psi(q) dq \\
& = & 0
\end{eqnarray*}
where (6.19) and
$ v^{\sigma}(x) = -(u^{\sigma})'(x) = \frac{x}{2\sigma^2} u^{\sigma}(x) $
have been taken into account.

From
\[
\Psi(q,p) = \frac{1}{\sqrt{2\pi}} \int \overline{u^{\sigma}_{qp}(x)} \psi(x) dx
\]
and (6.19) it follows that
\begin{eqnarray*}
\frac{\partial \Psi}{\partial q}(q,p) & = & \frac{1}{\sqrt{2\pi}}
      \int \frac{x-q}{2\sigma^2} \overline{u^{\sigma}_{qp}(x)} \psi(x) dx \\
\frac{\partial \Psi}{\partial p}(q,p) & = & \frac{1}{\sqrt{2\pi}}
      \int (-ix) \overline{u^{\sigma}_{qp}(x)} \psi(x) dx \ .
\end{eqnarray*}
Consequently, for every $ \Psi \in V{\cal H} $
\begin{equation}
\frac{\partial \Psi}{\partial p}(q,p) = -i \left( 2\sigma^2
               \frac{\partial \Psi}{\partial q}(q,p) + q\Psi(q,p) \right)
\end{equation}
holds. Proposition 6.5 and Eq.\ (6.23) imply the validity of
\[
VQV^{-1}\Psi = i\frac{\partial \Psi}{\partial p}
             = q\Psi + 2\sigma^2 \frac{\partial \Psi}{\partial q}
\]
for all $ \Psi \in VD(Q) $, i.e.\ the validity of Eq.\ (6.21), and similarly
that of Eq.~(6.22). \hfill $ \Box $\\

From Proposition 6.7 one can draw further conclusions. First, if
$ \Psi \in VD(Q) $ and $ \Psi \neq 0 $, then, by (6.20) and
$ VQV^{-1} = q + 2\sigma^2 \frac{\partial}{\partial q} $, $ q\Psi \in $
\linebreak $ L^{2}_{\C} \, (\R^2,dqdp) \setminus V{\cal H} $. Second,
if $ \Psi \in VD(P) $ and $ \Psi \neq 0 $, then, by (6.20) and
$ VPV^{-1} = p - i\frac{\partial}{\partial q} $,
$ p\Psi \in L^{2}_{\C} \, (\R^2,dqdp) \setminus V{\cal H} $. Third,
for $ \Phi \in V{\cal H} $ and $ \Psi \in VD(Q) $,
\begin{equation}
\langle \Phi | VQV^{-1}\Psi \rangle
    = i\left\langle \Phi \, \left| \, \frac{\partial \Psi}{\partial p}
                                                     \right. \right\rangle
    = \langle \Phi | q\Psi \rangle \ ;
\end{equation}
fourth, for $ \Phi \in V{\cal H} $ and $ \Psi \in VD(P) $,
\begin{equation}
\langle \Phi | VPV^{-1}\Psi \rangle = \langle \Phi | p\Psi \rangle \ .
\end{equation}

Finally, as an example, we apply the results (6.17), respectively, (6.21)
and (6.22) to the harmonic oscillator. The operator
\[
H := \frac{1}{2m}P^2 + \frac{m\omega^2}{2}Q^2
   = -\frac{1}{2m}\frac{d^2}{dx^2} + \frac{m\omega^2}{2}x^2
\]
defined on $ S_{\C} \, (\R) $, for instance, is essentially self-adjoint
(cf.\ the reasoning following (6.41)). Its self-adjoint extension is the
Hamiltonian of the harmonic oscillator and is also denoted by $H$. In
particular, $ D(H) \supseteq D(Q^2) \cap D(P^2) \supset S_{\C} \, (\R) $
where the Hamiltonian acts according to
$ H = \frac{1}{2m}P^2 + \frac{m\omega^2}{2}Q^2 $
precisely on $ D(Q^2) \cap D(P^2) $.

By Eqs.\ (6.17), we have
\begin{equation}
\begin{array}{ccl}
VHV^{-1} & = & \displaystyle{\frac{1}{2m}
               \left( p - i\frac{\partial}{\partial q} \right)^2
               - \frac{m\omega^2}{2} \frac{\partial^2}{\partial p^2}}
                                                       \vspace{2mm}\\
         & = & \displaystyle{-\left( \frac{1}{2m}
                                     \frac{\partial^2}{\partial q^2}
               + \frac{m\omega^2}{2} \frac{\partial^2}{\partial p^2} \right)
               -\frac{i}{m}p\frac{\partial}{\partial q} + \frac{1}{2m}p^2}
\end{array}
\end{equation}
on $ V(D(Q^2) \cap D(P^2)) $. Choosing $V$ with $ u = u^{\sigma} $ according
to (6.19), we obtain, using Eqs.\ (6.21) and (6.22),
\begin{eqnarray}
VHV^{-1} & = & \frac{1}{2m} \left( p - i\frac{\partial}{\partial q} \right)^2
               + \frac{m\omega^2}{2} \left( q + 2\sigma^2
                        \frac{\partial}{\partial q} \right)^2     \nonumber\\
         & = & \left( -\frac{1}{2m} + 2m\omega^2\sigma^4 \right)
                                        \frac{\partial^2}{\partial q^2}
               + \left( 2m\omega^2\sigma^2q - \frac{i}{m}p \right)
                          \frac{\partial}{\partial q} \hspace{1cm}\mbox{}  \\
         &   & \mbox{} + \frac{1}{2m}p^2 + \frac{m\omega^2}{2}q^2
                                        + m\omega^2\sigma^2 \ .   \nonumber
\end{eqnarray}
Moreover, one can choose $ \sigma = \frac{1}{\sqrt{2m\omega}} $. In this case
$ u^{\sigma} $ coincides with the ground state $ \phi_0 $ of the harmonic
oscillator and (6.27) becomes
\begin{equation}
VHV^{-1} = \left( \omega q - \frac{i}{m}p \right) \frac{\partial}{\partial q}
           + \frac{1}{2m}p^2 + \frac{m\omega^2}{2}q^2 + \frac{\omega}{2} \ .
\end{equation}
This is a funny result. The eigenfunctions of the operator (6.28) are
$ V\phi_n $, $ n = 0,1,2,\ldots \ $, where $ \phi_n $ are the eigenfunctions
of $H$. Using the result (6.46) of the next section, we can calculate the
eigenstates of $H$ in the present phase-space representation; we obtain
\begin{equation}
\begin{array}{crl}
\Phi_n(q,p) & := & (V\phi_n)(q,p)                   \vspace{2.4mm}\\
            &  = & \displaystyle{\frac{1}{\sqrt{2\pi n!}} \frac{1}{\sqrt{2^n}}
                    \left( \sqrt{m\omega}q - \frac{i}{\sqrt{m\omega}}p
                    \right)^n e^{-\frac{1}{2\omega} \left( \frac{1}{2m}p^2
                     + \frac{m\omega^2}{2}q^2 \right) - i\frac{qp}{2}}} \ .
\end{array}
\end{equation}
By an easy calculation, the equation
\[
VHV^{-1}\Phi_n = \omega \left( n + \frac{1}{2} \right) \Phi_n
\]
can directly be verified.

\section{The Relation to a Hilbert Space of Entire Functions}

If the function $u$ is a coherent state, the Hilbert space $ V{\cal H} $
defined by Eq.~(6.1) is closely related to a well-known Hilbert space of
entire functions (cf.\ V.~Bargmann, 1961, 1967). In this section, we first
introduce that Hilbert space, and then we investigate the relation of
$ V{\cal H} $ to it. Again, let $ N = 1 $ for simplicity, and let $u$
be given by Eq.\ (6.19).

Consider the Hilbert space $ L^{2}_{\C} \, (\R^2,\Xi(\R^2),\mu) $
where the measure $ \mu $ is \vspace{1pt} given by $ \mu(d(\xi,\eta))
:= \frac{1}{\pi} e^{-(\xi^2 + \eta^2)} \lambda^2(d(\xi,\eta))
= \frac{1}{\pi} e^{-(\xi^2 + \eta^2)} d\xi d\eta $; $ \mu(\R^2) = 1 $. We
write $ L^{2}_{\C} \, (\R^2,\Xi(\R^2),\mu) =: L^{2}_{\C} \, (\R^2,
\frac{1}{\pi} e^{-(\xi^2 + \eta^2)} d\xi d\eta) $. Identifying $ \R^2 $
with the complex plane $ \C \, $, let $ \hat{\cal H} $
be the subspace of all elements of \vspace{1pt}
$ L^{2}_{\C} \, (\R^2,\frac{1}{\pi} e^{-(\xi^2 + \eta^2)} d\xi d\eta) $
that can be represented by holomorphic functions. That is, $ \hat{\cal H} $
is the unitary space of all holomorphic functions
$ f \!\! : \C \, \rightarrow \C \, $ satisfying
\begin{equation}
\int |f(z)|^2 e^{-|z|^2} d\xi d\eta < \infty
\end{equation}
where $ z := \xi + i\eta $ and
\[
\langle f|g \rangle := \frac{1}{\pi} \int
                       \overline{f(z)} g(z) e^{-|z|^2} d\xi d\eta \ .
\]
We now show that $ \hat{\cal H} $ is even a Hilbert space,
i.e.\ a closed subspace of
$ L^{2}_{\C} \, (\R^2,\frac{1}{\pi} e^{-(\xi^2 + \eta^2)} d\xi d\eta) $.

Each $ f \in \hat{\cal H} $ is an entire function. Therefore, it can, for
all $ z \in \C \, $, be represented uniquely by a power series at $ z_0 = 0 $:
\begin{equation}
f(z) = \sum_{n=0}^{\infty} a_n z^n \ .
\end{equation}
We obtain
\begin{eqnarray*}
\|f\|^2 & = & \frac{1}{\pi} \int |f(z)|^2 e^{-|z|^2} d\xi d\eta \\
        & = & \frac{1}{\pi} \int_{0}^{\infty} \int_{0}^{2\pi}
        \left( \sum_{n=0}^{\infty} \overline{a_n} r^n e^{-in\phi} \right)
        \left( \sum_{m=0}^{\infty} a_m r^m e^{im\phi} \right)
                                           e^{-r^2} r d\phi dr \\
        & = & \frac{1}{\pi} \int_{0}^{\infty}
        \sum_{n=0}^{\infty} \sum_{m=0}^{\infty} \overline{a_n} a_m r^{n+m}
        \int_{0}^{2\pi} e^{i(m-n)\phi} d\phi \, e^{-r^2} r dr
\end{eqnarray*}
where we have introduced polar coordinates and used the fact that the
power series (6.31) converges uniformly on the circle $ \|z\| = r $. From
$ \int_{0}^{2\pi} e^{i(m-n)\phi} d\phi $ $ = 2\pi\delta_{mn} $ and the
monotone convergence theorem it follows that
\begin{eqnarray*}
\|f\|^2 & = & 2\int_{0}^{\infty} \sum_{n=0}^{\infty} |a_n|^2 r^{2n+1}
                                                             e^{-r^2} dr \\
        & = & 2\sum_{n=0}^{\infty} |a_n|^2 \int_{0}^{\infty} r^{2n+1}
                                                             e^{-r^2} dr \\
        & = & \sum_{n=0}^{\infty} |a_n|^2 n!
\end{eqnarray*}
holds. In particular, we have
\begin{equation}
\sum_{n=0}^{\infty} |a_n|^2 n! < \infty \ .
\end{equation}
Conversely, a power series with coefficients fulfilling (6.32) converges
everywhere, and the function $f$ defined by (6.31) satisfies (6.30). Hence,
the elements of $ \hat{\cal H} $ are just the entire functions
$ z \mapsto f(z) = \sum_{n=0}^{\infty} a_n z^n $ with the property (6.32). This
implies that $ \hat{\cal H} $ is, as a unitary space, isomorphic to the
Hilbert space of all sequences $ \{ a_n \}_{n \in \N} $ satisfying
$ \sum_{n=0}^{\infty} |a_n|^2 n! < \infty $; in consequence, $ \hat{\cal H} $
itself is a Hilbert space.

It is easily seen that the functions $ \hat{\phi}_n \in \hat{\cal H} $
given by
\[
\hat{\phi}_n(z) := \frac{1}{\sqrt{n!}} z^n
\]
form an orthonormal system. To show that $ \{ \hat{\phi}_n \}_{n \in \N_0} $
is complete, consider an arbitrary element $ f \in \hat{\cal H} $ with its
power series expansion $ f(z) = \sum_{n=0}^{\infty} a_n z^n $ and define
$ f_N(z) := \sum_{n=0}^{N} a_n z^n $. Analogously to the calculation
leading to (6.32), we obtain
\[
\| f_N - f \|^2 = \frac{1}{\pi} \int \left|
                  \sum_{n=N+1}^{\infty} a_n z^n \right|^2 e^{-|z|^2} d\xi d\eta
                = \sum_{n=N+1}^{\infty} |a_n|^2 n! \ .
\]
Because of (6.32) it follows that $ \| f_N - f \|^2 < \epsilon^2 $ for
almost all $N$. Hence,
\[
f = \| \, . \, \|\mbox{--} \! \lim_{N \rightarrow \infty} f_N
  = \sum_{n=0}^{\infty} a_n \sqrt{n!} \hat{\phi}_n
\]
where the sum converges with respect to the norm of $ \hat{\cal H} $. Thus,
$ \{ \hat{\phi}_n \}_{n \in \N_0} $ is a complete orthonormal system; in
particular, $ \langle \hat{\phi}_n | f \rangle = a_n \sqrt{n!} $ and
\begin{equation}
\|f\|^2 = \sum_{n=0}^{\infty} |\langle \hat{\phi}_n | f \rangle|^2
        = \sum_{n=0}^{\infty} |a_n|^2 n! \ .
\end{equation}

Next we define, for each $ \psi \in {\cal H} $, a function
$ f \!\! : \C \, \rightarrow \C \, $ by
\begin{eqnarray*}
f(z) & := & \sqrt{2\pi} e^{\frac{|z|^2}{2}} e^{i\frac{qp}{2}} (V\psi)(q,p)
        =   e^{\frac{|z|^2}{2}} e^{i\frac{qp}{2}}
                                    \langle u^{\sigma}_{qp}|\psi \rangle \\
     &  = & e^{\frac{|z|^2}{2}} \int
            \overline{e^{ip(x-\frac{q}{2})} u^{\sigma}(x-q)} \psi(x) dx
\end{eqnarray*}
where $u$ has been chosen according to Eq.\ (6.19) and
$ z = \xi + i\eta := \frac{1}{2} (\frac{q}{\sigma} - 2i\sigma p) $. Setting
$ V\psi = \Psi $, one can write
\begin{equation}
f(z) = f(\xi,\eta) = \sqrt{2\pi} e^{\frac{\xi^2 + \eta^2}{2} -i\xi\eta}
                                 \Psi (2\sigma\xi,-\frac{\eta}{\sigma}) \ .
\end{equation}
From Theorem 6.3 it follows that $ f \in C^{\infty}_{\C} (\R^2) $. Moreover,
$f$ is even holomorphic. Namely, from
\begin{eqnarray*}
\Psi(q,p) & = & \frac{1}{\sqrt{2\pi}} \int
                         \overline{u^{\sigma}_{qp}(x)} \psi(x) dx \\
\frac{\partial \Psi}{\partial q}(q,p) & = & \frac{1}{\sqrt{2\pi}} \int
   \frac{x-q}{2\sigma^2} \overline{u^{\sigma}_{qp}(x)} \psi(x) dx \\
\frac{\partial \Psi}{\partial p}(q,p) & = & \frac{1}{\sqrt{2\pi}} \int
                   (-ix) \overline{u^{\sigma}_{qp}(x)} \psi(x) dx \\
\end{eqnarray*}
we obtain
\begin{equation}
\frac{\partial \Psi}{\partial p}(q,p) = -i\left(
2\sigma^2 \frac{\partial \Psi}{\partial q}(q,p) + q\Psi(q,p) \right)
\end{equation}
(compare Eq.\ (6.23)). According to (6.34) we have
\begin{equation}
\frac{\partial f}{\partial \xi}(\xi,\eta)
= \sqrt{2\pi} e^{\frac{\xi^2 + \eta^2}{2} -i\xi\eta}
\left[ (\xi - i\eta) \Psi(2\sigma\xi,-\frac{\eta}{\sigma})
+ 2\sigma \frac{\partial \Psi}{\partial q} (2\sigma\xi,-\frac{\eta}{\sigma})
                                                                    \right]
\end{equation}
and
\begin{equation}
\frac{\partial f}{\partial \eta}(\xi,\eta)
= \sqrt{2\pi} e^{\frac{\xi^2 + \eta^2}{2} -i\xi\eta}
\left[ (\eta - i\xi) \Psi(2\sigma\xi,-\frac{\eta}{\sigma})
- \frac{1}{\sigma} \frac{\partial \Psi}{\partial p}
                                (2\sigma\xi,-\frac{\eta}{\sigma}) \right] \ .
\end{equation}
Eliminating $ \frac{\partial \Psi}{\partial p} $ in Eq.\ (6.37) by (6.35),
it follows
\begin{equation}
\frac{\partial f}{\partial \eta}(\xi,\eta)
= \sqrt{2\pi} e^{\frac{\xi^2 + \eta^2}{2} -i\xi\eta}
\left[ (\eta + i\xi) \Psi(2\sigma\xi,-\frac{\eta}{\sigma})
+ 2i\sigma \frac{\partial \Psi}{\partial q}
                              (2\sigma\xi,-\frac{\eta}{\sigma}) \right] \ .
\end{equation}
By comparison, Eqs.\ (6.36) and (6.38) imply
\[
\frac{\partial f}{\partial \eta} = i\frac{\partial f}{\partial \xi} \ ,
\]
the latter being equivalent to Cauchy--Riemann's differential
equations. \linebreak Hence, we have shown that $f$ is a holomorphic function.

Furthermore, $f$ satisfies (6.30). In fact,
\begin{equation}
\begin{array}{ccl}
\|f\|^2 & = & \displaystyle{\frac{1}{\pi} \int |f(z)|^2 e^{-|z|^2} d\xi d\eta
          =   2 \int e^{|z|^2} |\Psi (2\sigma\xi,-\frac{\eta}{\sigma})|^2
                                 e^{-|z|^2} d\xi d\eta}   \vspace{3mm}\\
        & = & \displaystyle{\int |\Psi(q,p)|^2 dqdp = \| \Psi \|^2
          =   \| V\psi \|^2 = \| \psi \|^2 \ .}
\end{array}
\end{equation}
Hence, $ f \in \hat{\cal H} $, and the operator
$ \hat{V} \!\! : {\cal H} \rightarrow \hat{\cal H} $ defined by
\begin{equation}
(\hat{V}\psi)(z) := f(z)
 = \sqrt{2\pi} e^{\frac{|z|^2}{2}} e^{i\frac{qp}{2}} (V\psi)(q,p)
 = e^{\frac{|z|^2}{2}} e^{i\frac{qp}{2}} \langle u^{\sigma}_{qp}|\psi \rangle
\end{equation}
where $ z = \xi + i\eta = \frac{1}{2} (\frac{q}{\sigma} - 2i\sigma p) $,
is an isometry. Thus, $ \hat{V}{\cal H} $ is a Hilbert space of
entire functions; we have
\[
\hat{V}{\cal H} \subseteq \hat{\cal H} \subset
L^{2}_{\C} \, (\R^2,\frac{1}{\pi} e^{-(\xi^2 + \eta^2)} d\xi d\eta) \ .
\]
Finally, $ \hat{V}{\cal H} $ coincides with $ \hat{\cal H} $, as we are
going to prove.

To that end, consider the harmonic oscillator. The operator
\begin{equation}
H := \frac{1}{2m}P^2 + \frac{m\omega^2}{2}Q^2
   = -\frac{1}{2m} \frac{d^2}{dx^2} + \frac{m\omega^2}{2} x^2
   = \omega \left( a^*a + \frac{1}{2}1 \right)
\end{equation}
as well as the operators $a$ and $ a^* $ can be defined on
$ S_{\C} \, (\R) \subset {\cal H} $, the latter two ones are given by
\begin{equation}
\begin{array}{rcl}
a   & := & \displaystyle{\frac{1}{\sqrt{2}} \left( \sqrt{m\omega}Q +
           \frac{i}{\sqrt{m\omega}}P \right)}  \vspace{0.3cm}\\
a^* & := & \displaystyle{\frac{1}{\sqrt{2}} \left( \sqrt{m\omega}Q -
           \frac{i}{\sqrt{m\omega}}P \right)} \ .
\end{array}
\end{equation}
On $ S_{\C} \, (\R) $, $H$ is a symmetric operator. As is well known,
the eigenvalue problem of $H$ reads
\[
H\phi_n = E_n \phi_n
\]
where $ E_n = \omega (n + \frac{1}{2}) $, $ n = 0,1,2,\ldots \ $, and the
functions $ \phi_n \in S_{\C} \, (\R) $ form a complete orthonormal system
of $ \cal H $. In consequence, $H$ is essentially self-adjoint. Its
self-adjoint extension, again denoted by $H$, is defined on
\[
D(H) := \left\{ \psi \in {\cal H} \, \left| \,
        \sum_{n=0}^{\infty} | \alpha_n |^2 E^2_n < \infty \right. \right\}
\]
by
\[
H\psi  = H \left( \sum_{n=0}^{\infty} \alpha_n \phi_n \right)
      := \sum_{n=0}^{\infty} \alpha_n H\phi_n
       = \sum_{n=0}^{\infty} E_n \alpha_n \phi_n
\]
where $ \alpha_n := \langle \phi_n | \psi \rangle $; this self-adjoint
operator $H$ is the Hamiltonian of the harmonic oscillator. Furthermore,
$a$ and $ a^* $ can also be defined on $ D(H) $, $ a^* $ is then the adjoint
of $a$. Finally, we recall that
\begin{equation}
\phi_n = \frac{1}{\sqrt{n!}} (a^*)^n \phi_0
\end{equation}
and
\begin{equation}
\phi_0 (x) = \frac{1}{\sqrt[4]{2\pi\sigma^2}} e^{-\frac{x^2}{4\sigma^2}}
           = u^{\sigma}(x)
\end{equation}
where $ \sigma := \frac{1}{\sqrt{2m\omega}} $.

An easy calculation shows that the coherent states
$ u^{\sigma}_{qp} = U_{qp} u^{\sigma} = $ \linebreak
$ e^{ipQ} e^{-iqP} u^{\sigma} $ with $ \sigma = \frac{1}{\sqrt{2m\omega}} $
are eigenfunctions of $a$. Namely, from
\[
a = \frac{1}{\sqrt{2}}
    \left( \sqrt{m\omega}Q + \frac{i}{\sqrt{m\omega}}P \right)
  = \frac{1}{2} \left( \frac{1}{\sigma} Q + 2i\sigma P \right)
  = \frac{1}{2} \left( \frac{1}{\sigma} x + 2\sigma \frac{d}{dx} \right) \ ,
\]
$ u^{\sigma}_{qp} (x) = e^{ipx} u^{\sigma} (x-q) $, and (6.19) it follows that
\begin{equation}
au^{\sigma}_{qp} = \overline{z} u^{\sigma}_{qp}
\end{equation}
holds with
\[
\overline{z} = \xi - i\eta
             = \frac{1}{2} \left( \frac{q}{\sigma} + 2i\sigma p \right) \ .
\]
From Eqs.\ (6.43), (6.45), and (6.44) we obtain
\begin{eqnarray*}
\langle \phi_n | u^{\sigma}_{qp} \rangle
& = & \frac{1}{\sqrt{n!}} \langle \phi_0 | a^n u^{\sigma}_{qp} \rangle
  =   \frac{\overline{z}^n}{\sqrt{n!}}
      \langle \phi_0 | u^{\sigma}_{qp} \rangle \\
& = & \frac{\overline{z}^n}{\sqrt{n!}} \frac{1}{\sqrt{2\pi\sigma^2}}
      \int e^{-\frac{x^2}{4\sigma^2} + ipx - \frac{(x-q)^2}{4\sigma^2}} dx \\
& = & \frac{\overline{z}^n}{\sqrt{n!}} \frac{1}{\sqrt{2\pi\sigma^2}}
      e^{-\frac{q^2}{8\sigma^2}}
      \int e^{ipx} e^{-\frac{(x - \frac{q}{2})^2}{2\sigma^2}} dx \\
& = & \frac{\overline{z}^n}{\sqrt{n!}} \frac{1}{\sqrt{2\pi\sigma^2}}
      e^{-\frac{q^2}{8\sigma^2}} e^{i\frac{qp}{2}}
      \int e^{ipx} e^{-\frac{x^2}{2\sigma^2}} dx \\
& = & \frac{\overline{z}^n}{\sqrt{n!}} e^{i\frac{qp}{2}}
      e^{-\left( \frac{q^2}{8\sigma^2} + \frac{\sigma^2 p^2}{2} \right)} \\
& = & \frac{\overline{z}^n}{\sqrt{n!}} e^{i\frac{qp}{2}}
      e^{-\frac{\xi^2 + \eta^2}{2}} \ ,
\end{eqnarray*}
that is,
\begin{equation}
\langle \phi_n | u^{\sigma}_{qp} \rangle
= e^{i\frac{qp}{2}} e^{-\frac{|z|^2}{2}} \frac{\overline{z}^n}{\sqrt{n!}} \ .
\end{equation}
Now, Eqs.\ (6.40) and (6.46) imply
\[
(\hat{V}\phi_n)(z) = e^{\frac{|z|^2}{2}} e^{i\frac{qp}{2}}
                        \langle u^{\sigma}_{qp} | \phi_n \rangle
 = \frac{z^n}{\sqrt{n!}} = \hat{\phi}_n (z) \ ,
\]
that is,
\begin{equation}
\hat{V}\phi_n = \hat{\phi}_n
\end{equation}
for all $ n \in \N_0 $. Since $ \{ \phi_n \}_{n \in \N_0} $ and
$ \{ \hat{\phi}_n \}_{n \in \N_0} $ are complete orthonormal systems in
$ \cal H $ and $ \hat{\cal H} $, respectively, $ \hat{V} $ is an isometry
from $ \cal H $ onto $ \hat{\cal H} $, i.e.\ a unitary map. In particular,
$ \hat{V}{\cal H} = \hat{\cal H} $.

Summarizing, we have proved the following theorem.

\begin{theorem}
The space $ \hat{\cal H} $ of all entire functions $f$ with the property
\linebreak $ \int |f(z)|^2 e^{-|z|^2} d\xi d\eta < \infty $
is a closed subspace of
$ L^{2}_{\C} \, (\R^2,\frac{1}{\pi} e^{-(\xi^2 + \eta^2)} d\xi d\eta) $
and hence itself a Hilbert space. An arbitrary function \vspace{2pt}
$ z \mapsto f(z) = \sum_{n=0}^{\infty} a_n z^n $ belongs to $ \hat{\cal H} $
if and only if $ \sum_{n=0}^{\infty} |a_n|^2 n! < \infty $. The functions
$ z \mapsto \hat{\phi}_n (z) = \frac{1}{\sqrt{n!}} z^n $, $ n \in \N_0 $,
constitute a complete orthonormal system of $ \hat{\cal H} $;
the representation $ f = \sum_{n=0}^{\infty} \alpha_n \hat{\phi}_n $
is related to the power-series expansion of $f$ \vspace{1pt} by
$ \alpha_n = \langle \hat{\phi}_n | f \rangle = a_n \sqrt{n!} $.

For $ N = 1 $ and $ u = u^{\sigma} $ according to (6.19), the isometry $V$
given by Eq.\ (6.1) relates the space $ {\cal H} = L^{2}_{\C} \, (\R,dx) $
with a Hilbert space of infinitely differentiable functions, $ V{\cal H}
\subset L^{2}_{\C} \, (\R^2,dqdp) \cap C^{\infty}_{\C} (\R^2) $. Moreover,
a further isometry $ \hat{V} $ is defined by Eq.\ (6.40); $ \hat{V} $
relates $ \cal H $ \vspace{1pt} with a Hilbert space of holomorphic functions,
$ \hat{V}{\cal H} = \hat{\cal H} \subset L^{2}_{\C} \,
(\R^2,\frac{1}{\pi} e^{-(\xi^2 + \eta^2)} d\xi d\eta) $. In particular,
$ \hat{V}\phi_n = \hat{\phi}_n $ holds where $ \phi_0,\phi_1,\phi_2,\ldots $
are the usual energy eigenfunctions of the harmonic oscillator with
$ m\omega = \frac{1}{2\sigma^2} $.
\end{theorem}

We remark that the proof of the fact that $ \hat{V} $ is a unitary map from
$ \cal H $ onto $ \hat{\cal H} $ can even be simplified. Namely, for
\[
f(z) := e^{\frac{|z|^2}{2}} e^{i\frac{qp}{2}}
                                \langle u^{\sigma}_{qp} | \psi \rangle
\]
we obtain, using $ \psi = \sum_{n=0}^{\infty} \alpha_n \phi_n $ and
Eq.\ (6.46), the power-series expansion
\[
f(z) = \sum_{n=0}^{\infty} \frac{\alpha_n}{\sqrt{n!}} z^n
     = \sum_{n=0}^{\infty} a_n z^n
\]
where $ a_n := \frac{\alpha_n}{\sqrt{n!}} $ and $ z \in \C \, $. In
consequence, $f$ is an entire function; more- \linebreak over,
\[
\sum_{n=0}^{\infty} |a_n|^2 n! = \sum_{n=0}^{\infty} |\alpha_n|^2 < \infty \ ,
\]
i.e., $ f \in \hat{\cal H} $. Hence, by (6.40) a linear map
$ \hat{V} \!\! : {\cal H} \rightarrow \hat{\cal H} $ is defined. From (6.33)
and $ \sum_{n=0}^{\infty} |a_n|^2 n! = \sum_{n=0}^{\infty} |\alpha_n|^2
= \|\psi\|^2 $ it follows that $ \hat{V} $ is an isometry, and (6.47) again
implies that it is unitary.---This proof does not make use of Eqs.\ (6.35)
and (6.39), the former one, however, shows explicitly the relationship of
$ \hat{\cal H} $ and $ \hat{V} $ to $ V{\cal H} $ and $V$.

We conclude with the transformation of our operators $Q$, $P$, $a$, $ a^* $,
\vspace{1pt} and $H$ under $ \hat{V} $. On their respective domains
$ \hat{V}D(Q) $, $ \hat{V}D(P) $, and $ \hat{V}D(H) $, the transformed
operators act according to
\begin{eqnarray*}
\hat{V} Q \hat{V}^{-1}   & = & \sigma \left( z + \frac{d}{dz} \right)  \\
\hat{V} P \hat{V}^{-1}   & = & \frac{i}{2\sigma}
                                      \left( z - \frac{d}{dz} \right)  \\
\hat{V} a \hat{V}^{-1}   & = & \frac{d}{dz}                            \\
\hat{V} a^* \hat{V}^{-1} & = & z                                       \\
\hat{V} H \hat{V}^{-1}   & = & \omega \left( z\frac{d}{dz}
                                            + \frac{1}{2} \right) \ ,  \\
\end{eqnarray*}
as follows from Proposition 6.5 and Eqs.\ (6.40) -- (6.42) by an easy
calculation.

\chapter{Classical Representations of Quantum Mechanics on Phase Space}

In this chapter, we specify the concepts introduced in Chapter 3 and discuss,
on the basis of statistically complete joint position-momentum observables,
the representation of quantum mechanics on phase space. In Section 7.1,
we first investigate the question of the statistical completeness of joint
position-momentum observables and consider then the corresponding classical
representations and dequantizations on phase space, the latter in the sense
of Section 3.2 for bounded observables. The dequantization of some
unbounded observables is the subject of Section 7.3, Section 7.2 deals
with a mathematical question occurring in the context of unbounded
observables. Finally, in Section 7.4 the reformulation of quantum dynamics
in terms of equations of motion on phase space is discussed.

\section{\sloppy Statistically Complete Joint Position-Momentum Observables}

The joint position-momentum observable $F$ according to Eq.\ (5.25) which is
a generalization of the observable (5.16), respectively, (6.2), can be used
to define a classical representation of quantum mechanics on phase space in
the sense of Definition 3.1, provided that $F$ is statistically complete. The
latter question was investigated by S. T. Ali and E. Prugove\v{c}ki (1977b)
with Theorem 7.4 below as result. To prove that theorem, we need some lemmata
which are interesting themselves.

Again, let $ {\cal H} = L^{2}_{\C} \, (\R^N,d^N \! x) $ and let $ U_{qp} $ be
the unitary operators (5.10). The first lemma generalizes Theorem 5.1.

\begin{lemma}
Let $ u,v \in {\cal H} $ be unit vectors and define $ u_{qp} := U_{qp}u $
and $ v_{qp} := U_{qp}v $. Then
\begin{equation}
\frac{1}{(2\pi)^N} \int |u_{qp}\rangle \langle v_{qp}| \, d^N \! q \, d^N \! p
                                     = \langle v|u \rangle  \, 1
\end{equation}
holds where the integral exists in the weak sense.
\end{lemma}
{\bf Proof:} Using Eq.\ (5.13) and the Fourier-Plancherel theorem, we obtain
\begin{eqnarray*}
\langle \psi | \, \langle v|u \rangle \, 1 \psi \rangle
& = & \|\psi\|^2 \langle v|u \rangle
  =   \int |\psi(q)|^2 \, d^N \! q \int \overline{v(x)} u(x) \, d^N \! x    \\
& = & \int \overline{v(x)} u(x) \int |\psi(x+q)|^2 \, d^N \! q \, d^N \! x  \\
& = & \int \int \overline{v(x)} u(x) |\psi(x+q)|^2 \, d^N \! x \, d^N \! q  \\
& = & \int \int \overline{\overline{u}(x-q)\psi(x)} \overline{v}(x-q)\psi(x)
                                                   \, d^N \! x \, d^N \! q  \\
& = & \int \int \overline{F(\overline{u}( \, . \, -q)\psi)(p)}
               F(\overline{v}( \, . \, -q)\psi)(p) \, d^N \! p \, d^N \! q  \\
& = & \int \int \frac{1}{(2\pi)^N} \, \langle \psi | u_{qp} \rangle
                     \langle v_{qp} | \psi \rangle \, d^N \! p \, d^N \! q  \\
& = & \frac{1}{(2\pi)^N} \int \langle \psi | u_{qp} \rangle
                     \langle v_{qp} | \psi \rangle \, d^N \! q \, d^N \! p
\end{eqnarray*}
where $ \psi \in {\cal H} $ is arbitrary. Hence, the lemma
is proved. \hfill $ \Box $\\

There is another, interesting proof of Lemma 7.1 which should be
presented. Since, as a consequence of Theorem 5.1, the functions
$ (q,p) \mapsto \langle u_{qp} | \phi \rangle $ and
$ (q,p) \mapsto \langle v_{qp} | \psi \rangle $ with
$ \phi,\psi \in {\cal H} $ are square-integrable with $ L^2 $-norm
$ \sqrt{(2\pi)^N} \|\phi\| $ and $ \sqrt{(2\pi)^N} \|\psi\| $, respectively,
it follows by means of the Schwarz inequality that
\[
\left| \int \langle \phi | u_{qp} \rangle
            \langle v_{qp} | \psi \rangle \, d^N \! q \, d^N \! p \right|
\leq (2\pi)^N \|\phi\| \, \|\psi\| \ .
\]
Hence, $ (\phi,\psi) \mapsto \int \langle \phi | u_{qp} \rangle
\langle v_{qp} | \psi \rangle \, d^N \! q \, d^N \! p $ is
a bounded sesquilinear functional, and the weak integral
$ A := \int |u_{qp}\rangle \langle v_{qp}| \, d^N \! q \, d^N \! p $
exists. It is easy to show that the operator $A$ commutes with all
$ U_{qp} $. The irreducibility of the representation $ (q,p) \mapsto U_{qp} $
then implies that $ A = \alpha 1 $ with $ \alpha \in \C \: $. It remains
to determine the value of $ \alpha $. According to Eq.\ (5.11), we obtain
\begin{eqnarray*}
\langle v|u \rangle
& = & \frac{1}{(2\pi)^N} \int \langle v|u_{qp} \rangle
                 \langle u_{qp}|u \rangle \, d^N \! q \, d^N \! p       \\
& = & \frac{1}{(2\pi)^N} \int \langle U_{-q,-p}v|u \rangle
                 \langle u|U_{-q,-p}u \rangle \, d^N \! q \, d^N \! p   \\
& = & \frac{1}{(2\pi)^N} \int \langle u|u_{qp} \rangle
                 \langle v_{qp}|u \rangle \, d^N \! q \, d^N \! p       \\
& = & \frac{1}{(2\pi)^N} \, \langle u|Au \rangle                        \\
& = & \frac{\alpha}{(2\pi)^N}
\end{eqnarray*}
where the invariance of Lebesgue measure under reflections has
been used. Hence, $ \alpha = (2\pi)^N \langle v|u \rangle $ and
$ A = (2\pi)^N \langle v|u \rangle \, 1 $, the latter being equivalent
to Eq.\ (7.1).

The next lemma is mainly technical.

\begin{lemma}
Let $ W = \sum_{i=1}^{\infty} \alpha_i P_{\chi_i} \in K({\cal H}) $
be any density operator where $ \alpha_i \geq 0 $,
$ \sum_{i=1}^{\infty} \alpha_i = 1 $, $ \|\chi_i\| = 1 $, and
$ P_{\chi_i} = |\chi_i\rangle \langle\chi_i| $. Then \vspace{1pt}
$ (q,p) \mapsto {\rm tr} \, WU_{qp} $ is a bounded continuous function
of $ L^{2}_{\C} \, (\R^{2N},d^N \! q \, d^N \! p) $. Moreover, in the
representation
\[
{\rm tr} \, WU_{qp}
       = \sum_{i=1}^{\infty} \alpha_i \, {\rm tr} \, P_{\chi_i}U_{qp}
       = \sum_{i=1}^{\infty} \alpha_i \langle \chi_i | U_{qp}\chi_i \rangle
\]
the series converges pointwise uniformly as well as with respect to
the $ L^2 $-norm.
\end{lemma}
{\bf Proof:} Since the functions
$ (q,p) \mapsto \langle \chi_i | U_{qp}\chi_i \rangle $
are continuous and bounded by $1$,
$ \sum_{i=1}^{\infty} \alpha_i \langle \chi_i | U_{qp}\chi_i \rangle $
is a uniformly convergent series of continuous functions. In consequence,
$ (q,p) \mapsto {\rm tr} \, WU_{qp} $ is also continuous, furthermore,
$ |{\rm tr} \, WU_{qp}| \leq 1 $. Now let $V$ be the isometry (6.1) induced
by any $ \chi_i $, i.e. $ u := \chi_i $. From
\[
(V\chi_i)(q,p) = \frac{1}{\sqrt{(2\pi)^N}}
                          \langle U_{qp}\chi_i | \chi_i \rangle
\]
it follows that $ (q,p) \mapsto \langle \chi_i | U_{qp}\chi_i \rangle $ is
square-integrable with $ L^2 $-norm $ \sqrt{(2\pi)^N} $. This implies
\begin{eqnarray*}
\int |{\rm tr} \, WU_{qp}|^2 \, d^N \! q \, d^N \! p
& \leq & \sum_{i,j=1}^{\infty} \alpha_i \alpha_j
         \int |\langle \chi_i | U_{qp}\chi_i \rangle| \,
              |\langle \chi_j | U_{qp}\chi_j \rangle|
                                             \, d^N \! q \, d^N \! p     \\
& \leq & \sum_{i,j=1}^{\infty} \alpha_i \alpha_j
                               \sqrt{(2\pi)^N} \sqrt{(2\pi)^N}           \\
&   =  & (2\pi)^N
\end{eqnarray*}
where the Schwarz inequality has been used. Hence,
$ (q,p) \mapsto {\rm tr} \, WU_{qp} $ is also square-integrable, and from
\[
\int \left| \sum_{i=n+1}^{\infty} \alpha_i
     \langle \chi_i | U_{qp}\chi_i \rangle \right|^2 d^N \! q \, d^N \! p
\leq \sum_{i,j=n+1}^{\infty} \alpha_i \alpha_j (2\pi)^N \rightarrow 0
\]
for $ n \rightarrow \infty $ we obtain that
$ \sum_{i=1}^{\infty} \alpha_i \langle \chi_i | U_{qp}\chi_i \rangle $
converges also in $ L^2 $-norm. \hfill $ \Box $\\

The third lemma asserts that the trace-class operators can, roughly speaking,
be represented as continuous linear combinations of the unitary operators
$ U^{\ast}_{qp} $.

\begin{lemma}
Every (not necessarily self-adjoint) trace-class operator
$ V \in {\cal T}({\cal H}) $ can be represented by a weak integral
over the phase space according to
\begin{equation}
V = \frac{1}{(2\pi)^N} \int ({\rm tr} \, VU_{qp}) \, U^{\ast}_{qp}
                                      \, d^N \! q \, d^N \! p \ .
\end{equation}
Moreover, we have
\begin{equation}
\langle V_1 | V_2 \rangle_{HS} = {\rm tr} \, V^{\ast}_{1}V_{2}
     = \frac{1}{(2\pi)^N} \int \overline{{\rm tr} \, V_{1}U_{qp}} \,
                 {\rm tr} \, V_{2}U_{qp} \, d^N \! q \, d^N \! p
\end{equation}
where $ \langle V_1 | V_2 \rangle_{HS} $ denotes the Hilbert-Schmidt scalar
product of the trace-class operators $ V_1,V_2 \in {\cal T}({\cal H}) $.
\end{lemma}
{\bf Proof:} Let $ W \in K({\cal H}) $ be a density operator and
$ \psi \in {\cal H} $ be an arbitrary vector. According to Lemma 7.2,
$ (q,p) \mapsto {\rm tr} \, WU_{qp} = \sum_{i=1}^{\infty} \alpha_i
\langle \chi_i | U_{qp}\chi_i \rangle $ and
$ (q,p) \mapsto \langle \psi | U^{\ast}_{qp}\psi \rangle $
are square-integrable functions where the sum converges in
$ L^2 $-norm. Therefore, we obtain, using Lemma 7.1,
\begin{eqnarray*}
\lefteqn{\frac{1}{(2\pi)^N} \int ({\rm tr} \, WU_{qp})
           \langle \psi | U^{\ast}_{qp}\psi \rangle \, d^N \! q \, d^N \! p}
                                                               \hspace{4cm} \\
& = & \frac{1}{(2\pi)^N} \sum_{i=1}^{\infty} \alpha_i
      \int \langle \chi_i | U_{qp}\chi_i \rangle
           \langle \psi | U^{\ast}_{qp}\psi \rangle \, d^N \! q \, d^N \! p \\
& = & \sum_{i=1}^{\infty} \alpha_i \langle \psi | \chi_i \rangle
                                   \langle \chi_i | \psi \rangle            \\
& = & \langle \psi | W\psi \rangle \ .
\end{eqnarray*}
Eq.\ (7.2) is now implied by linearity. For $ V \in {\cal T}({\cal H}) $ and
$ W \in K({\cal H}) $, it follows from Eq.\ (7.2) and Lemma 7.2 that
\begin{eqnarray*}
\langle V|W \rangle_{HS} & = & {\rm tr} \, V^{\ast}W
  =   \sum_{i=1}^{\infty} \alpha_i \langle \chi_i | V^{\ast}\chi_i \rangle  \\
& = & \frac{1}{(2\pi)^N} \sum_{i=1}^{\infty} \alpha_i
      \int ({\rm tr} \, V^{\ast}U_{qp})
      \langle \chi_i | U^{\ast}_{qp}\chi_i \rangle \, d^N \! q \, d^N \! p  \\
& = & \frac{1}{(2\pi)^N} \sum_{i=1}^{\infty} \alpha_i
      \int \overline{{\rm tr} \, VU_{-q,-p}} \,
      \langle \chi_i | U_{-q,-p}\chi_i \rangle \, d^N \! q \, d^N \! p      \\
& = & \frac{1}{(2\pi)^N} \int \overline{{\rm tr} \, VU_{qp}} \,
                        {\rm tr} \, WU_{qp} \, d^N \! q \, d^N \! p \ .
\end{eqnarray*}
Hence, Eq.\ (7.3) is also valid. \hfill $ \Box $\\

We add some remarks which go beyond what we need in the sequel. Eq.~(7.2)
motivates to consider the map assigning to each trace-class operator
$ V \in {\cal T}({\cal H}) $ the function
$ (q,p) \mapsto {\rm tr} \, VU_{qp} $ of
$ L^{2}_{\C} \, (\R^{2N},d^N \! q \, d^N \! p) $. According to (7.3),
this linear map is an isometry up to the factor $ \frac{1}{\sqrt{(2\pi)^N}} $,
provided that the space $ {\cal T}({\cal H}) $ is equipped with
the Hilbert-Schmidt norm. Since $ {\cal T}({\cal H}) $ is
$ \| \, . \, \|_{HS} $-dense in the space $ {\cal C}^{HS}({\cal H}) $
of all Hilbert-Schmidt operators, the considered map can uniquely be extended
to a map $ S \!\! : {\cal C}^{HS}({\cal H}) \rightarrow
L^{2}_{\C} \, (\R^{2N},d^N \! q \, d^N \! p) $ which is again an isometry
up to a factor. Now let $ f \in L^{2}_{\C} \, (\R^{2N},d^N \! q \, d^N \! p) $
be arbitrary and consider the map $ H \!\! :
L^{2}_{\C} \, (\R^{2N},d^N \! q \, d^N \! p) \rightarrow {\cal B}({\cal H}) $
defined by the weak integral
\[
Hf := \frac{1}{(2\pi)^N} \int f(q,p) U^{\ast}_{qp} \, d^N \! q \, d^N \! p \ ;
\]
by means of the Schwarz inequality it is easy to show that the weak integral
exists and that the linear map $H$ is bounded. According to (7.2), we have
for $ V \in {\cal T}({\cal H}) $ that $ HSV = V $. Writing
$ V \in {\cal C}^{HS}({\cal H}) $ as
$ V = \| \, . \, \|_{HS} $-$ \lim_{n \rightarrow \infty} V_n
= \| \, . \, \| $-$ \lim_{n \rightarrow \infty} V_n $
with some sequence of operators $ V_n \in {\cal T}({\cal H}) $ where
$ \| \, . \, \| $ is the usual operator norm of $ {\cal B}({\cal H}) $,
we obtain that $ HSV = V $ is valid for all
$ V \in {\cal C}^{HS}({\cal H}) $. Hence, $ \left. H \right|_{R(S)} = S^{-1} $
where $ \left. H \right|_{R(S)} $ denotes the restriction of $H$ to the range
of $S$, and $ \left. H \right|_{R(S)} $ is also an isometry up to a factor.

The range $ R(S) $ is the $ L^2 $-norm closure of the subspace
consisting of the functions $ (q,p) \mapsto {\rm tr} \, VU_{qp} $
with $ V \in {\cal T}({\cal H}) $. For $ g \in (R(S))^{\perp} $,
it follows that
\begin{eqnarray*}
\langle \psi | (Hg)\psi \rangle
& = & \frac{1}{(2\pi)^N} \int g(q,p) \langle U_{qp}\psi | \psi \rangle
                                     \, d^N\! q \, d^N \! p       \\
& = & \frac{1}{(2\pi)^N} \int \overline{{\rm tr} \, P_{\psi}U_{qp}} \, g(q,p)
                                     \, d^N\! q \, d^N \! p       \\
& = & 0
\end{eqnarray*}
where $ \psi \in {\cal H} $ is an arbitrary unit vector and
$ P_{\psi} = |\psi\rangle \langle\psi | $. Hence, \linebreak
$ Hg = 0 $. However, one can prove that even $ (R(S))^{\perp} = \{ 0 \} $
holds, i.e.\ $ R(S) = L^{2}_{\C} \, (\R^{2N},d^N \! q \, d^N \! p) $. In
particular, the functions $ (q,p) \mapsto {\rm tr} \, VU_{qp} $ with
$ V \in {\cal T}({\cal H}) $ are dense in
$ L^{2}_{\C} \, (\R^{2N},d^N \! q \, d^N \! p) $,
$S$ and $H$ are, up \vspace{1pt} to factors, unitary maps
between the Hilbert spaces $ {\cal C}^{HS}({\cal H}) $ and
$ L^{2}_{\C} \, (\R^{2N},d^N \! q \, d^N \! p) $, \linebreak and
$ H = S^{-1} $. The latter statements can be concluded from the fact
that our map $H$ is closely related to the Weyl correspondence (see
J. C. T. Pool, 1966). Similarly, our map $S$ is closely related to the
Wigner transformation (see E.~P.~Wigner, 1932; J. E. Moyal, 1949;
J. C. T. Pool, 1966; M. Hillery, R.~F.~O'Connell, M. O. Scully,
and E. P. Wigner, 1984).

After the preparing lemmata, we are able to present the proof of the
following theorem due to S. T. Ali and E. Prugove\v{c}ki (1977b).

\begin{theorem}
The joint position-momentum observable on $ (\R^{2N},\Xi(\R^{2N})) $ given by
\begin{equation}
F(B) = \frac{1}{(2\pi)^N} \int_{B} a_{qp} \, d^N \! q \, d^N \! p
\end{equation}
where $ a_{qp} = U_{qp}aU_{qp}^{\ast} $ and $ a \in K({\cal H}) $, is
statistically complete if
\begin{equation}
{\rm tr} \, aU_{qp} \neq 0
\end{equation}
holds for almost all $ (q,p) \in \R^{2N} $.
\end{theorem}
{\bf Proof:} The statistical completeness of the observable $F$ means that
\[
{\rm tr} \, W_1F(B) = {\rm tr} \, W_2F(B)
\]
for any $ W_1,W_2 \in K({\cal H}) $ and all $ B \in \Xi(\R^{2N}) $ implies
$ W_1 = W_ 2 $. According to the proof of Lemma 2.1, $F$ is statistically
complete if and only if from
\[
{\rm tr} \, VF(B) = 0
\]
for any $ V \in {\cal T}_s({\cal H}) $ and all $ B \in \Xi(\R^{2N}) $
it follows $ V = 0 $. In view of (7.4), $ {\rm tr} \, VF(B) = 0 $ is
equivalent to $ {\rm tr} \, Va_{qp} = 0 $ for \vspace{1pt} almost all
$ (q,p) \in \R^{2N} $, respectively, because of the continuity of
$ (q,p) \mapsto {\rm tr} \, Va_{qp} $, for all $ (q,p) $. Hence,
the observable (7.4) is statistically complete if and only if
\[
{\rm tr} \, Va_{qp} = 0
\]
for all $ (q,p) \in \R^{2N} $ implies $ V = 0 $.

Using Eq.\ (7.3) and $ U_{qp}U_{q'p'} = e^{-iqp'}U_{q+q',p+p'} $, we obtain
\begin{eqnarray*}
{\rm tr} \, Va_{qp} & = & {\rm tr} \, (U_{qp}a)^{\ast}VU_{qp}
  =   \langle U_{qp}a | VU_{qp} \rangle_{HS}                         \\
& = & \frac{1}{(2\pi)^N} \int \overline{{\rm tr} \, (aU_{q'p'}U_{qp})} \,
        {\rm tr} \, (VU_{qp}U_{q'p'}) \, d^N\! q' \, d^N \! p'       \\
& = & \frac{1}{(2\pi)^N} \int e^{i(q'p - qp')}
                              \overline{{\rm tr} \, aU_{q + q',p + p'}} \,
        {\rm tr} \, VU_{q + q',p + p'} \, d^N\! q' \, d^N \! p'      \\
& = & \frac{1}{(2\pi)^N} \int e^{i(q'p - p'q)}
                              \overline{{\rm tr} \, aU_{q'p'}} \,
        {\rm tr} \, VU_{q'p'} \, d^N\! q' \, d^N \! p' \ .
\end{eqnarray*}
We observe that
$ {\rm tr} \, Va_{qp} = \frac{1}{\sqrt{(2\pi)^N}} \tilde{h}(-p,q) $
holds where $ \tilde{h} $ is the Fourier transform of \vspace{1pt} the
$ L^1 $-function $ (q',p') \mapsto h(q',p')
:= \overline{{\rm tr} \, aU_{q'p'}} \, {\rm tr} \, VU_{q'p'} $. Since
the Fourier transformation also on
$ L^{1}_{\C} \, (\R^{2N},d^N \! q \, d^N \! p) $ is injective,
we have $ {\rm tr} \, Va_{qp} = 0 $ for all $ (q,p) \in \R^{2N} $,
respectively, $ \tilde{h} = 0 $ if and only if $ h = 0 $ is valid,
respectively,
\begin{equation}
\overline{{\rm tr} \, aU_{qp}} \, {\rm tr} \, VU_{qp} = 0
\end{equation}
for all $ (q,p) $. Hence, our observable $F$ is statistically complete
if and only if the validity of Eq.\ (7.6) for all $ (q,p) $ implies
$ V = 0 $.

Now assume that the inequality (7.5) is satisfied for almost all
$ (q,p) $. Then from Eq.\ (7.6) it follows that $ {\rm tr} \, VU_{qp} = 0 $
for all $ (q,p) $ and, by Eq.~(7.2), $ V = 0 $. Hence, $F$ is statistically
complete. \hfill $ \Box $\\

We have proved that the criterion (7.5) is sufficient for the statistical
completeness of the observable (7.4). S. T. Ali and E. Prugove\v{c}ki (1977b)
assert that condition (7.5) is also necessary for the statistical completeness
of (7.4); however, the author of this paper is not convinced of the
correctness of their argumentation. Further results on questions related
to Theorem 7.4 were obtained by R. Werner (1983, 1984).

For $ N = 1 $, consider the coherent state $ u^{\sigma} $ according to (6.19)
and $ a := |u^{\sigma}\rangle \langle u^{\sigma}| $. From Eqs.\ (6.44) and
(6.46) it then follows that
\begin{equation}
{\rm tr} \, aU_{qp} = \langle u^{\sigma} | U_{qp} u^{\sigma} \rangle
                    = \langle \phi_0 | u^{\sigma}_{qp} \rangle
                    = e^{i\frac{qp}{2}} e^{-\frac{|z|^2}{2}} \neq 0 \ .
\end{equation}
For $ N > 1 $, let
$ u(x) = u(x_1,\ldots,x_N) := u^{\sigma_1}(x_1) \ldots u^{\sigma_N}(x_N) $,
i.e.\ $ u = u^{\sigma_1} \otimes \ldots \otimes u^{\sigma_N} $, and
$ a := |u \rangle \langle u| $. From (7.7) we now obtain
\[
{\rm tr} \, aU_{qp} = \langle u | U_{qp} u \rangle
   = \prod_{j=1}^{N} \langle u^{\sigma_j} | U_{q_jp_j} u^{\sigma_j} \rangle
   \neq 0 \ .
\]
Hence, for every $N$, there exist statistically complete joint
position-momen- tum observables on the phase space $ \R^{2N} $, and
these observables may be of type (5.16). In particular, for $ N = 1 $,
the joint position-momentum observables generated by the coherent states
$ u^{\sigma} $ are statistically complete.

For a statistically complete joint position-momentum observable $F$,
the linear hull of $ F(\Xi(\R^{2N})) $ is, according to Lemma 2.1,
$ \sigma $-dense in the space $ {\cal B}_s({\cal H}) $. In contrast, the pair
of the PV-measures $ E^Q $ and $ E^P $ according to (5.6) and (5.7) is
not statistically complete (see, e.g., E. Prugove\v{c}ki, 1977a;
M. Singer and W. Stulpe, 1990), and consequently the linear hull of
$ E^Q(\Xi(\R^N)) \cup E^P(\Xi(\R^N)) $ is not $ \sigma $-dense in
$ {\cal B}_s({\cal H}) $. From the properties of the $ \sigma $-weak integral
(cf.\ W. Stulpe, 1986) it follows that the $ \sigma $-weak closure of the
linear hull of $ E^Q(\Xi(\R^N)) \cup E^P(\Xi(\R^N)) $ contains also the
operators $ G^Q(B) $ and $ G^P(B) $ where $ B \in \Xi(\R^N) $ and $ G^Q $
and $ G^P $ are the observables defined by (5.8) and (5.9). Hence, any pair
of an approximate position observable $ G^Q $ and an approximate momentum
observable $ G^P $ is, just as the pair of $ E^Q $ and $ E^P $, not
statistically complete. In particular, the pair of the marginal observables
$ F^Q $ and $ F^P $ of a joint position-momentum observable $F$ is not
statistically complete even if $F$ is statistically complete; by other
means, this result was already proved by S. T. Ali and E. Prugove\v{c}ki
(1977a).---Analogously to Lemma 2.1, the complex linear hull of
$ F(\Xi(\R^{2N})) $, $F$ being a statistically complete joint
position-momentum observable, is ultraweakly dense in the space
$ {\cal B}({\cal H}) $ of all bounded linear operators, whereas the complex
linear hull of $ E^Q(\Xi(\R^N)) \cup E^P(\Xi(\R^N)) $ is not ultraweakly dense
in $ {\cal B}({\cal H}) $; however, as is well known, the von Neumann algebra
generated by $ E^Q $ and $ E^P $ coincides with $ {\cal B}({\cal H}) $
(see, e.g., G. Emch, 1972).

Now let $F$ be a statistically complete joint position-momentum observable
according to (7.4) and $T$ the corresponding classical representation on
$ (\R^{2N},\Xi(\R^{2N})) $ according to Theorem 3.3. Since the integral (7.4)
exists in the $ \sigma $-weak sense, we have
\begin{equation}
\begin{array}{ccl}
(TV)(B) & = & {\rm tr} \, VF(B)
  =   \displaystyle{\frac{1}{(2\pi)^N} \int_{B} {\rm tr} \, Va_{qp} \,
                                 d^N \! q \, d^N \! p}   \vspace{3mm}\\
& = & \displaystyle{\int_{B} \rho(q,p) \, d^N \! q \, d^N \! p}
\end{array}
\end{equation}
where $ V \in {\cal T}_s({\cal H}) $, $ B \in \Xi(\R^{2N}) $, and
$ \rho(q,p) := \frac{1}{(2\pi)^N} \, {\rm tr} \, Va_{qp} $. In consequence,
a classical \vspace{1pt} representation $ \hat{T} $ on
$ (\R^{2N},\Xi(\R^{2N}),\lambda^{2N}) $, $ \hat{T} \!\! :
{\cal T}_s({\cal H}) \rightarrow L^{1}_{\R} (\R^{2N},d^N \! q \, d^N \! p) $,
is defined by
\begin{equation}
\hat{T}V := \rho \ .
\end{equation}
According to Eq.\ (3.6), the adjoint map $ \hat{T}' $ is given by the
$ \sigma $-weak integral
\[
\hat{T}'f = \int fdF
\]
where $ f \in L^{\infty}_{\R} (\R^{2N},d^N \! q \, d^N \! p) $. From (7.8)
it follows that
\begin{eqnarray*}
{\rm tr} \left( V\int fdF \right)
& = & \int f d({\rm tr} \, VF( \, . \, ))
  =   \frac{1}{(2\pi)^N} \int f(q,p) \, {\rm tr} \, Va_{qp} \,
                                             d^N \! q \, d^N \! p   \\
& = & {\rm tr} \left( V \frac{1}{(2\pi)^N} \int f(q,p) a_{qp} \,
                                             d^N \! q \, d^N \! p \right) \ .
\end{eqnarray*}
Hence,
\[
\int fdF = \frac{1}{(2\pi)^N} \int f(q,p) a_{qp} \, d^N \! q \, d^N \! p
\]
and
\begin{equation}
\hat{T}'f = \frac{1}{(2\pi)^N} \int f(q,p) a_{qp} \, d^N \! q \, d^N \! p
\end{equation}
where the latter integral exists also in the $ \sigma $-weak sense.

The next theorem restates Theorem 3.4, respectively, statement (3.8)
and concerns dequantizations.

\begin{theorem}
In the situation just described the following statements are valid:
\begin{enumerate}
\item[(a)] To each bounded self-adjoint \vspace{1pt} operator
$ A \in R(\hat{T}') = \hat{T}' L^{\infty}_{\R}(\R^{2N},d^N \! q \, d^N \! p) $
a function $ f \in L^{\infty}_{\R}(\R^{2N},d^N \! q \, d^N \! p) $
can be assigned such that for all states $ W \in K({\cal H}) $
\[
{\rm tr} \, WA = \int \rho(q,p) f(q,p) \, d^N \! q \, d^N \! p
\]
holds where $ \rho := \hat{T}W $ is the density of the probability
distribution $ P^F_W $.
\item[(b)] For every $ A \in {\cal B}_s({\cal H}) $, every $ \epsilon > 0 $,
and any finitely many states $ W_1,\ldots,W_n \in K({\cal H}) $ there exists
a function $ f \in L^{\infty}_{\R} (\R^{2N},d^N \! q \, d^N \! p) $ such that
\[
\left| {\rm tr} \, W_iA
   - \int \rho_i(q,p) f(q,p) \, d^N \! q \, d^N \! p \right| < \epsilon
\]
holds where $ \rho_i := \hat{T}W_i $ ($ i = 1,\ldots,n $).
\end{enumerate}
\end{theorem}

This result signifies a very far-reaching reformulation of the statistical
scheme of quantum mechanics in terms of the classical phase space. Namely,
probabilities and expectation values which appear in reality as relative
frequencies and mean values can be calculated on the basis of Hilbert space
and in principle also on the basis of phase space.

The proof of part (b) of Theorem 7.5 is based on the fact that, with respect
to the $ \sigma $-topology, every operator $ A \in {\cal B}_s({\cal H}) $ can
be \vspace{1pt} approximated by some operator of the form $ \hat{T}'f $ with
$ f \in L^{\infty}_{\R} (\R^{2N},d^N \! q \, d^N \! p) $ (compare the proof
of Theorem 3.4). In particular, the projections $ E^Q(b) $ and $ E^P(b) $
of the PV-measures for position and momentum, $ b \in \Xi(\R^N) $, can
be approximated physically arbitrarily well by operators of the form
(7.10). Roughly speaking, with suitable functions $ f^Q_b $ and $ f^P_b $
we have
\begin{equation}
\begin{array}{ccl}
E^Q(b) & \approx & {\displaystyle \frac{1}{(2\pi)^N}
         \int f^Q_b(q,p) a_{qp} \, d^N \! q \, d^N \! p} \vspace{3mm}\\
E^P(b) & \approx & {\displaystyle \frac{1}{(2\pi)^N}
         \int f^P_b(q,p) a_{qp} \, d^N \! q \, d^N \! p} \ .
\end{array}
\end{equation}
A similar approximation of $ E^Q(b) $ and $ E^P(b) $ is given by Eqs.\ (5.26):
\begin{equation}
\begin{array}{ccl}
F^Q (b) & = & {\displaystyle \frac{1}{(2\pi)^N}
               \int \chi_{b\times\R^N} a_{qp} \, d^N \! q \, d^N \! p}
                                                             \vspace{3mm}\\
        & = &  \displaystyle{\int \chi_b * \sum_{i=1}^{\infty}
               \lambda_i | u_i |^2 \, dE^Q} \approx E^Q(b)   \vspace{3mm}\\
F^P (b) & = & {\displaystyle \frac{1}{(2\pi)^N}
               \int \chi_{\R^N \times b} a_{qp} \, d^N \! q \, d^N \! p}
                                                             \vspace{3mm}\\
        & = &  \displaystyle{\int \chi_b * \sum_{i=1}^{\infty}
               \lambda_i | \tilde{u}_i |^2 \, dE^P} \approx E^P(b)
\end{array}
\end{equation}
($ a = \sum_{i=1}^{\infty} \lambda_i |u_i\rangle \langle u_i| $,
$ \lambda_i \geq 0 $, $ \sum_{i=1}^{\infty} \lambda_i = 1 $,
$ \|u_i\| = 1 $). However, it is intuitively clear by the properties of the
Fourier transformation that, if one of the approximations (7.12) is good, the
other one is bad. In contrast, both approximations (7.11) may be good. Hence,
in a certain sense, the observables of position and momentum both can be
approximated by functions of the joint position-momentum observable $F$
arbitrarily well.

\section{A Remark on the Trace}

Quantum observables described by PV-measures on $ \Xi(\R) $ correspond,
by the spectral theorem, uniquely to self-adjoint operators in $ \cal H $
which are generally unbounded. In the next section we discuss, in the context
of classical representations $ \hat{T} $ on phase space, dequantizations
for some unbounded self-adjoint operators $A$, i.e., we represent the
expressions $ {\rm tr} \, WA $ as integrals over the phase space. On this
occasion we present, in this section, a definition of the expression
$ {\rm tr} \, WA $ for unbounded self-adjoint operators; such traces are
used very often, but seldom rigorously defined.

Let $A$ be a (possibly unbounded) self-adjoint operator in $ \cal H $,
$ D(A) $ its domain, and $ E^A $ its spectral measure; let us interpret
$A$ as an observable. The expectation value of $A$ in a state
$ W \in K({\cal H}) $ is
\[
\langle A \rangle_W := \langle E^A \rangle_W = \int \xi P^A_W(d\xi)
\]
where
$ P^A_W(b) := P^{E^A}_{W}(b) = {\rm tr} \, WE^A(b) $ with $ b \in \Xi(\R) $,
provided that the integral exists. If $W$ is a pure state, i.e., if
$ W = P_{\psi} = |\psi\rangle \langle \psi | $, $ \psi \in {\cal H} $,
$ \|\psi\| = 1 $, then
\[
\langle A \rangle_{\psi} := \langle A \rangle_{P_{\psi}}
              = \int \xi \, \langle \psi | E^A(d\xi)\psi \rangle \ ,
\]
and the integral exists if and only if $ \psi \in D(|A|^{\frac{1}{2}}) $;
$ D(A) \subseteq D(|A|^{\frac{1}{2}}) \subseteq {\cal H} $. For
$ \psi \in D(A) $,
\begin{equation}
\langle A \rangle_{\psi} = \int \xi \, \langle \psi | E^A(d\xi)\psi \rangle
                         = \langle \psi | A\psi \rangle
\end{equation}
holds.

To obtain the analog of (7.13) for an arbitrary $ W \in K({\cal H}) $, assume
\begin{enumerate}
\item[{\rm (i)}] $ \int \xi \, {\rm tr} \, WE^A(d\xi) $ exists
\item[{\rm (ii)}] for some representation $ W = \sum_{i} \alpha_i P_{\chi_i} $
with $ \alpha_i \neq 0 $ and $ \langle \chi_i | \chi_j \rangle = \delta_{ij} $,
$ \chi_i \in D(A) $ holds.
\end{enumerate}
The vectors $ \chi_i $ form an orthonormal system of eigenvectors of $W$ that
is complete in $ (N(W))^{\perp} = \overline{R(W)} $, $ N(W) $ and $ R(W) $
denoting the kernel and the range of $W$, respectively. Since the eigenvalues
$ \alpha_i \neq 0 $ of $W$ have finite multiplicity, condition (ii) implies
that $ \chi_i \in D(A) $ is valid for every complete orthonormal system of
eigenvectors of $W$ in $ N(W))^{\bot} $. From (i) and (ii) it follows that
\begin{eqnarray*}
\langle A \rangle_W & = & \int \xi \, {\rm tr} \, WE^A(d\xi)
  =   \sum_{i} \alpha_i \int \xi \, {\rm tr} \, P_{\chi_i}E^A(d\xi)   \\
& = & \sum_{i} \alpha_i \int \xi \, \langle \chi_i | E^A(d\xi)\chi_i \rangle
                                                                      \\
& = & \sum_{i} \alpha_i \langle \chi_i | A\chi_i \rangle \ .
\end{eqnarray*}
If $A$ is a bounded self-adjoint operator, conditions (i) and (ii) are
always satisfied, and $ \sum_{i} \alpha_i \langle \chi_i | A\chi_i \rangle $
is just $ {\rm tr} \, WA $. If $A$ is an unbounded self-adjoint operator,
we suppose (i) and (ii) and define
\[
{\rm tr} \, WA := \sum_{i} \alpha_i \langle \chi_i | A\chi_i \rangle \ ;
\]
$ {\rm tr} \, WA $ does not depend on the complete orthonormal system of
eigenvectors of $W$ in $ (N(W))^{\perp} $. In particular,
\begin{equation}
\langle A \rangle_W = \int \xi \, {\rm tr} \, WE^A(d\xi) = {\rm tr} \, WA
\end{equation}
holds.

Finally, we remark that, if $A$ is unbounded, the above conditions (i) and
(ii) are independent of each other. Clearly, for $ W := P_{\psi} $ with
$ \psi \in D(|A|^{\frac{1}{2}}) \setminus D(A) $ condition (i) is fulfilled,
but not (ii). To show that condition (ii) does not imply (i), let $ A := H $
be the Hamiltonian of the harmonic oscillator,
$ H\phi_n = \omega (n + \frac{1}{2}) \phi_n $ with $ n = 0,1,2,\ldots \ $,
and $ W := \sum_{n=1}^{\infty} \frac{\alpha}{n^2} P_{\phi_n} $ with
$ \sum_{n=1}^{\infty} \frac{1}{n^2} =: \frac{1}{\alpha} $. Then
\begin{eqnarray*}
\langle H \rangle_W & = & \int \xi P^H_W(d\xi)
  =   \sum_{n=0}^{\infty} \omega \left( n + \frac{1}{2} \right)
                                {\rm tr} \, WP_{\phi_n}                     \\
& = & \sum_{n=0}^{\infty} \omega \left( n + \frac{1}{2} \right)
                                 \langle \phi_n | W\phi_n \rangle           \\
& = & \sum_{n=1}^{\infty} \omega \left( n + \frac{1}{2} \right)
                                 \frac{\alpha}{n^2}                         \\
& = & \frac{\omega}{2} + \alpha\omega \sum_{n=1}^{\infty} \frac{1}{n}       \\
& = & \infty \ .
\end{eqnarray*}
Hence, (i) does not hold, although (ii) does.

\section{\sloppy Dequantizations for Some Unbounded Self-Adjoint Operators}

Our next aim concerns the derivation of a classical expression for the
expectation values $ \langle A \rangle_W $ when $A$ is an unbounded
operator (cf.\ S. T. Ali and E.~Prugove\v{c}ki, 1977c; E. Prugove\v{c}ki,
1978, 1984). If $A$ is bounded, this problem is solved by Theorem 7.5. If $A$
is an unbounded self-adjoint operator, we have to look for an apparently
unbounded real-valued measurable function $f$ on phase space such that
\begin{equation}
\langle A \rangle_W = {\rm tr} \, WA
                    = \int \rho(q,p) f(q,p) \, d^N \! q \, d^N \! p
\end{equation}
holds for all $ W \in K({\cal H}) $ satisfying conditions (i) and (ii)
of Section 7.2 where $ \rho = \hat{T}W $, $ \hat{T} $ being a classical
representation according to (7.9).

Let $ N = 1 $, $ {\cal H} = L^{2}_{\C} \, (\R,dx) $, and
$ a := |u^{\sigma}\rangle \langle u^{\sigma}| $ with $ u^{\sigma} $
according to (6.19). Let $F$ then be the statistically complete observable
given by (7.4) and $ \hat{T} $ the corresponding classical representation on
$ (\R^2,\Xi(\R^2),\lambda^2) $ according to (7.9). Explicitly, $ \hat{T} $
is given by
\[
(\hat{T}W)(q,p) = \rho(q,p)
                = \frac{1}{2\pi} \, {\rm tr}
                  \, (W|u^{\sigma}_{qp} \rangle \langle u^{\sigma}_{qp}|)
                = \frac{1}{2\pi} \,
                  \langle u^{\sigma}_{qp} | W u^{\sigma}_{qp} \rangle
\]
where $ W \in K({\cal H}) $; in particular, for
$ W = P_{\psi} = |\psi\rangle \langle\psi | $, $ \psi \in {\cal H} $,
$ \|\psi\| = 1 $, we have
\begin{equation}
(\hat{T}P_{\psi})(q,p) = \rho(q,p)
         = \frac{1}{2\pi} \, |\langle u^{\sigma}_{qp} | \psi \rangle|^2
         = |(V\psi)(q,p)|^2 = |\Psi(q,p)|^2
\end{equation}
where $V$ and $ \Psi $ are defined by Eq.\ (6.1). For some unbounded operators
$A$ in $ \cal H $, we now calculate functions $f$ satisfying Eq.\ (7.15);
this calculation involves some ideas of E. Prugove\v{c}ki (1978, 1984).

First, we consider the position operator $Q$ in $ \cal H $. For
$ \psi \in D(Q) $ we obtain, using the unitary operator
$ V \!\! : {\cal H} \rightarrow V{\cal H} $, $ V\psi = \Psi $, and (6.21),
\[
\langle \psi | Q\psi \rangle = \langle \Psi | VQV^{-1}\Psi \rangle
= \left\langle \Psi \, \left| \, \left( q +
 2\sigma^2 \frac{\partial}{\partial q} \right) \Psi \right. \right\rangle \ .
\]
Taking account of Proposition 6.4, Corollary 6.6, and (6.20), it follows that
\[
\langle \psi | Q\psi \rangle = \langle \Psi | q\Psi \rangle
                             = \int q|\Psi(q,p)|^2 dqdp
\]
(compare also (6.24)). Hence, by (7.16),
\begin{equation}
\langle Q \rangle_{\psi} = \langle \psi | Q\psi \rangle = \int \rho(q,p)q dqdp
\end{equation}
holds for all $ \psi \in D(Q) $ with $ \|\psi\| = 1 $ where
$ \rho = \hat{T}P_{\psi} $. If $ W \in K({\cal H}) $ satisfies conditions
(i) and (ii) with respect to $Q$, Eq.\ (7.17) implies
\begin{eqnarray*}
\langle Q \rangle_W & = & {\rm tr} \, WQ = \sum_{i} \alpha_i
                                   \langle \chi_i | Q\chi_i \rangle       \\
& = & \sum_{i} \alpha_i \int (\hat{T}P_{\chi_i})(q,p)q dqdp \\
& = & \int \left( \sum_{i} \alpha_i \hat{T}P_{\chi_i} \right) (q,p) \, q dqdp
                                                                          \\
& = & \int \left( \hat{T} \left( \sum_{i} \alpha_i P_{\chi_i} \right) \right)
                                                        (q,p) \, q dqdp   \\
& = & \int (\hat{T}W)(q,p)q dqdp \ ;
\end{eqnarray*}
we have used that $ \sum_{i} \alpha_i P_{\chi_i} $, if it is an infinite sum,
converges in trace norm and thus $ \hat{T} (\sum_{i} \alpha_i P_{\chi_i})
= \sum_{i} \alpha_i \hat{T}P_{\chi_i} $, the latter sum converging in
$ L^1 $-norm as well as pointwise a.e. Hence,
\begin{equation}
\langle Q \rangle_W = {\rm tr} \, WQ = \int \rho(q,p)q dqdp
\end{equation}
where $ \rho = \hat{T}W $.

Second, we consider the momentum operator $P$ in $ \cal H $. For
$ \psi \in D(P) $ it follows from (6.16), Corollary 6.6, Proposition 6.4,
and (6.20) that
\begin{eqnarray*}
\langle \psi | P\psi \rangle & = & \langle \Psi | VPV^{-1}\Psi \rangle
  =   \left\langle \Psi \, \left| \, \left( p - i\frac{\partial}{\partial q}
                                     \right) \right. \Psi \right\rangle   \\
& = & \langle \Psi | p\Psi \rangle
  =   \int p|\Psi(q,p)|^2 dqdp
\end{eqnarray*}
(compare also (6.25)). Hence, by (7.16),
\[
\langle P \rangle_{\psi} = \langle \psi | P\psi \rangle = \int \rho(q,p)p dqdp
\]
holds for all $ \psi \in D(P) $ with $ \|\psi\| = 1 $ where
$ \rho = \hat{T}P_{\psi} $. Moreover,
\begin{equation}
\langle P \rangle_W = {\rm tr} \, WP = \int \rho(q,p)p dqdp
\end{equation}
is valid for all $ W \in K({\cal H}) $ satisfying conditions (i) and (ii)
with respect to $P$ where $ \rho = \hat{T}W $.

Third, we calculate a function on phase space for the operator $ Q^2 $. If
$ \psi \in D(Q^2) $, we obtain, applying Proposition 6.7 and Corollary 6.6,
\begin{equation}
\begin{array}{ccl}
\langle \psi | Q^2\psi \rangle
& = & \langle \Psi | VQ^2V^{-1}\Psi \rangle                    \vspace{3mm}\\
& = & \displaystyle{\left\langle \left. \left( q + 2\sigma^2
      \frac{\partial}{\partial q} \right) \Psi \, \right| \,
                                 VQV^{-1}\Psi \right\rangle}   \vspace{3mm}\\
& = & \displaystyle{\left\langle q\Psi \, \left| \,
      \left( q + 2\sigma^2 \frac{\partial}{\partial q} \right) \Psi \right.
                                           \right\rangle}      \vspace{3mm}\\
& = & \displaystyle{\langle q\Psi | q\Psi \rangle + 2\sigma^2
      \left\langle q\Psi \, \left| \, \frac{\partial}{\partial q}\Psi
                                              \right. \right\rangle \ .}
\end{array}
\end{equation}
Eq.\ (6.20) does not apply to the last term since $ q\Psi \not\in V{\cal H} $
for $ \Psi \neq 0 $. However, it can be recasted as follows. Integration
by parts yields
\begin{eqnarray}
\left\langle q\Psi \, \left| \, \frac{\partial}{\partial q}\Psi
                                                \right. \right\rangle
& = & \int q\overline{\Psi(q,p)} \frac{\partial \Psi}{\partial q}(q,p) dqdp
                                                                   \nonumber\\
& = & -\int \left( \overline{\Psi(q,p)}
 + q\overline{\frac{\partial \Psi}{\partial q}(q,p)} \right) \Psi(q,p) dqdp \\
& = & -\|\Psi\|^2 - \left\langle \left. \frac{\partial}{\partial q}\Psi
                         \, \right| \, q\Psi \right\rangle \ .     \nonumber
\end{eqnarray}
Here we have assumed that, if $ \Psi \in VD(Q^2) $,
$ \lim_{|q| \rightarrow \infty} q|\Psi(q,p)|^2 = 0 $ for almost all $p$;
actually this assumption had to be proved. From (7.21) it follows that
\[
2 \, {\rm Re} \left\langle q\Psi \, \left| \, \frac{\partial}{\partial q}\Psi
      \right. \right\rangle = -\|\Psi\|^2 \ ,
\]
whereas (7.20) implies that
$ \langle q\Psi | \frac{\partial}{\partial q}\Psi \rangle $
is real. In consequence,
\begin{equation}
\left\langle q\Psi \, \left| \, \frac{\partial}{\partial q}\Psi \right.
\right\rangle = -\frac{1}{2} \|\Psi\|^2 \ .
\end{equation}
Inserting (7.22) into (7.20), we obtain
\[
\langle \psi | Q^2\psi \rangle = \int (q^2 - \sigma^2) |\Psi(q,p)|^2 dqdp \ .
\]
Hence,
\[
\langle Q^2 \rangle_{\psi} = \langle \psi | Q^2\psi \rangle
                           = \int \rho(q,p) (q^2 - \sigma^2) dqdp
\]
holds for all $ \psi \in D(Q^2) $ with $ \|\psi\| = 1 $ where
$ \rho = \hat{T}P_{\psi} $. Moreover, for suitable $ W \in K({\cal H}) $
we have
\begin{equation}
\langle Q^2 \rangle_W = {\rm tr} \, WQ^2 = \int \rho(q,p) (q^2 - \sigma^2) dqdp
\end{equation}
where $ \rho = \hat{T}W $.

Finally, for the operator $ P^2 $ and $ \psi \in D(P^2) $ it follows from
Proposition~6.7 and Corollary 6.6 that
\begin{eqnarray*}
\langle \psi | P^2\psi \rangle & = & \langle \Psi | VP^2V^{-1}\Psi \rangle  \\
& = & \left\langle \left. \left( p - i\frac{\partial}{\partial q} \right) \Psi
      \, \right| \, VPV^{-1}\Psi \right\rangle                              \\
& = & \left\langle p\Psi \, \left| \, \left( p + \frac{i}{2\sigma^2}q
      + \frac{1}{2\sigma^2} \frac{\partial}{\partial p} \right) \Psi
                                                \right. \right\rangle \ .
\end{eqnarray*}
If, in addition, $ \psi \in D(Q) $, one can conclude that
\begin{eqnarray}
\langle \psi | P^2\psi \rangle & = & \langle p\Psi | p\Psi \rangle
      + \frac{i}{2\sigma^2} \, \langle p\Psi | q\Psi \rangle
      + \frac{1}{2\sigma^2} \left\langle p\Psi \, \left| \,
        \frac{\partial}{\partial p}\Psi \right. \right\rangle     \nonumber\\
& = & \| p\Psi \|^2 + \frac{1}{2\sigma^2} \, {\rm Re} \left\langle p\Psi
      \, \left| \, \frac{\partial}{\partial p}\Psi \right. \right\rangle   \\
&   & \mbox{} + \frac{i}{2\sigma^2} \left( \langle p\Psi | q\Psi \rangle
      + {\rm Im} \left\langle p\Psi \, \left| \,
      \frac{\partial}{\partial p}\Psi \right. \right\rangle \right) \ .
                                                                  \nonumber
\end{eqnarray}
Integration by parts again yields
\[
\left\langle p\Psi \, \left| \, \frac{\partial}{\partial p}\Psi
                                                \right. \right\rangle
 = -\|\Psi\|^2 - \left\langle \left. \frac{\partial}{\partial p}\Psi
                                     \, \right| \, p\Psi \right\rangle
\]
which implies
\begin{equation}
{\rm Re} \left\langle p\Psi \, \left| \, \frac{\partial}{\partial p}\Psi
                                               \right. \right\rangle
 = -\frac{1}{2} \|\Psi\|^2 \ ;
\end{equation}
we have assumed that $ \lim_{|p| \rightarrow \infty} p|\Psi(q,p)|^2 = 0 $
for almost all $q$ and every $ \Psi \in VD(P^2) \cap VD(Q) $. Inserting (7.25)
into (7.24), we obtain, since the last term of (7.24) is purely imaginary and
consequently equal to zero,
\[
\langle \psi | P^2\psi \rangle
        = \int (p^2 - \frac{1}{4\sigma^2}) |\Psi(q,p)|^2 dqdp \ .
\]
Hence,
\[
\langle P^2 \rangle_{\psi} = \langle \psi | P^2\psi \rangle
        = \int \rho(q,p) (p^2 - \frac{1}{4\sigma^2}) dqdp
\]
holds for $ \psi \in D(P^2) \cap D(Q) $, $ \|\psi\| = 1 $, and
$ \rho = \hat{T}P_{\psi} $. Furthermore,
\begin{equation}
\langle P^2 \rangle_W = {\rm tr} \, WP^2
        = \int \rho(q,p) (p^2 - \frac{1}{4\sigma^2}) dqdp
\end{equation}
is valid for suitable $ W \in K({\cal H}) $ and $ \rho = \hat{T}W $.

There is another way due to the author to derive Eqs.\ (7.18), (7.19),
(7.23), and (7.26) and even to generalize them, which is the subject
of the following proposition and its proof.

\begin{proposition}
Let $ N = 1 $, $ {\cal H} = L^{2}_{\C} \, (\R,dx) $, and
$ a \in K({\cal H}) $, and let $F$ be the joint position-momentum
observable given by (7.4). Suppose that $a$ satisfies the condition (7.5),
i.e., $F$ is statistically complete; let $ \hat{T} $ be the corresponding
classical representation on $ (\R^2,\Xi(\R^2),\lambda^2) $ according to
(7.9). Finally, let $ \eta^Q $ and $ \eta^P $ be the confidence functions
of the marginal observables $ F^Q $ and $ F^P $ of $F$ according to
Eqs.\ (5.8), (5.9), and (5.26); assume that the expectation values and
variances of $ \eta^Q $ and $ \eta^P $ exist.

If, for $ W \in K({\cal H}) $, the expectation value and the variance of $Q$
exist, then $ \langle Q^2 \rangle_W = \int q P^{Q^2}_{W}(dq) $ exists also,
and
\begin{eqnarray}
\langle Q \rangle_W   & = & \int \rho(q,p) (q - \langle \eta^Q \rangle) dqdp \\
\langle Q^2 \rangle_W & = & \int \rho(q,p) ((q - \langle \eta^Q \rangle)^2
                                                 - {\rm var} \, \eta^Q) dqdp
\end{eqnarray}
holds where $ \rho = \hat{T}W $. Furthermore, condition (ii) of
Section 7.2 is satisfied for $W$ with respect to $Q$, and
$ \langle Q \rangle_W = {\rm tr} \, WQ $ holds; the validity of
$ \langle Q^2 \rangle_W = {\rm tr} \, WQ^2 $ requires the additional
assumption that $W$ satisfies condition (ii) also with respect to $ Q^2 $.

The analogous statements for $P$ and $ P^2 $ are valid under analogous
suppositions; in particular,
\begin{eqnarray}
\langle P \rangle_W   & = & \int \rho(q,p) (p - \langle \eta^P \rangle) dqdp \\
\langle P^2 \rangle_W & = & \int \rho(q,p) ((p - \langle \eta^P \rangle)^2
                                                 - {\rm var} \, \eta^P) dqdp
\end{eqnarray}
holds where $ \rho = \hat{T}W $.
\end{proposition}
{\bf Proof:} From the existence of $ \langle Q \rangle_W $ and
\[
{\rm var}_W Q := {\rm var}_W E^Q = \int (q - \langle Q \rangle_W)^2 P^Q_W(dq)
\]
it follows that $ \int q^2 P^Q_W(dq) $ exists. Since the spectral measures
of $Q$ and $ Q^2 $ are related by $ E^{Q^2} = E^Q \circ f^{-1} $ where $f$
is the function $ q \mapsto f(q) := q^2 $, we have
\[
P^{Q^2}_{W}(b) = {\rm tr} \, WE^{Q^2}(b) = {\rm tr} \, WE^Q(f^{-1}(b))
                                      = P^Q_W(f^{-1}(b))
\]
where $ b \in \Xi(\R) $. In consequence, $ P^{Q^2}_{W} = P^Q_W \circ f^{-1} $
and
\[
\int q^2 P^Q_W(dq) = \int f dP^Q_W = \int {\rm id}_{\R} d(P^Q_W \circ f^{-1})
                   = \int q P^{Q^2}_W(dq) = \langle Q^2 \rangle_W \ .
\]
In particular, $ \langle Q^2 \rangle_W $ exists, and
\begin{equation}
{\rm var}_W Q = \int q^2 P^Q_W(dq) - \langle Q \rangle^2_W
              = \langle Q^2 \rangle_W - \langle Q \rangle^2_W
\end{equation}
holds.

According to Eqs.\ (5.4) and (5.5), we have
\begin{equation}
\langle F^Q \rangle_W = \langle Q \rangle_W + \langle \eta^Q \rangle
\end{equation}
and
\begin{equation}
{\rm var}_W F^Q = {\rm var}_W Q + {\rm var} \, \eta^Q \ .
\end{equation}
Since the probability distribution $ P^F_W $ of $F$ has the density
$ \rho = \hat{T}W $, the probability distribution $ P^{F^Q}_{W} $ of $ F^Q $
has the density $ q \mapsto \rho^Q(q) := \int \rho(q,p) dp $. The latter
implies
\[
\langle F^Q \rangle_W = \int q P^{F^Q}_{W}(dq) = \int q\rho^Q(q) dq
                      = \int \int q \rho(q,p) dpdq \ ,
\]
that is,
\begin{equation}
\langle F^Q \rangle_W = \int q\rho(q,p) dqdp \ .
\end{equation}
Hence, from (7.32) and (7.34) we obtain
\[
\langle Q \rangle_W = \int q\rho(q,p) dqdp - \langle \eta^Q \rangle
                    = \int \rho(q,p)(q - \langle \eta^Q \rangle) dqdp \ ,
\]
i.e.\ Eq.\ (7.27). From (7.31) -- (7.33) it follows that
\begin{equation}
\langle Q^2 \rangle_W = {\rm var}_W Q + \langle Q \rangle^2_W
 = {\rm var}_W F^Q - {\rm var} \, \eta^Q
   + (\langle F^Q \rangle_W - \langle \eta^Q \rangle)^2 \ .
\end{equation}
Inserting
\begin{eqnarray*}
{\rm var}_W F^Q & = & \int (q - \langle F^Q \rangle_W)^2 P^{F^Q}_{W}(dq) \\
                & = & \int q^2 P^{F^Q}_{W}(dq) - \langle F^Q \rangle^2_W \\
                & = & \int q^2\rho(q,p) dqdp - \langle F^Q \rangle^2_W
\end{eqnarray*}
into (7.35) and taking account of (7.34), we obtain
\begin{eqnarray*}
\langle Q^2 \rangle & = & \int q^2\rho(q,p) dqdp
                          - 2\langle F^Q \rangle_W \langle \eta^Q \rangle
                          + \langle \eta^Q \rangle^2 - {\rm var} \, \eta^Q \\
                    & = & \int \rho(q,p) (q^2 - 2\langle \eta^Q \rangle q
                          + \langle \eta^Q \rangle^2 - {\rm var} \, \eta^Q)
                                                                      dqdp \\
                    & = & \int \rho(q,p) ((q - \langle \eta^Q \rangle)^2
                          - {\rm var} \, \eta^Q) dqdp \ ,
\end{eqnarray*}
i.e.\ Eq.\ (7.28).

Let $ W = \sum_{i} \alpha_i P_{\chi_i} $, $ \alpha_i \neq 0 $, and
$ \langle \chi_i | \chi_j \rangle = \delta_{ij} $. From
$ \langle Q^2 \rangle_W < \infty $ and
\begin{eqnarray*}
\langle Q^2 \rangle_W & = & \int q P^{Q^2}_{W}(dq) = \int q^2 P^Q_W(dq)   \\
& = & \int q^2 \, {\rm tr} \, WE^Q(dq)                                    \\
& = & \sum_{i} \alpha_i \int q^2 \, \langle \chi_i | E^Q(dq)\chi_i \rangle
\end{eqnarray*}
it follows that $ \int q^2 \, \|E^Q(dq)\chi_i\|^2 < \infty $ or,
equivalently, $ \chi_i \in D(Q) $. That is, $W$ satisfies condition (ii)
of Section 7.2 with respect to $Q$. Since condition~(i) just means
the existence of $ \langle Q \rangle_W $, we have
$ \langle Q \rangle_W = {\rm tr} \, WQ $.---The proof of
the remaining statements is evident. \hfill $ \Box $\\

\begin{corollary}
If $ \langle Q^2 \rangle_W $ and $ \langle P^2 \rangle_W $ exist and
$ \chi_i \in D(Q^2) \cap D(P^2) $ for $ W = \sum_{i} \alpha_i P_{\chi_i} $
with $ \alpha_i \neq 0 $ and $ \langle \chi_i | \chi_j \rangle = \delta_{ij} $,
then $ \langle Q \rangle_W $ and $ \langle P \rangle_W $ exist also, and
$ \langle Q \rangle_W $, $ \langle P \rangle_W $, $ \langle Q^2 \rangle_W $,
and $ \langle P^2 \rangle_W $ can be expressed as traces. Moreover,
the representations (7.27) -- (7.30) are valid with $ \rho = \hat{T}W $.
\end{corollary}

For the special case that $ F$ and $ \hat{T} $ are generated by the coherent
states (6.19), Eqs.\ (7.27) -- (7.30) reduce to (7.18), (7.19), (7.23), and
(7.26). For this particular case we now investigate how the energy expectation
values of the harmonic oscillator can be expressed classically.

\begin{proposition}
Let the observable $F$ be generated by
$ a := |u^{\sigma}\rangle \langle u^{\sigma}| $ with $ u^{\sigma} $
according to (6.19), and let $ \hat{T} $ be the corresponding
classical representation. Let $H$ be the Hamiltonian
of the harmonic oscillator; that is, $H$ with domain $ D(H) $
is the self-adjoint extension of the symmetric operator
$ -\frac{1}{2m} \frac{d^2}{dx^2} + \frac{m\omega^2}{2} x^2 $,
the latter one being defined on $ S_{\C} \, (\R) $, for instance. If, for
$ W = \sum_{i} \alpha_i P_{\chi_i} $ with $ \alpha_i \neq 0 $ and
$ \langle \chi_i | \chi_j \rangle = \delta_{ij} $, $ \langle H \rangle_W $
exists and $ \chi_i \in D(Q^2) \cap D(P^2) $, then
\begin{equation}
\begin{array}{ccl}
\langle H \rangle_W & = & {\rm tr} \, WH                      \vspace{4mm}\\
& = & \displaystyle{\int \rho(q,p) \left( H(q,p) - \left(
      \frac{1}{8m\sigma^2} + \frac{m\omega^2\sigma^2}{2} \right) \right) dqdp}
                                                              \vspace{4mm}\\
& = & \displaystyle{\int \rho(q,p) H(q,p) dqdp - \left( \frac{1}{8m\sigma^2}
                                     + \frac{m\omega^2\sigma^2}{2} \right)}
\end{array}
\end{equation}
holds where $ \rho = \hat{T}W $ and the function
$ (q,p) \mapsto H(q,p) $ is the classical Hamiltonian,
$ H(q,p) = \frac{p^2}{2m} + \frac{m\omega^2}{2}q^2 $.
\end{proposition}
{\bf Proof:} Since the operator $ \frac{1}{2m} P^2 + \frac{m\omega^2}{2} Q^2 $
defined on $ D(Q^2) \cap D(P^2) $ \vspace{1pt} is a symmetric extension of
$ -\frac{1}{2m} \frac{d^2}{dx^2} + \frac{m\omega^2}{2} x^2 $, we have
\[
-\frac{1}{2m} \frac{d^2}{dx^2} + \frac{m\omega^2}{2} x^2
\subseteq \frac{1}{2m} P^2 + \frac{m\omega^2}{2} Q^2 \subseteq H
\]
and $ S_{\C} \, (\R) \subset D(Q^2) \cap D(P^2) \subseteq D(H) $. In
consequence, $W$ satisfies conditions (i) and (ii) with respect to $H$,
i.e., $ \langle H \rangle_W = {\rm tr} \, WH $. From
\begin{eqnarray*}
\langle H \rangle_W & = & {\rm tr} \, WH
  =   \sum_{i} \alpha_i \langle \chi_i | H\chi_i \rangle                   \\
& = & \sum_{i} \alpha_i \left( \frac{1}{2m} \langle \chi_i | P^2\chi_i \rangle
      + \frac{m\omega^2}{2} \langle \chi_i | Q^2\chi_i \rangle \right)     \\
& = & \sum_{i} \alpha_i \left( \frac{1}{2m} \int \xi^2 \,
                        \langle \chi_i | E^P(d\xi)\chi_i \rangle
      + \frac{m\omega^2}{2} \int \xi^2 \,
                        \langle \chi_i | E^Q(d\xi)\chi_i \rangle \right)   \\
& = & \frac{1}{2m} \int \xi^2 \, {\rm tr} \, WE^P(d\xi)
      + \frac{m\omega^2}{2} \int \xi^2 \, {\rm tr} \, WE^Q(d\xi)
\end{eqnarray*}
it follows that $ \langle Q^2 \rangle_W $ and $ \langle P^2 \rangle_W $
exist. Moreover, by Corollary 7.7, respectively, Eqs.\ (7.28) and (7.30),
we obtain
\begin{eqnarray*}
\langle H \rangle_W & = & \frac{1}{2m} \langle P^2 \rangle_W
                          + \frac{m\omega^2}{2} \langle Q^2 \rangle_W  \\
& = & \int \rho(q,p) \left( H(q,p) - \left( \frac{1}{8m\sigma^2}
                        + \frac{m\omega^2\sigma^2}{2} \right) \right) dqdp \ .
\end{eqnarray*}
\hspace*{\fill} $ \Box $\\

The additive constant $ \frac{1}{8m\sigma^2} + \frac{m\omega^2\sigma^2}{2} $
depends on the parameter $ \sigma $ and takes its minimum for
$ \sigma = \frac{1}{\sqrt{2m\omega}} $, in that case $ u^{\sigma} $ coincides
with the ground state $ \phi_0 $ of the harmonic oscillator. Therefore,
suppose finally that $ \hat{T} $ is the classical representation induced by
$ u^{\sigma} = \phi_0 $. Eq.\ (7.36) then reads
\begin{equation}
\langle H \rangle_W = {\rm tr} \, WH
                    = \int \rho(q,p) H(q,p) dqdp - \frac{\omega}{2}
                    = \langle H \rangle_{\rho} - \frac{\omega}{2}
\end{equation}
where $ \langle H \rangle_{\rho} = \int \rho(q,p) H(q,p) dqdp $ is
the classical energy expectation value.

As an explicit example, we discuss the phase-space densities for the
energy eigenstates $ \phi_n $, $ n = 0,1,2,\ldots \ $, and verify (7.37)
directly. According to (7.16) we have
\[
\rho_n(q,p) := (\hat{T}P_{\phi_n})(q,p)
 = \frac{1}{2\pi} \, |\langle u^{\sigma}_{qp} | \phi_n \rangle|^2 \ ;
\]
using our former result (6.46), we obtain
\[
\rho_n(q,p) = \frac{1}{2\pi n!} \, |z|^{2n} e^{-|z|^2}
\]
where
\[
z = \xi + i\eta = \frac{1}{2} \left( \frac{q}{\sigma} - 2i\sigma p \right)
                = \frac{1}{\sqrt{2}} \left( \sqrt{m\omega}q
                  - \frac{i}{\sqrt{m\omega}}p \right) \ .
\]
Hence,
\[
|z|^2 = \frac{q^2}{4\sigma^2} + \sigma^2 p^2
      = \frac{p^2}{2m\omega} + \frac{m\omega}{2} q^2 = \frac{H(q,p)}{\omega}
\]
and
\begin{equation}
\rho_n(q,p) = \frac{1}{2\pi n!} \left( \frac{H(q,p)}{\omega} \right)^n
                                   e^{-\frac{H(q,p)}{\omega}} \ ;
\end{equation}
note also $ \rho_n = |\Phi_n|^2 $ and the result (6.29) for
$ \Phi_n = V\phi_n $. In particular, $ \rho_n $ is of the form
$ \rho_n = \tilde{\rho}_n \circ H $. Introducing the new coordinates
$E$ and $ \phi $ defined by
\begin{equation}
\begin{array}{ccl}
q & = & \displaystyle{\sqrt{\frac{2E}{m\omega^2}} \cos\phi} \vspace{3mm}\\
p & = & \displaystyle{\sqrt{2mE} \sin\phi}
\end{array}
\end{equation}
where $ E > 0 $ and $ 0 < \phi < 2\pi $, we obtain $ H(q,p) = E $ and
\begin{eqnarray*}
\langle H \rangle_{\rho_n} & = & \int \rho_n(q,p) H(q,p) dqdp             \\
& = & \frac{1}{2\pi n!} \int \frac{(H(q,p))^{n+1}}{\omega^n}
                                 e^{-\frac{H(q,p)}{\omega}} dqdp          \\
& = & \frac{\omega}{2\pi n!} \int_{0}^{2\pi} \int_{0}^{\infty}
      \left( \frac{E}{\omega} \right)^{n+1} e^{-\frac{E}{\omega}}
                                                \frac{1}{\omega} dEd\phi  \\
& = & \frac{\omega}{n!} \int_{0}^{\infty} x^{n+1} e^{-x} dx \ .
\end{eqnarray*}
Hence,
\[
\langle H \rangle_{\rho_n} = \omega (n + 1) \ ,
\]
this result being in accordance with Eq.\ (7.37) which gives, for
$ W = P_{\phi_n} $,
\begin{equation}
\langle H \rangle_{\phi_n} = \omega \left( n + \frac{1}{2} \right)
        = \langle H \rangle_{\rho_n} - \frac{\omega}{2} \ ,
\end{equation}
$ n = 0,1,2,\ldots \ $.

We also compare the quantum variance $ {\rm var}_{\phi_n} H $ with
the corresponding classical one, $ {\rm var}_{\rho_n} H $. Whereas
$ {\rm var}_{\phi_n} H = 0 $, we obtain
\begin{eqnarray*}
{\rm var}_{\rho_n} H
& = & \int \rho_n(q,p) (H(q,p) - \langle H \rangle_{\rho_n})^2 dqdp  \\
& = & \langle H^2 \rangle_{\rho_n} - \langle H \rangle^{2}_{\rho_n}  \\
& = & \omega^2 (n + 1)(n + 2) - \omega^2 (n + 1)^2                   \\
& = & \omega^2 (n + 1) \ .
\end{eqnarray*}
In particular, for the relative standard deviation we have
\[
\frac{\Delta_{\rho_n} H}{\langle H \rangle_{\rho_n}}
= \frac{\sqrt{{\rm var}_{\rho_n} H}}{\langle H \rangle_{\rho_n}}
= \frac{1}{\sqrt{n + 1}}
\]
which converges to zero for $ n \rightarrow \infty $. This indicates
a correspondence between the quantum probability distributions of energy
in the states $ \phi_n $ and the classical ones in $ \rho_n $, in addition
to that given by Eq.\ (7.40).

We conclude our example with the calculation of the classical probability
distributions of energy in the states $ \rho_n $. The latter ones are given
by $ P^{H}_{\rho_n} := P^{H}_{\mu_n} = \mu_n \circ H^{-1} $ where $ \mu_n $
are the probability measures corresponding to $ \rho_n $. Taking account of
$ \rho_n = \tilde{\rho}_n \circ H $ and (7.39), one obtains
\begin{eqnarray*}
P^{H}_{\rho_n}(b) & = & (\mu_n \circ H^{-1})(b) = \mu_n(H^{-1}(b))         \\
& = & \int_{H^{-1}(b)} \rho_n(q,p) dqdp                                    \\
& = & \int \chi_b(H(q,p)) \tilde{\rho}_n(H(q,p)) dqdp                      \\
& = & \int_{0}^{2\pi} \int_{0}^{\infty} \chi_b(E) \tilde{\rho}_n(E)
                                                  \frac{1}{\omega}dEd\phi  \\
& = & \int_{b \cap \R^+_0} \frac{2\pi}{\omega} \tilde{\rho}_n(E) dE
\end{eqnarray*}
where $ b \in \Xi(\R) $. Hence, there exists also a probability density
$ \hat{\rho}_n $ for $ P^{H}_{\rho_n} $ which is given by
\[
\hat{\rho}_n(E) := \left\{
\begin{array}{cll}
0 & {\rm for} & E < 0 \vspace{3mm}\\
\displaystyle{\frac{2\pi}{\omega} \tilde{\rho}_n(E)} & {\rm for} & E \geq 0 \ .
\end{array}
\right.
\]
According to (7.38) we have
\[
\hat{\rho}_n(E) = \frac{1}{n!} \frac{E^n}{\omega^{n+1}} e^{-\frac{E}{\omega}}
\]
for $ E \geq 0 $.

In conclusion, our example demonstrates two aspects of our classical
representations. First, $ \hat{T} $ gives rise to a {\it dequantization}
in the sense of Eq.\ (7.15): the statistical scheme of quantum mechanics
is reformulated on phase space, and quantum observables given as
self-adjoint operators $A$ in Hilbert space are represented by
functions $f$ on phase space (in the example we have $ A = H $
and $ f(q,p) = H(q,p) - \frac{\omega}{2} $; cf.\ also Theorem 7.5,
Proposition 7.6, Corollary~7.7, and Proposition 7.8). Second, one can
consider $ \rho = \hat{T}W $ as a classical analog of $ W \in K({\cal H}) $
also in the following sense: if the operator $A$ corresponds to the classical
observable $ (q,p) \mapsto g(q,p) $ by {\it quantization}, then there should
be some correspondence between the probability distribution, the expectation
value, the variance, etc.\ of $A$ in $W$ and the probability distribution
etc.\ of $g$ in $ \rho $ (in the example we have $ A = H $ and
$ g(q,p) = H(q,p) $). Note that the usual quantization of $g$ yields $A$ and
the dequantization of $A$ according to (7.15) $f$; $f$ is related to $g$,
but in general not equal.---In particular, we see that classical
representations can be applied in the context of performing classical limits
of quantum mechanical results.

\section{Quantum Dynamics on Phase Space}

The reformulation of quantum dynamics on phase space is based on the
following theorem which is essentially a restatement of the Theorem 3.7
due to the author, respectively, a repetition of Eq.\ (3.17).

\begin{theorem}
Let $ \hat{T} $ be a classical representation on phase space in the sense of
Eq.\ (7.9), $ \{ \tau_t \}_{t \in \R} $ the strongly continuous one-parameter
group of norm-automorphisms of $ {\cal T}_s({\cal H}) $ corresponding to some
Hamiltonian $H$ according to Eq.\ (3.9), and $Z$ its infinitesimal generator
which is given by Eq.\ (3.11). Then
\[
t \mapsto \rho_t := \hat{\delta}_t \rho := \hat{T}\tau_t\hat{T}^{-1}\rho
\]
where $ t \in \R $, $ \rho \in \hat{T}K({\cal H}) $, and
$ \hat{T}^{-1} \!\! : R(\hat{T}) \rightarrow {\cal T}_s({\cal H}) $,
satisfies the equation
\begin{equation}
\dot{\rho}_t = \hat{L} \rho_t \ ,
\end{equation}
provided that in addition $ \rho \in D(\hat{L}) = \hat{T}D(Z) $ holds;
the operator $ \hat{L} $ is given by $ \hat{L} = \hat{T}Z\hat{T}^{-1} $,
and the derivative $ \dot{\rho}_t $ can be taken with respect to the norm
in $ L^{1}_{\R}(\R^{2N},d^N \! q \, d^N \! p) $.
\end{theorem}

For $ \hat{L} $ we obtain, by Eqs.\ (3.11) and (7.9), the representation
\begin{equation}
(\hat{L}\rho)(q,p) = \frac{-i}{(2\pi)^N} \, {\rm tr} \, (HW - WH)a_{qp}
                   = \frac{2}{(2\pi)^N} \, {\rm Im} \, {\rm tr} \, HWa_{qp}
\end{equation}
where $ \rho \in \hat{T}K({\cal H}) \cap D(\hat{L})
= \hat{T}(K({\cal H}) \cap D(Z)) $ and $ \rho = \hat{T}W $. However,
the last term of (7.42) has possibly only a formal meaning since, according to
(3.11), $ HW - WH $ can be extended to a trace-class operator on $ \cal H $,
but not necessarily $ HW $ and $ WH $. In the special case that
$ \rho \in D(\hat{L}) $ corresponds to a pure state
$ W = P_{\psi} = |\psi\rangle \langle\psi | $, $ \psi \in {\cal H} $,
$ \|\psi\| = 1 $, a rigorous meaning of the last term of (7.42) is
guaranteed. The condition $ W = P_{\psi} \in D(Z) $ is equivalent to
$ \psi \in D(H) $, as is easily seen. Moreover, for $ \psi \in D(H) $,
$ HP_{\psi} $ and $ P_{\psi}H $ can be considered as trace-class
operators. Hence, for $ \rho = \hat{T}P_{\psi} \in D(\hat{L}) $,
Eq.\ (7.42) reads
\begin{equation}
\begin{array}{ccl}
(\hat{L}\rho)(q,p)
& = & \displaystyle{\frac{-i}{(2\pi)^N} \,
                   (\langle a_{qp}\psi | H\psi \rangle
                  - \langle H\psi | a_{qp}\psi \rangle)}   \vspace{3mm}\\
& = & \displaystyle{\frac{2}{(2\pi)^N} \, {\rm Im} \,
                    \langle a_{qp}\psi | H\psi \rangle \ .}
\end{array}
\end{equation}
If, in particular, the classical representation $ \hat{T} $ is induced by
$ a := |u\rangle \langle u | $, $ u \in {\cal H} $, $ \|u\| = 1 $,
(7.42) becomes
\begin{equation}
\begin{array}{ccl}
(\hat{L}\rho)(q,p)
& = & \displaystyle{\frac{-i}{(2\pi)^N} \,
                    \langle u_{qp} | (HW - WH)u_{qp} \rangle}   \vspace{3mm}\\
& = & \displaystyle{\frac{2}{(2\pi)^N} \, {\rm Im} \,
                    \langle u_{qp} | HWu_{qp} \rangle}
\end{array}
\end{equation}
and (7.43)
\begin{equation}
\begin{array}{ccl}
(\hat{L}\rho)(q,p)
& = & \displaystyle{\frac{2}{(2\pi)^N} \,
       {\rm Im} \, (\langle u_{qp} | H\psi \rangle
                    \langle \psi | u_{qp} \rangle)}   \vspace{3mm}\\
& = & 2 \, {\rm Im} \, ((VHV^{-1}\Psi)(q,p)\overline{\Psi(q,p)})
\end{array}
\end{equation}
where $V$ and $ \Psi $ are defined by Eq.\ (6.1). Again, the last term of
(7.44) does not necessarily make sense, whereas Eq.\ (7.45) is completely
rigorous.

In order to derive a partial differential equation from (7.41),
we need a representation for $ \hat{L} $ that is more explicit than
(7.42) -- (7.45). Such representations were derived for concrete Hamiltonians
by S. T. Ali and E. Prugove\v{c}ki (1977c; E. Prugove\v{c}ki, 1978, 1984)
on a nonrigorous level. We consider only a particular case which, adopting
some ideas of S. T. Ali and E. Prugove\v{c}ki, we treat rigorously. This
particular case is again the one-dimensional harmonic oscillator where
$ \hat{T} $ is induced by $ a := |u^{\sigma}\rangle \langle u^{\sigma} | $
with $ u^{\sigma} $ according to (6.19). For our procedure, the next
proposition is useful which sharpens the validity of Eqs.\ (6.26) and (6.27).

\begin{proposition}
Let, for $ N = 1 $ and a function $ u \in S_{\C} \, (\R) $ of norm $1$,
$V$ be the isometry given by Eq.\ (6.1). Then the Hamiltonian $ VHV^{-1} $
of the harmonic oscillator in phase-space representation acts as the
differential operator (6.26) on the entire domain $ VD(H) $. In the case
$ u := u^{\sigma} $ according to (6.19), a second representation of
$ VHV^{-1} $ on its entire domain $ VD(H) $ is given by Eq.\ (6.27).
\end{proposition}
{\bf Proof:} For $ \Psi \in VD(H) $ and $ \psi := V^{-1}\Psi $, we have
\begin{eqnarray*}
\lefteqn{(VHV^{-1}\Psi)(q,p) = (VH\psi)(q,p)
  =   \frac{1}{\sqrt{2\pi}} \langle Hu_{qp} | \psi \rangle}\hspace{1cm}     \\
& = & \frac{1}{\sqrt{2\pi}} \int \overline{\left(
              -\frac{1}{2m} \frac{d^2}{dx^2} + \frac{m\omega^2}{2} x^2 \right)
                                               u_{qp}(x)} \psi(x) dx        \\
& = & \frac{1}{\sqrt{2\pi}} \int \overline{\left(
              -\frac{1}{2m} \frac{\partial^2}{\partial q^2}
             - \frac{m\omega^2}{2} \frac{\partial^2}{\partial p^2}
            + i\frac{p}{m} \frac{\partial}{\partial q}
             + \frac{p^2}{2m} \right) u_{qp}(x)} \psi(x) dx                 \\
& = & \left( \left( -\frac{1}{2m} \frac{\partial^2}{\partial q^2}
           - \frac{m\omega^2}{2} \frac{\partial^2}{\partial p^2}
          - i\frac{p}{m} \frac{\partial}{\partial q}
           + \frac{p^2}{2m} \right) \Psi \right) (q,p)
\end{eqnarray*}
where Theorem 6.3 has been used. Hence, we have proved the first statement
of the proposition. From this the second one follows by means of
Eq.\ (6.23). \hspace*{\fill} $ \Box $\\

The proof of the proposition is analogous to the proof of Proposition 6.5. We
emphasize once more that Proposition 7.10 extends the validity of Eqs.\ (6.26)
and (6.27) for $ \Psi \in V(D(Q^2) \cap D(P^2)) $ to $ \Psi \in VD(H) $.

Now, we come back to Eq.\ (7.45) where $ u = u^{\sigma} $, $H$ is
\vspace{1pt} the Hamiltonian of the harmonic oscillator, and
$ \rho = \hat{T}P_{\psi} $ with $ \psi \in D(H) $. Taking account of
Proposition 7.10 and inserting (6.27) into (7.45), we obtain
\begin{eqnarray*}
\hat{L}\rho & = & 2 \, {\rm Im} \, ((VHV^{-1}\Psi) \overline{\Psi})     \\
& = & 2 \, {\rm Im} \left[ \left( \left( -\frac{1}{2m} + 2m\omega^2\sigma^4
                                  \right) \frac{\partial^2 \Psi}{\partial q^2}
                         + \left( 2m\omega^2\sigma^2q - \frac{i}{m}p \right)
                                    \frac{\partial \Psi}{\partial q} \right)
                                          \overline{\Psi} \right] \ .
\end{eqnarray*}
Using Eq.\ (6.23), it follows that
\begin{eqnarray*}
\hat{L}\rho & = & -2\frac{p}{m} \, {\rm Re} \,
                    \frac{\partial \Psi}{\partial q} \overline{\Psi}
              + 4m\omega^2\sigma^2q \, {\rm Im} \left( \frac{1}{2\sigma^2}
                  \left( i\frac{\partial \Psi}{\partial p} - q\Psi \right)
                                         \overline{\Psi} \right)            \\
&   & \mbox{} + 2\left( -\frac{1}{2m} + 2m\omega^2\sigma^4 \right)
                {\rm Im} \left( \frac{1}{2\sigma^2} \left(
                        i\frac{\partial^2 \Psi}{\partial q \partial p}
                      - q\frac{\partial \Psi}{\partial q} - \Psi \right)
                                              \overline{\Psi} \right)       \\
& = & -2\frac{p}{m} \, {\rm Re} \, \frac{\partial \Psi}{\partial q}
                                                  \overline{\Psi}
      + 2m\omega^2q \, {\rm Re} \, \frac{\partial \Psi}{\partial p}
                                                  \overline{\Psi}           \\
&   & \mbox{} + 2\left( -\frac{1}{4m\sigma^2} + m\omega^2\sigma^2 \right)
        {\rm Im} \left( i\frac{\partial^2 \Psi}{\partial q \partial p}
                                                \overline{\Psi}
        + \frac{\partial \Psi}{\partial q}
        \left( 2\sigma^2 \frac{\partial \overline{\Psi}}{\partial q}
        + i\frac{\partial \overline{\Psi}}{\partial p} \right) \right)      \\
& = & -2\frac{p}{m} \, {\rm Re} \, \frac{\partial \Psi}{\partial q}
                                                        \overline{\Psi}
      + 2m\omega^2q \, {\rm Re} \, \frac{\partial \Psi}{\partial p}
                                                        \overline{\Psi}     \\
&   & \mbox{} + 2\left( -\frac{1}{4m\sigma^2} + m\omega^2\sigma^2 \right)
        {\rm Re} \left( \frac{\partial^2 \Psi}{\partial q \partial p}
                                               \overline{\Psi}
         + \frac{\partial \Psi}{\partial q}
                 \frac{\partial \overline{\Psi}}{\partial p} \right) \ .
\end{eqnarray*}
According to (7.16) and Theorem 6.3,
$ \rho = \hat{T}P_{\psi} = |\Psi|^2 \in C^{\infty}_{\R}(\R^2) $
holds which implies
\[
\frac{\partial \rho}{\partial q}
  = 2 \, {\rm Re} \, \frac{\partial \Psi}{\partial q} \overline{\Psi} \ ,
\]
\[
\frac{\partial \rho}{\partial p}
  = 2 \, {\rm Re} \, \frac{\partial \Psi}{\partial p} \overline{\Psi} \ ,
\]
and
\[
\frac{\partial^2 \rho}{\partial q \partial p}
  = 2 \, {\rm Re} \left( \frac{\partial^2 \Psi}{\partial q \partial p}
                  \overline{\Psi} + \frac{\partial \Psi}{\partial q}
                  \frac{\partial \overline{\Psi}}{\partial p} \right) \ .
\]
Hence,
\begin{equation}
\hat{L}\rho = -\frac{p}{m} \frac{\partial \rho}{\partial q}
              + m\omega^2q \frac{\partial \rho}{\partial p}
              + \left( -\frac{1}{4m\sigma^2} + m\omega^2\sigma^2 \right)
                        \frac{\partial^2 \rho}{\partial q \partial p}
\end{equation}
where $ \rho \in D(\hat{L}) $ is of the form $ \rho = \hat{T}P_{\psi} $.

One might try to conclude the validity of Eq.\ (7.46) for arbitrary
$ \rho \in D(\hat{L}) \cap \hat{T}K({\cal H}) $ from its validity for
$ \rho = \hat{T}P_{\psi} \in D(\hat{L}) $. However, there are some
difficulties. If $ \rho = \hat{T}W = \sum_{i} \alpha_i \rho_i $ with
$ W = \sum_{i} \alpha_i P_{\chi_i} \in K({\cal H}) $, $ \alpha_i > 0 $,
$ \|\chi_i\| = 1 $, and $ \rho_i := \hat{T}P_{\chi_i} $ is an infinite sum,
it converges in $ L^1 $-norm and, as a consequence of the monotone convergence
theorem, also pointwise a.e.; moreover, since the functions $ \rho_i $ are
bounded by $ \frac{1}{2\pi} $, $ \sum_{i} \alpha_i \rho_i $ converges even
uniformly. Furthermore, the functions $ \rho_i $ are infinitely differentiable
and $ \rho $ is continuous. Because of Eq.\ (6.14), the partial derivatives
of $ \rho_i $ need not be bounded; thus
$ \sum_{i} \alpha_i \frac{\partial \rho_i}{\partial p} $,
for instance, need not converge uniformly. Hence, in view of these arguments,
we can neither conclude that $ \rho = \sum_{i} \alpha_i \rho_i $ can be
differentiated term by term nor that it is a differentiable function
at all. This is one of two difficulties in the derivation of (7.46)
for a general $ \rho $. The other one is that, for
$ \rho = \hat{T}W \in D(\hat{L}) \cap \hat{T}K({\cal H}) $, we do not know
whether there exists a representation $ W = \sum_{i} \alpha_i P_{\chi_i} $
with $ \rho_i = \hat{T}P_{\chi_i} \in D(\hat{L}) $, i.e.\ $ \chi_i \in D(H) $.

On the level of formal calculations, however, we have shown by (7.46) that
\begin{equation}
\hat{L}\rho = -\{ H,\rho \}
              + \left( -\frac{1}{4m\sigma^2} + m\omega^2\sigma^2 \right)
                        \frac{\partial^2 \rho}{\partial q \partial p}
\end{equation}
holds for all $ \rho \in D(\hat{L}) $ where $H$ here is the classical
Hamiltonian and $ \{ \, . \, , \, . \, \} $ the Poisson bracket. Eq.\ (7.41)
now yields
\begin{equation}
\dot{\rho}_t = -\{ H,\rho_t \}
               + \left( -\frac{1}{4m\sigma^2} + m\omega^2\sigma^2 \right)
                         \frac{\partial^2 \rho_t}{\partial q \partial p}
\end{equation}
where the time derivative is understood in $ L^1 $-norm. It is suggestive
to replace the derivative $ \dot{\rho}_t $ by the usual derivative
$ \frac{\partial \rho_t}{\partial t} $ which is performed pointwise
for every $ (q,p) $. Eq.\ (7.48) then reads
\[
\frac{\partial \rho_t}{\partial t}
 = -\{ H,\rho_t \} + \left( -\frac{1}{4m\sigma^2} + m\omega^2\sigma^2 \right)
                             \frac{\partial^2 \rho_t}{\partial q \partial p}
\]
which is the classical Liouville equation with a correction term. Again, for
$ \sigma = \frac{1}{\sqrt{2m\omega}} $, i.e., if $ u^{\sigma} $ coincides with
the ground state of the harmonic oscillator, the correction term vanishes
(cf.\ Eqs.\ (6.27) and (7.36)), and $ t \mapsto \rho_t $ is a solution
of the classical Liouville equation
\begin{equation}
\frac{\partial \rho_t}{\partial t} = -\{ H,\rho_t \} \ .
\end{equation}
For this result there is another, rigorous proof. That proof is prepared by
the following lemma which states a result of elementary quantum mechanics.

\begin{lemma}
Let $H$ be the Hamiltonian of the one-dimensional harmonic oscillator and
$ \sigma = \frac{1}{\sqrt{2m\omega}} $. Then
\[
\psi_t := e^{-iHt}u^{\sigma}_{qp}
        = e^{\frac{i}{2}(qp - q_tp_t - \omega t)} u^{\sigma}_{q_tp_t}
\]
holds where $ (q_t,p_t) := \Phi_t(q,p) $ and $ \{\Phi_t\}_{t \in \R} $ is
the classical Hamiltonian flow of the harmonic oscillator.
\end{lemma}
{\bf Proof:} From
\[
\psi_t = e^{-iHt}u^{\sigma}_{qp}
       = \sum_{n=0}^{\infty} e^{-iE_nt}
                     \langle \phi_n | u^{\sigma}_{qp} \rangle \phi_n
\]
and Eq.\ (6.46) we obtain
\begin{eqnarray*}
\psi_t & = & e^{-i\frac{\omega}{2}t} \sum_{n=0}^{\infty} (e^{-i\omega t})^n
                    e^{i\frac{qp}{2}} e^{-\frac{|z|^2}{2}}
                        \frac{\overline{z}^n}{\sqrt{n!}} \phi_n         \\
& = & e^{-i\frac{\omega}{2}t} e^{i\frac{qp - q_tp_t}{2}}
           \sum_{n=0}^{\infty} e^{i\frac{q_tp_t}{2}} e^{-\frac{|z_t|^2}{2}}
                                   \frac{\overline{z_t}^n}{\sqrt{n!}} \phi_n
\end{eqnarray*}
where $ (q_t,p_t) = \Phi_t(q,p) $ and
\begin{eqnarray*}
z_t & := & e^{i\omega t}z = \frac{1}{2} (\cos\omega t + i\sin\omega t)
                            \left( \frac{q}{\sigma} - 2i\sigma p \right)   \\
& = & \frac{1}{2} \left( \frac{1}{\sigma}
                       (q\cos\omega t + 2\sigma^2p\sin\omega t)
            - 2i\sigma (p\cos\omega t - \frac{q}{2\sigma^2}\sin\omega t)
                                                           \right)         \\
& = & \frac{1}{2} \left( \frac{1}{\sigma}
                       (q\cos\omega t + \frac{p}{m\omega}\sin\omega t)
            - 2i\sigma (p\cos\omega t - m\omega q\sin\omega t) \right)     \\
& = & \frac{1}{2} \left( \frac{q_t}{\sigma} - 2i\sigma p_t \right) \ .
\end{eqnarray*}
Hence,
\begin{eqnarray*}
\psi_t & = & e^{\frac{i}{2}(qp - q_tp_t - \omega t)} \sum_{n=0}^{\infty}
                \langle \phi_n | u^{\sigma}_{q_tp_t} \rangle \phi_n        \\
       & = & e^{\frac{i}{2}(qp - q_tp_t - \omega t)} u^{\sigma}_{q_tp_t} \ .
\end{eqnarray*}
\hspace*{\fill} $ \Box $\\

The proposition now presents our rigorous results on the explicit form of the
operator $ \hat{L} $, respectively, Eq.\ (7.41) as well as on Eq.\ (7.49).
\begin{proposition}
Let, for $ N = 1 $, $ \hat{T} $ be the classical representation induced by
the coherent state $ u^{\sigma} $, $ \sigma > 0 $, and let $H$ be the
Hamiltonian of the harmonic oscillator. Then the following statements
are valid:
\begin{enumerate}
\item[(a)] If $ W \in K({\cal H}) $ is a finite convex linear combination
of pure states $ P_{\psi} $ with $ \psi \in D(H) $, then
$ \rho = \hat{T}W \in D(\hat{L}) \cap C^{\infty}_{\R}(\R^2) $
and $ \hat{L}\rho $ is given by Eq.~(7.47). Moreover, $ \rho_t
:= \hat{T}\tau_t\hat{T}^{-1}\rho \in D(\hat{L}) \cap C^{\infty}_{\R}(\R^2) $
for all $ t \in \R $, and $ t \mapsto \rho_t $ is a solution of Eq.\ (7.48)
where $ \dot{\rho}_t $ is understood in $ L^1 $-norm.
\item[(b)] If $ \sigma = \frac{1}{\sqrt{2m\omega}} $, then
\[
\rho_t = \hat{T}\tau_t\hat{T}^{-1}\rho = \rho \circ \Phi^{-1}_{t}
\]
holds for every $ \rho \in \hat{T}K({\cal H}) $ and
\[
\frac{\partial \rho_t}{\partial t} = -\{ H,\rho_t \}
\]
for every $ \rho \in \hat{T}K({\cal H}) \cap C^{\infty}_{\R}(\R^2) $ where
$ \frac{\partial \rho_t}{\partial t} $ is the usual partial derivative and
$H$ the classical Hamiltonian.
\end{enumerate}
\end{proposition}
{\bf Proof:} For $ W = P_{\psi} $ with $ \psi \in D(H) $ or, equivalently,
$ \rho = \hat{T}P_{\psi} \in D(\hat{L}) $, Eq.~(7.47) has been derived
rigorously; in particular, $ \rho = |\Psi|^2 \in C^{\infty}_{\R}(\R^2) $. From
$ \rho _t = \hat{T}\tau_tP_{\psi} = \hat{T}P_{\psi_t} $ with
$ \psi_t := e^{-iHt}\psi \in D(H) $ it follows that
$ \rho_t \in D(\hat{L}) \cap C^{\infty}_{\R}(\R^2) $ and
\[
\dot{\rho}_t = \hat{L}\rho_t = -\{ H,\rho_t \}
               + \left( -\frac{1}{4m\sigma^2} + m\omega^2\sigma^2 \right)
                         \frac{\partial^2 \rho_t}{\partial q \partial p} \ ,
\]
i.e.\ Eq.\ (7.48). Clearly, these statements on $ \rho $ and $ \rho_t $
are also valid if $W$ is a finite convex linear combination of pure states
$ P_{\psi} $ with $ \psi \in D(H) $.

Lemma 7.11 implies for $ \sigma = \frac{1}{\sqrt{2m\omega}} $ and every
$ \rho = \hat{T}W \in \hat{T}K({\cal H}) $ that
\begin{eqnarray*}
\rho_t(q,p) & = & (\hat{T}\tau_t\hat{T}^{-1}\rho)(q,p)                     \\
            & = & \frac{1}{2\pi} \, \langle u^{\sigma}_{qp} |
                           e^{-iHt}We^{iHt} u^{\sigma}_{qp} \rangle        \\
            & = & \frac{1}{2\pi} \, \langle u^{\sigma}_{\Phi_{-t}(q,p)} |
                                   W u^{\sigma}_{\Phi_{-t}(q,p)} \rangle   \\
            & = & (\hat{T}W)(\Phi_{-t}(q,p))                               \\
            & = & \rho(\Phi^{-1}_{t}(q,p)) \ ,
\end{eqnarray*}
i.e., $ \rho_t = \rho \circ \Phi^{-1}_{t} $. If $ \rho $ is
sufficiently differentiable, we obtain furthermore, differentiating
$ \rho_t \circ \Phi_t = \rho $ with respect to time and using Hamilton's
equations, $ \frac{\partial \rho_t}{\partial t} = -\{ H,\rho_t \} $.
\hfill $ \Box $\\

The two parts of the proposition were proved independently of
each other. If $ \sigma = \frac{1}{\sqrt{2m\omega}} $ and
$ \rho \in \hat{T}K({\cal H}) \cap D(\hat{L}) \cap C^{\infty}_{\R}(\R^2) $,
we have
\[
\dot{\rho}_t = \hat{L}\rho_t
\]
as well as
\[
\frac{\partial \rho_t}{\partial t} = -\{ H,\rho_t \} \ .
\]
However, we have not proved that $ \hat{L}\rho = -\{ H,\rho \} $ and
$ \dot{\rho}_t = \frac{\partial \rho_t}{\partial t} $ hold for all
$ \rho \in \hat{T}K({\cal H}) \cap D(\hat{L}) \cap C^{\infty}_{\R}(\R^2) $. The
concluding corollary presents some information on this problem.

\begin{corollary}
If $ \sigma = \frac{1}{\sqrt{2m\omega}} $ and
$ W = \sum_{i=1}^{n} \alpha_i P_{\psi_i} $ with $ \alpha_i > 0 $,
$ \sum_{i=1}^{n} \alpha_i = 1 $, $ \|\psi_i\| = 1 $, and $ \psi_i \in D(H) $,
then \vspace{2pt} $ \rho = \hat{T}W \in \hat{T}K({\cal H}) \cap D(\hat{L})
\cap C^{\infty}_{\R}(\R^2) $, $ \hat{L}\rho = -\{ H,\rho \} $, and
$ \dot{\rho}_t = \frac{\partial \rho_t}{\partial t} $ for all $ t \in \R $.
\end{corollary}
{\bf Proof:} In view of part (a) of Proposition 7.12 and specially
of Eq.\ (7.47), one only has to show
$ \dot{\rho}_t = \frac{\partial \rho_t}{\partial t} $. In fact,
taking account of part (b) of the proposition, we obtain
\[
\dot{\rho}_t = \hat{L}\rho_t = -\{ H,\rho_t \}
             = \frac{\partial \rho_t}{\partial t} \ .
\]
\hspace*{\fill} $ \Box $\\

It may be that all probability densities $ \rho \in D(\hat{L}) $ are
infinitely differentiable, i.e., $ \hat{T}K({\cal H}) \cap D(\hat{L}) \cap
C^{\infty}_{\R}(\R^2) = \hat{T}K({\cal H}) \cap D(\hat{L}) $. The \vspace{1pt}
set $ \hat{T}K({\cal H}) \cap C^{\infty}_{\R}(\R^2) $, however,
is properly larger \vspace{1pt} than
$ \hat{T}K({\cal H}) \cap D(\hat{L}) \cap C^{\infty}_{\R}(\R^2) $,
as the example $ \psi \not\in D(H) $, $ \|\psi\| = 1 $,
$ \rho := \hat{T}P_{\psi} = |\Psi|^2 $ shows. Moreover, since
$ \rho \not\in D(\hat{L}) $, $ \left. \dot{\rho}_t \right|_{t=0} $
does not exist, but, since $ \rho \in C^{\infty}_{\R}(\R^2) $,
$ \left. \frac{\partial \rho_t}{\partial t} \right|_{t=0} $ does. Hence,
there are cases in which the time derivatives $ \dot{\rho}_t $ and
$ \frac{\partial \rho_t}{\partial t} $ do not coincide.

Summarizing, we have shown by Theorem 7.9 that quantum dynamics can always
be reformulated on phase space, and we have demonstrated by the particular
example of the harmonic oscillator that the corresponding equations of motion
are related to the classical Liouville equation.

\chapter*{References}
\begin{description}
\item[{\rm Ali, S. T. (1985):}]
           Stochastic Localization, Quantum Mechanics on
           Phase \linebreak Space, and Quantum Space-Time,
           {\it Riv.\ Nuovo Cimento} {\bf 8} (11), \vspace{-3mm} 1--128.
\item[{\rm Ali, S. T., and E. Prugove\v{c}ki (1977a):}]
           Systems of Imprimitivity and Representations
           of Quantum Mechanics on Fuzzy Phase Spaces,
           {\it J. Math.\ Phys.}\ {\bf 18}, 219--228. \vspace{-3mm}
\item[{\rm Ali, S. T., and E. Prugove\v{c}ki (1977b):}]
           Classical and Quantum Statistical \linebreak Mechanics
           in a Common Liouville Space,
           {\it Physica} {\bf 89A}, 501--521. \vspace{-3mm}
\item[{\rm Ali, S. T., and E. Prugove\v{c}ki (1977c):}]
           Quantum Statistical Mechanics on \linebreak
           Stochastic Phase Space,
           {\it Int.\ J. Theor.\ Phys.}\ {\bf 16}, 689--706. \vspace{-3mm}
\item[{\rm Bargmann, V. (1961):}]
           On a Hilbert Space of Analytic Functions and an Associated
           Integral Transform, Part I,
           {\it Comm.\ Pure Appl.\ Math.}\ {\bf 14}, 187--214.
           \vspace{-3mm}
\item[{\rm Bargmann, V. (1967):}]
           On a Hilbert Space of Analytic Functions and an Associated
           Integral Transform, Part II,
           {\it Comm.\ Pure Appl.\ Math.}\ {\bf 20}, 1--101.
           \vspace{-3mm}
\item[{\rm Beltrametti, E. G., and S. Bugajski (1995):}]
           A Classical Extension of Quantum Mechanics,
           {\it J. Phys.\ A:\ Math.\ Gen.}\ {\bf 28}, 3329--3343.
           \vspace{-3mm}
\item[{\rm Bugajski, S. (1993a):}]
           Classical Frames for a Quantum Theory---A Bird's-Eye View,
           {\it Int.\ J. Theor.\ Phys.}\ {\bf 32}, 969--977. \vspace{-3mm}
\item[{\rm Bugajski, S. (1993b):}]
           On Classical Representations of Convex Descriptions,
           {\it Z. Naturforsch.}\ {\bf 48a}, 469--470. \vspace{-3mm}
\item[{\rm Busch, P. (1985):}]
           Indeterminacy Relations and Simultaneous Measurements
           in Quantum Theory,
           {\it Int.\ J. Theor.\ Phys.}\ {\bf 24}, 63--92. \vspace{-3mm}
\item[{\rm Busch, P. (1986):}]
           Unsharp Reality and Joint Measurements for Spin Observables,
           {\it Phys.\ Rev.\ D} {\bf 33}, 2253--2261. \vspace{-3mm}
\item[{\rm Busch, P. (1987):}]
           Some Realizable Joint Measurements of Complementary Observables,
           {\it Found.\ Phys.}\ {\bf 17}, 905--937. \vspace{-3mm}
\item[{\rm Busch, P.,}] M. Grabowski, and P. J. Lahti (1995):\
           {\it Operational Quantum \linebreak Physics} (Lecture Notes in
           Physics, Vol.\ m 31), Springer-Verlag, \vspace{-3mm} Berlin.
\item[{\rm Busch, P., K.-E. Hellwig, and W. Stulpe (1993):}]
           On Classical Representa- \linebreak tions
           of Finite-Dimensional Quantum Mechanics,
           {\it Int.\ J. Theor.\ Phys.}\ {\bf 32}, 399--405. \vspace{-3mm}
\item[{\rm Busch, P., and P. J. Lahti (1984):}]
           On Various Joint Measurements of Position
           and Momentum Observables in Quantum Theory,
           {\it Phys.\ Rev.\ D} {\bf 29}, 1634--1646. \vspace{-3mm}
\item[{\rm Busch, P., and P. J. Lahti (1989):}]
           The Determination of the Past and the Future
           of a Physical System in Quantum Mechanics,
           {\it Found.\ Phys.}\ {\bf 19}, 633--678. \vspace{-3mm}
\item[{\rm Cohen, L. (1966):}]
           Generalized Phase-Space Distribution Functions,
           {\it J. Math.\ Phys.}\ {\bf 7}, 781--786. \vspace{-3mm}
\item[{\rm Davies, E. B. (1970):}]
           On the Repeated Measurement of Continuous Observables
           in Quantum Mechanics,
           {\it J. Funct.\ Anal.}\ {\bf 6}, 318--346. \vspace{-3mm}
\item[{\rm Davies, E. B. (1976):}]
           {\it Quantum Theory of Open Systems},
           Academic Press, London. \vspace{-3mm}
\item[{\rm Dunford, N., and J. T. Schwartz (1958):}]
           {\it Linear Operators}, Part I,
           Inter- \linebreak science Publishers, New York. \vspace{-3mm}
\item[{\rm Emch, G. (1972):}]
           {\it Algebraic Methods in Statistical Mechanics
           and Quantum Field Theory},
           Wiley-Interscience, New York. \vspace{-3mm}
\item[{\rm Gudder, S. (1984):}]
           Probability Manifolds,
           {\it J. Math.\ Phys.}\ {\bf 25}, 2397--2401. \vspace{-3mm}
\item[{\rm Gudder, S. (1985):}]
           Amplitude Phase-Space Model for Quantum Mechanics,
           {\it Int.\ J. Theor.\ Phys.}\ {\bf 24}, 343--353. \vspace{-3mm}
\item[{\rm Gudder, S. (1988):}]
           {\it Quantum Probability},
           Academic Press, Boston. \vspace{-3mm}
\item[{\rm Gudder, S., J. Hagler, and W. Stulpe (1988):}]
           An Uncertainty Relation for Joint Position-Momentum Measurements,
           {\it Found.\ Phys.\ Lett.}\ {\bf 1}, 287--292. \vspace{-3mm}
\item[{\rm Guz, W. (1984):}]
           Foundations of Phase-Space Quantum Mechanics,
           {\it Int.\ J. Theor.\ Phys.}\ {\bf 23}, 157--184. \vspace{-3mm}
\item[{\rm Haag, R., and D. Kastler (1964):}]
           An Algebraic Approach to Quantum Field Theory,
           {\it J. Math.\ Phys.}\ {\bf 5}, 849--861. \vspace{-3mm}
\item[{\rm Hellwig, K.-E. (1993):}]
           Quantum Measurements and Information Theory, \linebreak
           {\it Int.\ J. Theor.\ Phys.}\ {\bf 32}, 2401--2411. \vspace{-3mm}
\item[{\rm Hellwig, K.-E., and W. Stulpe (1993):}]
           A Classical Reformulation of Finite-Dimensional Quantum Mechanics,
           in {\it Symposium on the Foundations of Modern Physics 1993},
           P. Busch, P. J. Lahti, and P. Mittelstaedt (eds.), 209--214,
           World Scientific, Singapore. \vspace{-3mm}
\item[{\rm Hillery, M.,}] R. F. O'Connell, M. O. Scully,
           and E. P. Wigner (1984):
           Dis- tribution Functions in Physics:\ Fundamentals,
           {\it Phys.\ Rep.}\ {\bf 106}, 121--167. \vspace{-3mm}
\item[{\rm Holevo, A. S. (1973):}]
           Statistical Decision Theory for Quantum Systems,
           {\it J. Multivariate Anal.}\ {\bf 3}, 337--394. \vspace{-3mm}
\item[{\rm Hudson, R. L. (1974):}]
           When Is the Wigner Quasi-Probability Density Non-Negative?,
           {\it Rep.\ Math.\ Phys.}\ {\bf 6}, 249--252. \vspace{-3mm}
\item[{\rm Lanz, L.,}] O. Melsheimer, and E. Wacker (1985):
           Introduction of a Boltz- \linebreak mann Observable and
           Boltzmann Equation, {\it Physica} {\bf 131A}, 520--544.
           \vspace{-3mm}
\item[{\rm Ludwig, G. (1970):}]
           {\it Deutung des Begriffs ``physikalische Theorie'' und
           axiomatische Grundlegung der Hilbertraumstruktur der
           Quantenmechanik durch Haupts\"{a}tze des Messens},
           (Lecture Notes in Physics, Vol.\ 4),
           Springer-Verlag, Berlin. \vspace{-3mm}
\item[{\rm Ludwig, G. (1983):}]
           {\it Foundations of Quantum Mechanics I},
           Springer-Verlag, New York. \vspace{-3mm}
\item[{\rm Ludwig, G. (1985):}]
           {\it An Axiomatic Basis for Quantum Mechanics}, Vol.\ I,
           Springer-Verlag, Berlin. \vspace{-3mm}
\item[{\rm Ludwig, G. (1987):}]
           {\it An Axiomatic Basis for Quantum Mechanics}, Vol.\ II,
           Springer-Verlag, Berlin. \vspace{-3mm}
\item[{\rm Moyal, J. E. (1949):}]
           Quantum Mechanics as a Statistical Theory,
           {\it Proc. \linebreak Cambridge Philos.\ Soc.}\ {\bf 45},
           99--124. \vspace{-3mm}
\item[{\rm Perelomov, A. M. (1986):}]
           {\it Generalized Coherent States and their Applica- \linebreak
           tions}, Springer-Verlag, Berlin. \vspace{-3mm}
\item[{\rm Pitowsky, I. (1983):}]
           Deterministic Model of Spin and Statistics,
           {\it Phys.\ Rev.\ D} {\bf 27}, 2316--2335. \vspace{-3mm}
\item[{\rm Pitowsky, I. (1989):}]
           {\it Quantum Probability---Quantum Logic}
           (Lecture Notes in Physics, Vol.\ 321),
           Springer-Verlag, Berlin. \vspace{-3mm}
\item[{\rm Pool, J. C. T. (1966):}]
           Mathematical Aspects of the Weyl-Correspondence,
           {\it J. Math.\ Phys.}\ {\bf 7}, 66--77. \vspace{-3mm}
\item[{\rm Prugove\v{c}ki, E. (1977a):}]
           Information-Theoretical Aspects of Quantum Measurement,
           {\it Int.\ J. Theor.\ Phys.}\ {\bf 16}, 321--331. \vspace{-3mm}
\item[{\rm Prugove\v{c}ki, E. (1977b):}]
           On Fuzzy Spin Spaces,
           {\it J. Phys.\ A:\ Math.\ Gen.}\ {\bf 10},
           543--549. \vspace{-3mm}
\item[{\rm Prugove\v{c}ki, E. (1978):}]
           A Unified Treatment of Dynamics and Scattering in Classical
           and Quantum Statistical Mechanics,
           {\it Physica} {\bf 91A}, 202--228. \vspace{-3mm}
\item[{\rm Prugove\v{c}ki, E. (1984):}]
           {\it Stochastic Quantum Mechanics and Quantum Spacetime},
           Reidel, Dordrecht. \vspace{-3mm}
\item[{\rm Schroeck, F. E., Jr.\ (1981):}]
           A Model of a Quantum Mechanical Treatment of Measurement
           with a Physical Interpretation,
           {\it J. Math.\ Phys.}\ {\bf 22}, 2562--2572. \vspace{-3mm}
\item[{\rm Schroeck, F. E., Jr.\ (1982a):}]
           On the Stochastic Measurement of Incompatible Spin Components,
           {\it Found.\ Phys.}\ {\bf 12}, 479--497. \vspace{-3mm}
\item[{\rm Schroeck, F. E., Jr.\ (1982b):}]
           The Transitions among Classical Mechanics, \linebreak
           Quantum Mechanics, and Stochastic Quantum Mechanics,
           {\it Found. \linebreak Phys.}\ {\bf 12}, 825--841. \vspace{-3mm}
\item[{\rm Singer, M., and W. Stulpe (1990):}]
           Some Remarks on the Determination of Quantum States by Measurements,
           {\it Found.\ Phys.\ Lett.}\ {\bf 3}, 153--166. \vspace{-3mm}
\item[{\rm Singer, M., and W. Stulpe (1992):}]
           Phase-Space Representations of General
           Statistical Physical Theories,
           {\it J. Math.\ Phys.}\ {\bf 33}, 131--142. \vspace{-3mm}
\item[{\rm Srinivas, M. D., and E. Wolf (1975):}]
           Some Nonclassical Features of Phase-Space Representations
           of Quantum Mechanics,
           {\it Phys.\ Rev.\ D} {\bf 11}, 1477--1485. \vspace{-3mm}
\item[{\rm Stulpe, W. (1986):}]
           {\it Bedingte Erwartungen und stochastische Prozesse in der
           generalisierten Wahrscheinlichkeitstheorie -- Beschreibung
           sukzessiver Messungen mit zuf\"alligem Ausgang},
           Doctoral Dissertation, Berlin. \vspace{-3mm}
\item[{\rm Stulpe, W. (1988):}]
           Conditional Expectations, Conditional Distributions, \linebreak
           and {\it A Posteriori} Ensembles in Generalized Probability Theory,
           {\it Int.\ J. Theor.\ Phys.}\ {\bf 27}, 587--611. \vspace{-3mm}
\item[{\rm Stulpe, W. (1992):}]
           On the Representation of Quantum Mechanics on Phase Space,
           {\it Int.\ J. Theor.\ Phys.}\ {\bf 31}, 1785--1795. \vspace{-3mm}
\item[{\rm Stulpe, W. (1994):}]
           Some Remarks on Classical Representations of Quantum Mechanics,
           {\it Found.\ Phys.}\ {\bf 24}, 1089--1094. \vspace{-3mm}
\item[{\rm Varadarajan, V. S. (1970):}]
           {\it Geometry of Quantum Theory}, Vol.\ II,
           Van Nostrand Reinhold, New York. \vspace{-3mm}
\item[{\rm Werner, R. (1983):}]
           Physical Uniformities on the State Space
           of Nonrelativistic Quantum Mechanics,
           {\it Found.\ Phys.}\ {\bf 13}, 859--881. \vspace{-3mm}
\item[{\rm Werner, R. (1984):}]
           Quantum Harmonic Analysis on Phase Space,
           {\it J. Math.\ Phys.}\ {\bf 25}, 1404--1411. \vspace{-3mm}
\item[{\rm Wigner, E. P. (1932):}]
           On the Quantum Correction for Thermodynamic \linebreak Equilibrium,
           {\it Phys.\ Rev.}\ {\bf 40}, 749--759. \vspace{-3mm}
\item[{\rm Wigner, E. P. (1971):}]
           Quantum Mechanical Distribution Functions Revisited,
           in {\it Perspectives in Quantum Theory}, W. Yourgrau and
           A. van der Merwe (eds.), 25--36, MIT Press, Cambridge, Mass.
\end{description}

%
\end{document}